\documentclass[english,preprint,aps,prd,showpacs,superscriptaddress,nofootinbib,tightenlines]{revtex4}
\usepackage[T1]{fontenc}
\usepackage[latin9]{inputenc}
\usepackage{float}
\usepackage{amsmath}
\usepackage{esint}
\usepackage{ulem}
\usepackage{amsfonts}
\usepackage{multirow}
\usepackage{mathrsfs}
\usepackage{subfig}
\usepackage{amsmath}
\usepackage{amssymb}
\usepackage{bm}
\usepackage{bbm}
\usepackage{slashbox}
\usepackage{xcolor}
\usepackage{babel}
\usepackage{slashed}
\usepackage{footmisc}
\usepackage{graphicx}
\usepackage{caption}
\usepackage{simplewick}
\newcommand {\bseq}{\begin{subequations}}
\newcommand {\eseq}{\end{subequations}}
\newcommand {\pvint}[2]{{\int\!\!\!\!\!\!-}_{\!\!\!\!#1}^{#2}}
\newcommand*{\rom}[1]{\uppercase\expandafter{\romannumeral #1\relax}}
\def\Xint#1{\mathchoice
{\XXint\displaystyle\textstyle{#1}}%
{\XXint\textstyle\scriptstyle{#1}}%
{\XXint\scriptstyle\scriptscriptstyle{#1}}%
{\XXint\scriptscriptstyle\scriptscriptstyle{#1}}%
\!\int}
\def\XXint#1#2#3{{\setbox0=\hbox{$#1{#2#3}{\int}$ }
\vcenter{\hbox{$#2#3$ }}\kern-.6\wd0}}

\def\dashint{\Xint-}

\newcommand{\nn}{\nonumber}

\newcommand{\beq}{\begin{equation}}
\newcommand{\eeq}{\end{equation}}
\newcommand{\bqa}{\begin{eqnarray}}
\newcommand{\eqa}{\end{eqnarray}}

\begin{document}

\title{Partonic quasidistributions in two-dimensional QCD}

\author{Yu Jia\footnote{jiay@ihep.ac.cn}}
\affiliation{Institute of High Energy Physics,  Chinese Academy of Sciences, Beijing 100049, China\vspace{0.2cm}}
\affiliation{School of Physics, University of Chinese Academy of Sciences,
Beijing 100049, China\vspace{0.2cm}}

\author{Shuangran Liang\footnote{liangsr@ihep.ac.cn}}
\affiliation{Institute of High Energy Physics, Chinese Academy of Sciences, Beijing 100049, China\vspace{0.2cm}}
\affiliation{School of Physics, University of Chinese Academy of Sciences,
Beijing 100049, China\vspace{0.2cm}}

\author{Xiaonu Xiong\footnote{x.xiong@fz-juelich.de}}
\affiliation{Institute for Advanced Simulation,
Institut f\"ur Kernphysik and J\"ulich Center for Hadron Physics,
Forschungszentrum J\"ulich, D-52425 J\"ulich, Germany}

\author{Rui Yu\footnote{yurui@ihep.ac.cn}}
\affiliation{Institute of High Energy Physics, Chinese Academy of Sciences, Beijing 100049, China\vspace{0.2cm}}
\affiliation{School of Physics, University of Chinese Academy of Sciences,
Beijing 100049, China\vspace{0.2cm}}

\date{\today}

\begin{abstract}
As a sequel of our preceding work~\cite{Jia:2017uul}, we carry out a comprehensive
comparative study between the quasi parton distribution functions (PDFs),
distribution amplitudes (DAs) and their light-cone counterparts
for various flavor-neutral mesons,
in the context of the 't Hooft model, that is, the two-dimensional QCD in the large $N$ limit.
In contrast to the original derivation via diagrammatic techniques exemplified by
Dyson-Schwinger and Bethe-Salpeter equations,
here we employ the Hamiltonian operator approach to reconstruct the celebrated 't Hooft
equation in light-front quantization,
and Bars-Green equations in equal-time quantization.
The novelty of our derivation is to employ the soft momentum cutoff as the IR regulator.
As a virtue of this operator approach,  the functional form of the quasi distributions
can be transparently built out of the Bars-Green wave functions and the Bogoliubov angle with the aid of
bosonization technique.
Equipped with various bound-state wave functions numerically inferred in Ref.~\cite{Jia:2017uul}, we then
investigate how rapidly the quasi distributions approach their light-cone counterparts with the increasing meson momentum.
We observe that, light mesons' quasi distributions approach the light-cone distributions in a slower pace
than the heavy quarkonia. Curiously, lattice simulations of quasi distributions in four-dimensional QCD also discover
this feature.
Furthermore, we also compute the partonic light-cone PDF and quasi-PDF to one-loop order in perturbation theory,
again employing the momentum cutoff as the IR regulator.
We explicitly verify one of the backbones underlying the large momentum effective field theory (LaMET),
namely, both quasi-PDFs and light-cone PDFs in ${\rm QCD}_2$ indeed possess the same IR behavior at leading
order in $1/P^z$.
\end{abstract}










\maketitle

\section{Introduction}

Parton distributions functions (PDFs) and distribution amplitudes (DAs) encapsulate
the nonperturbative structures of quarks and gluons inside a hadron.
Parton distributions are the key ingredient for making predictions for any hard process in the high-energy hadron
collision experiments.
Undoubtedly, the most promising approach of calculating the parton distributions from the first principle of QCD is
lattice simulation.
Nevertheless, due to their intrinsic Minkowski nature, it is very difficult to directly deduce
the parton distributions as functions of $x$ on Euclidean lattice.
Until recently,  only the first few Mellin moments of parton distributions can be accessible to lattice study~\cite{Detmold:2001dv,Hagler:2007xi,Musch:2011er,Alexandrou:2015qia,Braun:2015lfa}.

A breakthrough occurred several years ago, exemplified by the introduction of
quasi distributions and the Large Momentum Effective field Theory (LaMET)~\cite{Ji:2013dva,Ji:2014gla}.
This novel approach principally paves the way for directly calculating the $x$ dependence of parton distributions
on Euclidean lattice.
Ref.~\cite{Xiong:2013bka} explicitly shows that, the infrared structures of the quark
quasi-PDF and light-cone PDF are identical at one-loop level,
consequently, the matching factor linking these two sets of PDFs were derived to this perturbative order.
The factorization theorem conjectured in \cite{Ji:2013dva} that links the quasi-PDF and ordinary PDF is later
proved to all orders in $\alpha_s$~\cite{Ma:2014jla}. The renormalization of quasi quark PDFs to all orders in $\alpha_s$
is addressed in Refs.~\cite{Ishikawa:2017faj,Ji:2017oey}.
The nonperturbative matching program has also been discussed~\cite{Stewart:2017tvs,Alexandrou:2017huk}.
A plethora of exploratory lattice simulations of quark quasi-PDFs, DAs have been available recently~\cite{Lin:2014zya,Chen:2016utp,Alexandrou:2016jqi,
Alexandrou:2015rja,Zhang:2017bzy,Chen:2017mzz,Lin:2017ani,Chen:2017lnm,Chen:2017gck,Chen:2018xof}.
Moreover, there also appear studies based on lattice perturbation theory for quasi-PDF~\cite{Ishikawa:2016znu,Carlson:2017gpk,Xiong:2017jtn} and some improvement of quasi-PDF are suggested in Ref.~\cite{Chen:2016fxx,Monahan:2016bvm,Monahan:2017hpu}.
We also notice that one-loop matching for the gluon quasi-PDF has also recently been explored~\cite{Wang:2017qyg}.

Solving the realistic 4-dimensional QCD is a notoriously difficult mission.
Conceivably, there is still a long way to proceed before obtaining the phenomenologically
competitive parton distributions
from the angle of lattice simulations.
In the meanwhile, it may also look attractive if we can learn something useful about
partonic quasi distributions from much more tractable model field theories. To date,
most solvable field theories live in $1+1$ dimensional spacetime.
In this paper, we will utilize the two-dimensional QCD (hereafter abbreviated ${\rm QCD}_2$) as a specific toy model,
to unravel various aspects of (quasi) partonic distributions.
Curiously, some qualitative features observed in this work,
especially when regarding the behavior of quasi distributions under boost, is not specific to
${\rm QCD}_2$ only, instead are also captured by realistic ${\rm QCD}_4$.

$1/N$ expansion has historically served a powerful nonperturbative tool of QCD,
since the theory becomes considerably simpler in the large $N$ limit~\cite{tHooft:1973alw,Witten:1979kh,Coleman:1985}.
Some essential nonperturbative features of strong interaction, are impressively captured in this limit.
In a similar vein, ${\rm QCD}_2$ in the large $N$ limit, often referred to as 't Hooft model~\cite{tHooft:1974pnl},
turns out to become an exactly solvable model.
Via diagramatically-based Dyson-Schwinger and Bethe-Salpeter methods,
't Hooft was able to resum the planar diagrams to arrive at the bound state equation in the context of
light-cone quantization and light-cone gauge.
The resulting bound state wave function can be readily interpreted as the light-cone wave function (LCWFs).
Light-cone correlations can thus be naturally constructed out of the 't Hooft wave function.
For instance, PDF and GPD in ${\rm QCD}_2$ have been studied long ago by Burkardt~\cite{Burkardt:2000uu}.

The motif of this work is to carefully investigate the nature and characteristics of
quasi-PDFs and DAs for various flavor-neutral mesons in the 't Hooft model.
To this purpose, a reformulation of ${\rm QCD}_2$ in the equal-time quantization,
looks much more appropriate.
Theoretical foundation along this line was first laid down by Bars and Green in 1978~\cite{Bars:1977ud}.
We will illustrate how to express the quasi distributions in terms of more
fundamental building blocks in 't Hooft model, a pair of bound-state wave functions first introduced in \cite{Bars:1977ud}.
We will be particularly interested in examining how the quasi distributions approach
the light-cone distributions as the meson gets boosted.

All in all, we hope this work can provide some valuable insight on unravelling some
gross features of quasi distributions. Moreover, ${\rm QCD}_2$ may also
serve as a benchmark to examine the efficiency among different approaches, {\it e.g.},
quasi distribution approach versus pseudo PDF~\cite{Radyushkin:2017cyf,Orginos:2017kos}
and lattice cross section approach~\cite{Ma:2017pxb}.

As mentioned earlier, meson spectra of ${\rm QCD}_2$ in the large $N$ limit can be
obtained in two equivalent ways.
One is through solving the 't Hooft equation~\cite{tHooft:1974pnl}, derived from light-cone quantization
flavored with light-cone gauge,
while the other is through solving the Bars-Green equations~\cite{Bars:1977ud},
which are inferred from ordinary equal-time quantization
combined with axial gauge. The solutions of 't Hooft equation correspond to meson's LCWFs,
denoted by $\varphi^n(x)$, where $x$ denotes the light-cone momentum fraction carried by the quark relative to that by the meson.
$n=0,1,\cdots$ denotes the discrete quantum number,
which resembles the principal quantum number
$n$ arising from the solution of Schr\"{o}dinger equation in one-dimensional space.
The dimensionless momentum ratio $x$ is restricted in the interval $[0,1]$.
In contrast, the solutions of Bars-Green equations are represented by a pair of bound-state
wave functions $\varphi^n_{\pm}(k,P)$, where $k = xP$ is the spatial component of momentum carried by the quark and $P$ denotes the meson momentum's spatial component. Here the dimensionless ratio $x$ is completely unbounded, $-\infty < x < \infty$.

In \cite{Jia:2017uul}, we have numerically solved the Bars-Green equations for a variety of quark mass, and
with several different meson momenta. We have explicitly verified the Poincar\'e invariance of the 't Hooft model
in the equal-time quantization, in the sense that meson spectra
do not depend on the reference frame. We have also numerically confirmed that,
in the infinite momentum frame (IMF), {\it i.e.}, $P\rightarrow\infty$,
the Bars-Green wave functions approach asymptotically
\beq
\lim_{P\rightarrow\infty} \varphi^n_+\left(xP,P\right) = \varphi^{n}(x),\qquad
\lim_{P\rightarrow\infty} \varphi^n_-\left(xP,P\right) = 0.
\eeq
Note the ``negative energy'' (backward motion in time) component of the wave functions, $\varphi_-$
fades away as $P\rightarrow\infty$, while the ``positive energy'' (forward motion in time) component of
the wave functions, $\varphi_+$ recovers the 't Hooft wave function in the IMF.

Our primary achievement in this work is to construct the light-cone (quasi) distributions out of
the 't Hooft wave function (Bars-Green) wave functions,
thus develop a concrete feel about the nature of the quasi distributions.
Based on the numerical solutions of the wave functions
reported in \cite{Jia:2017uul}, we then quantitatively compare the quasi parton distributions and
their light-cone counterparts accordingly.
Rather than Wick-rotate into Euclidean spacetime,
we stay in the Minkowski spacetime to compute the quasi distributions.

Apart from looking into nonperturbative aspects, we also study the quasi distributions in ${\rm QCD}_2$
from the angle of perturbation theory. By replacing a meson by a quark (or quark-antiquark pair), we
compute these fictitious ``mesonic''  light-cone PDF and PDF (LCDA and quasi-DA),
to one-loop order, explicitly verify both of them share the identical IR behavior at the leading order in $1/P_z$.
Hence, one of the backbones of LaMET is explicitly validated, in this novel theoretical setting.

The rest of the paper is structured as follows.
In Section~\ref{Sec:Setup:Theory}, we set up the theory of ${\rm QCD}_2$ and introduce our notations.
In Section~\ref{Sec:LC_H_method}, we recapitulate the Hamiltonian operator approach to derive the 't Hooft equation,
in the context of light-cone quantization flavored with light-cone gauge.
Though being an old subject, we feel that there still exists some notable innovation in our derivation.
In Section~\ref{Sec:H_Axl}, in the framework of equal-time quantization flavored with axial gauge,
we revisit the derivation of the Bars-Green equations by employing Hamiltonian operator approach,
as well as Bogoliubov transformation.
The novelty of our derivation is that, we adopt momentum cutoff rather than principle
value prescription as an alternative means to regularize IR singularities encountered in the intermediate stage.
In Section~\ref{Sec:PDF:Quasi:PDF}, with the aid of the bosonization procedure,
we present an analytic expressions for the light-cone and the quasi-PDFs
in terms of the bound-state wave functions.
We stress that quasi distributions also depend on an essential quantity, the Bogoliubov-Chiral angle.
In Section~\ref{Sec:LCDA:Quasi:DA}, following the same bosonization program,
we proceed to present the analytical expressions for the LCDA and quasi-DA.
In Section~\ref{Sec:numerical}, we proceed to conduct a comprehensive numerical study for
light-cone and quasi-PDFs, DAs related to a variety of meson species:
chiral $\pi$, physical pion, a fictitious ``strangeonium'', and charmonium.
For the quasi distributions, we choose several different reference frames for each meson species,
to illuminate how their profiles evolve with the increasing meson momentum.
In Section~\ref{Sec:Matching}, we conduct the one-loop perturbative calculation for both
light-cone and quasi-PDFs, as well as LCDA and quasi-DA, associated with a fictitious meson,
using both covariant and time-ordered perturbation theory.
Again with the IR singularities regularized by a soft momentum cutoff,
we explicitly examine the IR cutoff dependence of the light-cone and quasi distributions.
Finally, we summarize in Section~\ref{Sec:Summary}.
In Appendix~\ref{Sec:g0_gz},we conduct a comparative study between one
variant of the quasi-PDF (DA) and the canonical quasi-PDF (DA),
examining which version of quasi distributions evolve to their light-cone counterpart
with a faster pace under Lorentz boost.
In Appendix~\ref{Sec:distribution:identities}, we present some distribution identities
that are useful to express the perturbative LCDA and quasi-DA in terms of the so-called ``4-plus" function
in Section~\ref{Sec:Matching}.

\section{Setup of the notations}
\label{Sec:Setup:Theory}

For simplicity, throughout this work we will only consider single flavor of quark.
Consequently, we will be interested only in flavor-singlet mesons (quarkonia).
Adding more flavors does not pose any principal difficulty and we will avoid this unnecessary complication.
Bearing a local color $SU(N)$ symmetry, the lagrangian density of the ${\rm QCD}_2$ reads
\beq
\label{QCD2:lagrangian}
{\mathcal L}_{{\rm QCD}_2} = -\frac{1}{4}\left(F^a_{\mu\nu}\right)^2
+\bar{\psi}\left(i\slashed D -m\right)\psi,
\eeq
where $m$ denotes the quark mass. $\psi$ represents the quark field,
which contains two components in Dirac spinor space and $N$ components in the color space.
$F^a_{\mu\nu} \equiv \partial_\mu A^a_\nu-\partial_\nu A^a_\mu + g_s f^{abc}A^b_\mu A^c_\nu$ is
the gluon field strength tensor, with $A^a_\mu$ denoting the gluon field.
$D_\mu=\partial_\mu-ig_s A^a_\mu T^a$ denotes the color covariant derivative.
Here $T^a$ represent the generators in the fundamental representation of the color $SU(N)$ group,
which are $N\times N$ Hermitian matrices satisfying
\bseq\label{eq:T_nor}
\bqa
\mathrm{tr}(T^aT^b) &=&\frac{\delta^{ab}}{2},
\\
\sum_a T^a_{i j}T^a_{k l} &=&
\frac{1}{2}\left(\delta_{i l}\delta_{j k}-\frac{1}{N}\delta_{ij}\delta_{kl}\right).
\eqa
\eseq
where $a, b =1,2,\ldots,N^2-1$.

Throughout this work, we define the Lorentz two-vector as $x^\mu=\left(x^0,x^z\right)$, with the superscript $0$ indicating
the temporal component and $z$ indicating the spatial component~\footnote{It may appear unwieldy to label the spatial index by ``$z$'' in a $1+1$-dimensional field theory.
The reason we choose the superscript $z$ instead of $1$ or $x$ is to keep conformity with the convention adopted by majority of the literature about
quasi distributions in four dimensions.}. Moreover, we will adhere to the Weyl-Chiral representation
for the Dirac $\gamma$-matrices:
\beq
\gamma^0=\sigma_1,\quad \gamma^z=-i\sigma_2,\quad \gamma_5\equiv \gamma^0\gamma^z=\sigma_3,
\label{gamma:matrices:Weyl:Chiral}
\eeq
where $\sigma_i$ ($i=1,2,3$) are the familiar Pauli matrices. The advantage of choosing this specific representation
is to make chirality projection simpler, since $\gamma_5$
(which coincides with the Lorentz boost generator in 2 dimensions) becomes diagonal in this basis.

In this paper, we will specify the large $N$ limit of the ${\rm QCD}_2$ (the `t Hooft model) as
\beq
\label{eqn:lrgN_weak}
N\rightarrow\infty,\qquad \lambda \equiv \frac{g_s^2N}{4\pi}\text{ fixed},\qquad m\gg g_s \sim {1\over \sqrt{N}}.
\eeq
where $\lambda$ is dubbed the {\it 't Hooft coupling constant}.
The last condition in Eq.~\eqref{eqn:lrgN_weak} specifies the so-called
weak coupling phase~\cite{Zhitnitsky:1985um}.
It is necessary to state this clearly in the outset, since the chiral limit and large $N$ limit
do not generally commute. It is only in the weak coupling regime, {\it i.e.},
the $m\to 0$ limit imposed {\it after} taking the $N\to \infty$ limit,
that the massless ``Goldstone'' boson (chiral pion) can arise.

\section{Hamiltonian approach in light-front quantization in light-cone gauge}\label{Sec:LC_H_method}

The bound state equation for ${\rm QCD}_2$ in light-cone framework was originally derived by 't Hooft in 1974, based on Feynman
diagrammatic approach~\cite{tHooft:1974pnl}. In the following years, the same equation was also reproduced in the
light-cone hamiltonian formalism~\cite{Kikkawa:1980dc,Nakamura:1981zi,Rajeev:1994tr,Dhar:1994ib,Dhar:1994aw,Cavicchi:1993jh,Barbon:1994au,Itakura:1996bk}.
In this section, we rederive 't Hooft equation from the angle of light-cone quantization once again.
The novelty of our derivation is that, we adopt a soft momentum cutoff to regularize severe IR divergence encountered in the
intermediate stage, rather than the principal value prescription used in most of the preceding literature.
Of course, in the end of the day, we will recover the celebrated 't Hooft equation,
which is no longer plagued with infrared singularity.

\subsection{The light-front Hamiltonian}
\label{Sec:H_LC}

We adopt the widely-used Kogut-Soper convention~\cite{Kogut:1969xa}
that the light-cone coordinates are defined through $x^\pm=\left(x^0\pm x^z\right)/\sqrt{2}$.
Consequently, only the off-diagonal components of the metric tensor survive,
$g_{+-}=g^{+-}=1$, and $x^\pm = x_\mp$.

It is convenient to decompose the quark Dirac field $\psi$ into the right-handed (``good'') component $\psi_{R}$
and left-handed (``bad'') component  $\psi_{L}$,
by acting the chirality projectors $\psi_{R,L}= {1\pm \gamma_5 \over 2}\psi$.
Owing to the diagonal form of $\gamma_5$ in Weyl representation,
as specified in (\ref{gamma:matrices:Weyl:Chiral}),
one can explicitly decompose
\begin{align}
\psi =&  2^{-{1\over 4}} \left(\begin{array}{c}
\psi_R
\\
\psi_L
\end{array} \right),
\end{align}
where $\psi_{R,L}$ represent the single-component Grassmann variables.

To facilitate the light-front quantization, it is convenient to
reexpress the ${\rm QCD}_2$ lagrangian (\ref{QCD2:lagrangian}) in terms of light-cone coordinates.
Furthermore, the theory gets significantly simplified once imposing the light-cone gauge $A^{+a}= 0$:
\beq
{\mathcal L}_{{\rm QCD}_2}  = \frac{1}{2}\left(\partial_-A^{-a}\right)^2+i\left(\psi_R^\dagger D_+ \psi_R +\psi_L^\dagger\partial_-\psi_L\right)-\frac{m}{\sqrt{2}}\left(\psi_L^\dagger\psi_R+\psi_R^\dagger\psi_L\right).
\label{eq:LF_Lgrgn}
\eeq
As an exhilarating virtue of non-covariant gauge, the characteristic complication of QCD,
the triple and quartic gluon self-interactions are absent in ${\rm QCD}_2$.

Regarding $x^+$ as the light-front time, one observes that only the
right-handed quark field constitutes the dynamical degree of freedom.
From \eqref{eq:LF_Lgrgn}, one then derives the Euler-Lagrange equation for the left-handed fermion
field and the gluon field:
\bseq\label{eq:lc_ctr}
\bqa
&& \partial_-^2 A^{-a}(x) -g_s\psi_R^\dagger(x) T^a\psi_R(x) = 0,
\\
&& i\partial_-\psi_L(x) - \frac{m}{\sqrt{2}}\psi_R(x) = 0.
\eqa
\eseq
Hence $\psi_L$ and $A^{-a}$ are non-propagating (constrained) degrees of freedom,
rather than the canonical variables.
Solutions to the equations of motion (\ref{eq:lc_ctr}) are
\bseq
\label{eq:thesolution}
\bqa
&& \psi_L(x^+,x^-) =  \frac{m}{\sqrt{2}i}\int dy^-\, G_\rho^{(1)}(x^--y^-)\psi_R(x^+,y^-),
\\
&& A^{-a}(x^+,x^-) = g_s \int dy^- \,G_\rho^{(2)}(x^--y^-)\,\psi_R^\dagger(x^+,y^-)\,T^a\,\psi_R(x^+,y^-),
\eqa
\eseq
where $G^{(1)}$ and $G^{(2)}$ correspond to the Green functions associated with the differential
operators $\partial_-$ and $\partial_{-}^2$, respectively:
\bseq
\bqa
&& G_\rho^{(1)}(x^--y^-)= i\int_{-\infty}^{+\infty}\frac{dk^+}{2\pi}\Theta\left(|k^+|\!-\!\rho\right)
\frac{e^{-ik^+(x^-\! -\! y^-)}}{k^+},
\\
&& G_\rho^{(2)}(x^--y^-)= -\int_{-\infty}^{+\infty}\frac{dk^+}{2\pi}\Theta\left(|k^+|\!-\!\rho\right)\frac{e^{-ik^+(x^-\! -\! y^-)}}{\left(k^+\right)^2}.
\eqa
\eseq
where $\Theta$ signifies the Heaviside step function,
and the sharp momentum cutoff $\rho\rightarrow 0^+$ is introduced as an IR regulator.
We put a subscript $\rho$ in the coordinate-space Green function
to stress its implicit dependence on the IR cutoff upon Fourier transform.
We note that this peculiar regularization scheme has already been used by
't Hooft~\cite{tHooft:1974pnl} and Callan, Coote and Gross~\cite{Callan:1975ps}.
Einhorn even interpreted this IR regulator as a gauge parameter~\cite{Einhorn:1976uz}.
This IR regulator $\rho$ may linger around the intermediate steps, but must drop away in the physical
observables such as meson-meson
scattering amplitude~\cite{Callan:1975ps}.

Identifying the light-front Hamiltonian from the Lagrangian \eqref{eq:LF_Lgrgn} through the stress tensor $T^{+-}$,
eliminating the occurrences of $\psi_L$ and $A^{a\mu}$ in line with \eqref{eq:thesolution},
we finally end with the desired form of the light-front Hamiltonian~\footnote{Note that
our light-front Hamiltonian differs from the light-cone Hamiltonian defined in some influential paper~\cite{Brodsky:1997de},
which are connected via $H_{\rm LC} \equiv 2 P^+ H_{\rm LF}$.
Our $H_{\rm LF}$ is frame-dependent, while $H_{\rm LC}$ in \cite{Brodsky:1997de} is not, yet carrying mass dimension two.}:
\bqa
H_{\rm LF} &\equiv  P^- = &\int_{x^+=\mathrm{const.}}\!\!\!\!\!\!\!\!\!dx^-\left\{\frac{m^2}{2i}\psi_R^\dagger(x^-)\int dy^-G_\rho^{(1)}(x^--y^-)\psi_R(y^-)\right.
\nn \\
&& { \left.  - {g_s^2 \over 2}\sum_a\psi_R^\dagger(x^-) T^a\psi_R(x^-)  \int dy^- \,G_\rho^{(2)}(x^- - y^-)
\psi_R^\dagger(y^-) T^a\psi_R(y^-)  \right\}  }.
\label{LF:Hamilton:final:form}
\eqa
Note only the canonical variable $\psi_R$ survives in the light-front Hamiltonian
(\ref{LF:Hamilton:final:form})~\footnote{When concentrating on the color-singlet sectors
of Fock space, one is allowed to drop the boundary term in the light-cone Hamiltonian,
once the spatial size of the system extends to infinity.
One can refer to Hornbostel's thesis for a comprehensive and lucid discussion on this issue~\cite{Hornbostel:1988ne}.}.

The canonical quantization rules in equal light-front time are then
\bseq
\bqa
&& \left \{\psi^i_R(x^+, x^-),{\psi^j_R}^\dagger(y^+, y^-) \right\}_{x^+=y^+} =\delta^{ij}\delta (x^--y^- ),
\\
&& \left  \{\psi^i_R(x^+, x^-),\psi^j_R(y^+, y^-) \right \}_{x^+=y^+} = \left\{ {\psi^i_R}^\dagger(x^+, x^-),
{\psi^j_R}^\dagger(y^+, y^-)\right\}_{x^+=y^+}=0.
\eqa
\eseq
For clarity, we attach the color indices $i,j=1,\ldots, N$ to the $\psi_R$ field explicitly.

\subsection{Bosonization}

To quantize \eqref{LF:Hamilton:final:form}, one may expand the $\psi_R$ field in terms of the annihilation and creation operators:
\beq
\psi_R^i(x^-)=\int^\infty_0\frac{dk^+}{2\pi}\Big(b^i(k^+)e^{-ik^+x^-}+{d^i}^\dagger(k^+)e^{ik^+x^-}\Big),
\label{PsiL:Fourier:expansion}
\eeq
where $i$ is the color index. The Fock vacuum $\vert 0 \rangle$ is defined to satisfy
\begin{align}
b^i({k^+})\left|0\right\rangle = d^i({k^+})\left|0\right\rangle =0
\end{align}
for any nonnegative $k^+$.

Following the bosonization procedure~\cite{Kikkawa:1980dc,Nakamura:1981zi,Rajeev:1994tr,Dhar:1994ib,Dhar:1994aw,Cavicchi:1993jh,Barbon:1994au,Itakura:1996bk}, we
define the following four color-singlet compound operators:
\bqa
&& \displaystyle{ M\left({k^+,p^+}\right)\equiv\frac{1}{\sqrt{N}}\sum_i d^{i}(k^+) b^{i}(p^+)},\qquad
M^{ \dagger}\left({k^+,p^+}\right)\equiv\frac{1}{\sqrt{N}}\sum_{i} b^{i\dagger}(p^+)d^{i\dagger}(k^+),
\nn \\
&& \displaystyle{ B\left({k^+,p^+}\right)\equiv\frac{1}{\sqrt{N}}\sum_i b^{i\dagger}(k^+)b^i(p^+)},\qquad
D\left({k^+,p^+}\right)\equiv\frac{1}{\sqrt{N}}\sum_i d^{i\dagger}(k^+)d^i(p^+).
\label{Four:Bosonic:operators}
\eqa

The commutation relations among $B$, $D$, $M$ and $M^\dagger$ are
\bseq
\bqa
&& \left[M\left(k_1^+,p_1^+\right),M^\dagger\left(k_2^+,p_2^+\right)\right]=
\left(2\pi\right)^2\delta(k_1^+\!-\!k_2^+)\delta(p_1^+\!-\!p_2^+)+\,\mathcal{O}\left(\frac{1}{\sqrt{N}}\right),
\\
&& \left[M\left(k_1^+,p_1^+\right),B\left(k_2^+,p_2^+\right)\right]=
\left[M^\dagger\!\left(k_1^+,p_1^+\right),B\left(k_2^+,p_2^+\right)\right]=\mathcal{O}\left(\frac{1}{\sqrt{N}}\right),
\\
&&\left[M\left(k_1^+,p_1^+\right),D\left(k_2^+,p_2^+\right)\right]=\left[M^\dagger\!\left(k_1^+,p_1^+\right),D\left(k_2^+,p_2^+\right)\right]=
\mathcal{O}\left(\frac{1}{\sqrt{N}}\right),
\\
&& \left[B\left(k_1^+,p_1^+\right),B\left(k_2^+,p_2^+\right)\right]=\left[D\,\left(k_1^+,\,p_1^+\right),
D\left(k_2^+,\,p_2^+\right)\right]=\mathcal{O}\left(\frac{1}{\sqrt{N}}\right),
\\
&& \left[B\left(k_1^+,p_1^+\right),D\left(k_2^+,p_2^+\right)\right]=\,0.
\eqa
\label{eq:MBD_cmmtr}
\eseq

Substituting the Fourier expansion of $\psi_R$, (\ref{PsiL:Fourier:expansion}), into the LF Hamiltonian (\ref{LF:Hamilton:final:form}), then expressing everything in terms of the bosonic compound
operators introduced in (\ref{Four:Bosonic:operators}), dropping terms that are suppressed by powers of $1/N$
(with the aid of the $SU(N)$ identity \eqref{eq:T_nor}), we can decompose
the light-front Hamiltonian into three pieces:
\begin{align}
H_{\rm LF}= H_{{\rm LF};0}+:H_{{\rm LF};2}:+:H_{{\rm LF};4}:,
\end{align}
where $:\;:$ denotes the standard normal ordering.
Organized by the frequency of occurrences of the
bosonic compound operators, these three parts turn out to be
\bseq
\bqa
&& H_{{\rm LF};0} = N \int \frac{dx^-}{2\pi}  \left[{\lambda\over 2}+
{\lambda-m^2 \over 2} \int^\infty_\rho \!{dk^+\over k^+}\right],
\\
&& :\!H_{{\rm LF};2}\!: \;\; =\sqrt{N}\int^\infty_\rho \frac{dk^+}{2\pi}\left[{m^2-2\lambda\over 2}{1\over k^+}+ {\lambda\over \rho}
\right]\left[B\left({k^+,k^+}\right)+D\left({k^+,k^+}\right)\right],
\\
&& :\!H_{{\rm LF};4}\!: \;\;=
\frac{\lambda}{8\pi^2}\int_0^\infty dk^+_{1}\int_0^\infty dk^+_{2}\int_0^\infty dk^+_{3}\int_0^\infty dk^+_{4}
\\
& & \bigg\{\Theta(|k^+_1\!-\!k^+_2|\!-\!\rho)\frac{M^{ \dagger}({k^+_{2},k^+_{3}})D({k^+_{4},k^+_{1}})-B({k^+_{1},k^+_{4}})M({k^+_{3},k^+_{2}})}{(k^+_{1}-
k^+_{2})^2}\delta\left(k^+_{1}\!-\!k^+_{2}-k^+_{3}\!-\!k^+_{4}\right)
\nn\\
&&+\Theta(|k^+_1\!-\!k^+_2|\!-\!\rho)\frac{D({k^+_{2},k^+_{3}})M({k^+_{1},k^+_{4}})-M^\dagger({k^+_{4},k^+_{1}})B({k^+_{3},k^+_{2}})}
{(k^+_{1}-k^+_{2})^2}\delta\left(k^+_{1}\!-\!k^+_{2}\!+\!k^+_{3}\!+\!k^+_{4}\right)
\nn\\
 &&-\Theta(|k^+_1\!-\!k^+_2|\!-\!\rho)\frac{M^{\dagger}({k^+_{1},k^+_{4}})M({k^+_{2},k^+_{3}})+
 M^{\dagger}({k^+_{2},k^+_{3}})M({k^+_{1},k^+_{4}})}{(k^+_{1}-k^+_{2})^2}\delta\left(k^+_{1}\!-\!k^+_{2}\!-\!k^+_{3}\!+\!k^+_{4}
 \right)
\nn \\
&& +\Theta(|k^+_1\!+\!k^+_2|\!-\!\rho)\frac{D({k^+_{4},k^+_{1}})B({k^+_{3},k^+_{2}})+B({k^+_{1},k^+_{4}})D({k^+_{2},k^+_{3}})}
{(k^+_{1}+k^+_{2})^2}\delta\left(k^+_{1}\!+\!k^+_{2}\!-\!k^+_{3}\!-\!k^+_{4}\right)
\nn\\
&&  +\Theta(|k^+_1\!+\!k^+_2|\!-\!\rho)\frac{M^\dagger({k^+_{4},k^+_{1}})D({k^+_{2},k^+_{3}})-B({k^+_{3},k^+_{2}})M({k^+_{1},k^+_{4}})}
{(k^+_{1}+k^+_{2})^2}\delta\left(k^+_{1}\!+\!k^+_{2}\!-\!k^+_{3}\!+\!k^+_{4}\right)
\nn\\
&&+\Theta(|k^+_1\!+\!k^+_2|\!-\!\rho)\frac{D({k^+_{4},k^+_{1}})M({k^+_{3},k^+_{2}})-M^\dagger({k^+_{2},k^+_{3}})
B({k^+_{1},k^+_{4}})}{(k^+_{1}+k^+_{2})^2}\delta\left(k^+_{1}\!+\!k^+_{2}\!+\!k^+_{3}\!-\!k^+_{4}\right)
\nn\\
&&-\Theta(|k^+_1\!-\!k^+_2|\!-\!\rho)\frac{D({k^+_{2},k^+_{3}})D({k^+_{4},k^+_{1}})+B({k^+_{3},k^+_{2}})B({k^+_{1},
k^+_{4}})}{(k^+_{1}-k^+_{2})^2}\delta\left(k^+_{1}\!-\!k^+_{2}\!+\!k^+_{3}\!-\!k^+_{4}\right)\bigg\},
\nn
\eqa
\label{eq:LF_H}
\eseq
where $\lambda={g_s^2 N}/{4\pi}$ is the dimensional 't Hooft coupling constant, and
those terms suppressed by $1/N$ have been suppressed.
$H_{{\rm LF};0}$ can be interpreted as the vacuum light-cone energy, which is both UV and IR divergent~\cite{Lenz:1991sa}.
This constant is irrelevant for our purpose, so will be neglected henceforth.
Note the soft momentum cutoff $\rho$
has been introduced in (\ref{eq:LF_H}) to regularize the IR divergence.

A key observation is that the ${\rm QCD}_2$ is a confining theory, and one cannot create or annihilate
isolated quarks and antiquarks. Therefore, to create a quark, one has to create an accompanying antiquark;
{\it vice versa}, to annihilate a quark, one has to annihilate an accompanying antiquark.
Only the color-singlet $q\bar{q}$ pair can be created or annihilated.
The consequence is that the operators in (\ref{Four:Bosonic:operators}) cannot be all independent.
Rather one finds that the compound operators $B$ and $D$ can be built out of $M$ and $M^\dagger$~\cite{Itakura:1996bk}:
\bseq
\begin{align}
&B(k^+,p^+)\rightarrow\frac{1}{\sqrt{N}}\int_0^\infty\frac{dq^+}{2\pi}M^\dagger(q^+,k^+)M(q^+,p^+),
\\
&D(k^+,p^+)\rightarrow\frac{1}{\sqrt{N}}\int_0^\infty\frac{dq^+}{2\pi}M^\dagger(k^+,q^+)M(p^+,q^+).
\end{align}
\label{Replace:B:D}
\eseq
As can be readily verified, these relations are compatible with the commutation relation (\ref{eq:MBD_cmmtr}).

Substituting (\ref{Replace:B:D}) into \eqref{eq:LF_H}, relabelling the momenta $p^+=x P^+$ and $k^+=(1-x)P^+$, keeping only the
leading order terms in $1/N$, one finds that the $:H_{{\rm LF};2}:$ and $:H_{{\rm LF};4}:$ components now read
\bseq
\bqa
&& :H_{{\rm LF};2}\!: \;=\frac{1}{(2\pi)^2}\int_\rho^\infty dP^+\int_0^1 dxM^\dagger((1-x)P^+,xP^+)M((1-x)P^+,xP^+)
\\
&& \times\left\{\!\left[\left(\frac{m^2}{2}\!-\!\lambda\right)\frac{1}{x}\!+\!\frac{P^+\lambda}{\rho}\right]
\Theta\left(x\!-\!\frac{\rho}{P^+}\right)\!+\!\left[\left(\frac{m^2}{2}\!-\!\lambda\right)\frac{1}{1-x}\!+
\!\frac{P^+\lambda}{\rho}\right]\Theta\left(1\!-\!\frac{\rho}{P^+}\!-\!x\right)\!\right\},
\nn\\
&& :H_{{\rm LF};4}\!: \;=-\!\frac{\lambda}{(2\pi)^2}\!\int_\rho^\infty\! \!\!dP^+{\int_0^1\!\!\int_0^1}dxdy\Theta\left(|x\!-\!y|\!-\!\frac{\rho}{P^+}\right)
\frac{1}{(x\!-\!y)^2}
\nn\\
&& \times M^\dagger((1\!-\!x)P^+,xP^+)M((1\!-\!y)P^+,yP^+).
\eqa
\label{LC:Hamilton:M:Mdagger}
\eseq

\subsection{Diagonalization, principal value prescription, and the 't Hooft equation}

Our goal is to diagonalize the light-front hamiltonian (\ref{LC:Hamilton:M:Mdagger}).
To this purpose, it is convenient to introduce an infinite set of meson annihilnation/creation operators: $m_n(P^+)$/$m^\dagger_n(P^+)$,
where $n$ stands for the principal quantum number, and $P^+$ represents the light-cone momentum of the corresponding meson.
We postulate that the $M$ operator basis is connected to the $m_n$ basis through
\bseq
\label{m:connected:to:M}
\begin{align}
&M ((1-x)P^+,xP^+)=\sqrt{\frac{2\pi}{P^+}} \sum_{n=0}^\infty \varphi_n(x) m_n(P^+)\label{m:connected:to:M_1},
\\
& m_n(P^+)=\sqrt{\frac{P^+}{2\pi}}\int_0^1 dx\,\varphi_n(x) M\left((1-x)P^+,xP^+\right),
\end{align}
\eseq
where $\varphi_n(x)$ is understood to be the $n$-th coefficient function.
The physical picture is clear, since confinement nature of 't Hooft model,
exciting a quark-antiquark pair from the Fock vacuum would eventually lead to the
formation of a meson, in the large $N$ limit.
To the best of our knowledge, the explicit writing of the decomposition formula
\eqref{m:connected:to:M} is new.

We postulate that the mesonic annihilation and creation operators $m_n$ and $m_n^\dagger$ obey
the canonical commutation relations:
\beq\label{LC:m:mdagger:commutation}
\left[m_n(P^+_1), m_r^\dagger(P^+_2)\right]= 2\pi \delta_{nr} \delta(P^+_1-P^+_2),
\eeq
and all other commutators vanish.
It is straightforward to check that, in order to satisfy these commutation
relations, the coefficient functions $\varphi_n(x)$ must be subject to the following
orthogonality and completeness conditions:
\bseq
\begin{align}
\int_0^1\!\! dx\, \varphi_n(x)\varphi_m(x)&=\delta_{nm},
\\
\sum_n\varphi_n(x)\varphi_n(y)&=\delta(x-y),
\end{align}
\eseq

We wish that the light-front Hamiltonian in the basis of $m_n$ and $m^\dagger_n$ operators is in
a diagonal form,
\beq
\label{eq:dicon}
H_{\mathrm{LF}}=  H_{{\rm LF};0} +\int \frac{dP^+}{2\pi} P_n^- m^\dagger_n(P^+)m_n(P^+),
\eeq
where $P^-$ is the light-cone energy of the $n$-th mesonic state, $P^-_n = M_n^2/(2P^+)$.

If the light-front Hamiltonian can be diagonalized in the new $m_n$ operator basis,
the $n$-th mesonic state in the large $N$ limit
can be directly constructed via
\beq
\label{eq:msn_st}
\left| P_n^-, P^+ \right\rangle = \sqrt{2P^+} m_n^\dagger(P^+)\left| 0 \right\rangle.
\eeq

In order to reach the desired form (\ref{eq:dicon}), one should enforce the condition that
the coefficients of all the unwanted operators of the form $m_n^\dagger m_r$ ($n\neq r$)
vanish. This requirement leads to the following equations which must
be satisfied by $\varphi_n(x)$ in different intervals of $x$:
\begin{subequations}
\begin{align}
&\frac{m^2}{1-x}\varphi_n(x)-2\lambda\int_{x+\frac{\rho}{P^+}}^1dy\frac{\varphi_n(y)-\varphi_n(x)}{(x-y)^2}=M_n^2\varphi_n(x)& 0<x<\frac{\rho}{P^+},\\
&\left(\!\frac{m^2}{x}\!+\!\frac{m^2}{1\!-\!x}\!\right)\varphi_n(x)\!-\!2\lambda\int_{0}^{1} dy\, \Theta\left(|x\!-\!y|\!-\!\rho\right)\frac{\varphi_n(y)\!-\!\varphi_n(x)}{(x\!-\!y)^2}=M_n^2\varphi_n(x)& \frac{\rho}{P^+}<x<1-\frac{\rho}{P^+},\\
&\frac{m^2}{x}\varphi_n(x)-2\lambda\int^{x-\frac{\rho}{P^+}}_0dy\frac{\varphi_n(y)-\varphi_n(x)}{(x-y)^2}=M_n^2\varphi_n(x)& 1-\frac{\rho}{P^+}<x<1,
\end{align}
\end{subequations}
In the $\rho\rightarrow 0^+$ limit, these equations merge into a single equation:
\beq\label{eq:tHooft_eq}
\left({m^2\over x}+ {m^2\over 1-x}\right) \varphi_n(x)-
2\lambda\pvint{0}{1} dy {\varphi_n(y)-\varphi_n(x)\over (x-y)^2} =  M_n^2 \varphi_n(x),\qquad 0<x<1,
\eeq
where the dashed integral $\dashint$ in (\ref{eq:tHooft_eq}) denotes the principal value (PV) prescription:
\beq
\pvint{}{}dy\, \frac{f(y)}{(x-y)^2} = \lim_{\epsilon\rightarrow0^+}\int_{}^{} dy  \Theta(|x-y|-\epsilon)
{f(y)-f(x)\over(x-y)^2}.
\label{PV:pres:defined:this:work}
\eeq
with $f(y)$  a test function that is regular at $y=x$.

Eq.~\eqref{eq:tHooft_eq} is nothing but the celebrated 't Hooft equation.
Now the coefficient functions $\varphi_n(x)$, first introduced in (\ref{m:connected:to:M}),
can be interpreted as the 't Hooft wave function, or the light-cone wave function
of the $n$-th mesonic state.

We emphasize that the PV prescription as specified in \eqref{eq:tHooft_eq}
needs not be unique. Here we just list two additional popular PV prescriptions:
\bseq
\label{TWO:popular:PV:schemes}
\bqa
 \pvint{}{}dy\, \frac{f(y)}{(x-y)^2} = && \lim_{\epsilon\rightarrow 0} \int dy\,
\frac{f(y)}{2}\left[\frac{1}{(x-y+i\epsilon)^2}+\frac{1}{(x-y-i\epsilon)^2}\right],
\\
 \pvint{}{}dy\, \frac{f(y)}{(x-y)^2} =&&  \lim_{\epsilon \rightarrow 0^+}\int
dy \, \Theta(|x-y|-\epsilon)
{f(y) \over (x-y)^2} - {2 f(x) \over \epsilon},
\eqa
\eseq
where the first one was adopted in \cite{Einhorn:1976uz} (also referred to as Mandelstam-Leibbrandt prescription~\cite{Mandelstam:1982cb,Leibbrandt:1987qv}),
and the second one was introduced by Hadamard long ago~\cite{Hadamard:1923}.
All the aforementioned PV prescriptions are mathematically equivalent,
but may practically differ in efficiency upon numerical implementation.

\section{Hamiltonian approach in equal-time  quantization in axial gauge}
\label{Sec:H_Axl}

The bound-state equations in ${\rm QCD}_2$ in equal-time quantization and in axial gauge
were originally derived by Bars and Green in 1978~\cite{Bars:1977ud},
largely utilizing Feynman diagrammatic techniques.
In 2001 Kalashnikova and Nefediev presented an elegant derivation based on
the Hamiltonian operator approach~\cite{Kalashnikova:2001df}.
A nice feature of this method is that, through introducing
Bogoliubov transformation, the physical meaning of Bars-Green wave functions get
greatly clarified.

It appears rather obscure to link the Bars-Green bound-state wave functions with the quasi distributions
based on the diagrammatic methods. On the contrary, it is quite transparent to achieve this goal
with the aid of bosonaization technique. Therefore, it is rewarding
to recapitulate the derivation of the Bars-Green equations in this section,
again within the Hamiltonian approach.

Ref.~\cite{Kalashnikova:2001df} employs the PV prescription to sweep away the potential IR divergences.
Nevertheless, to be compatible with our treatment in light-front quantization
in Sec.~\ref{Sec:LC_H_method} as well as in the perturbative one-loop computation
for quasi distributions in Sec.~\ref{Sec:Matching},
here we adopt the same momentum cutoff as the IR regulator.
We will explicitly verify that, though differing in intermediate steps,
after the IR momentum cutoff is removed in the end,
the famous mass gap equation and Bars-Green equations will be recovered.

\subsection{The Hamiltonian in the axial gauge}

Enforcing the axial gauge condition $A^z=0$, the ${\rm QCD}_2$ lagrangian in (\ref{QCD2:lagrangian}) reduces to
\beq
{\mathcal L}_{{\rm QCD}_2} = \frac{1}{2}(\partial_zA^{a}_0)^2+i\psi^\dagger
(D_0 + \gamma^5\partial_z)\psi-m\bar{\psi}\psi.
\eeq

Unlike the light-cone case, both components of the quark Dirac field $\psi$ remain as the
propagating degrees of freedom. The equation of motion for $A^{0a}$'s turns to be
\beq
\partial^2_z A^{0a} =  g_s\psi^\dagger T^a\psi,
\label{EL:eq:for:A0a}
\eeq
thus $A^{0a}$ is a constrained rather than dynamical variable.

The solution of Eq.~(\ref{EL:eq:for:A0a}) is
\beq
\label{eq:A0_axial}
A^{0a}(t,z)=g_s\int dz^\prime {\widetilde G}_\rho^{(2)}(z-z^\prime)\psi^\dagger(t,z^\prime)T^a\psi(t,z^\prime),
\eeq
where $\widetilde{G}^{(2)}_\rho$ denotes the Green function associated with the operator $\partial_z^2$~\footnote{For
notational brevity, in this section we have suppressed the superscript ``$z$'' for the spatial component of a 2-vector,
so $k$ should be understood as $k^z$ if no confusion arises.}:
\beq
\widetilde{G}^{(2)}_\rho (z-z') =
-\int_{-\infty}^{+\infty}\frac{dk}{2\pi}\Theta\left(|k|\!-\!\rho\right)
\frac{e^{ik(z\! -\! z')}}{k^2}.
\eeq
This Fourier integral is ill-defined due to the singularity caused by $k\to 0$.
For consistency with the rest of the paper, here we again employ a
momentum cutoff $\rho\rightarrow 0^+$ to regularize the IR divergence.

The equal-time Hamiltonian in the axial gauge, when expressed in terms of the canonical variables,
is~\footnote{Bars and Green expounded why one is allowed to drop the boundary terms in
the color-singlet sector, in the context of
equal-time quantization and axial gauge~\cite{Bars:1977ud}. \label{foot:bndry} }
\bqa
H && \equiv P^0 = \int_{t=\mathrm{const}} \!\!\!\!dz \bigg\{ \psi^\dagger(z)\,\left(-i \gamma^5 \partial_z +m\gamma^0\right)\,\psi(z)
\nn \\
&&   - {g_s^2\over 2} \sum_a\int \!\! dz'\,\psi^\dagger(z)T^a\psi(z)\,\widetilde{G}^{(2)}_\rho(z-z')\,\psi^\dagger(z')T^a\psi(z')\bigg\},
\label{eq:H_orgnl}
\eqa

The equal-time canonical quantization rule is then
\bseq
\bqa
&& \left \{\psi^i(t,z),{\psi^j}^\dagger(t',z')\right \}_{t=t'} = \delta_{ij}\delta (z-z'),
\\
&& \left \{\psi^i(t,z),\psi^j(t',z')\right \}_{t=t'} =\left\{{\psi^i}^\dagger(z),{\psi^j}^\dagger(z')\right\}_{t=t'}=0,
\eqa
\eseq
where $i,j=1,\ldots, N$ denote the color indices carried by $\psi$, and the spinor indices have been suppressed for simplicity.

\subsection{Dressed quark basis and mass-gap equation}
\label{mass:gap}

To proceed, we expand the Dirac $\psi$ field in terms of the quark annihilation and creation operators:
\beq
\label{Psi:Fourier:expansion}
\psi^i(z)= \int \frac{dp}{2\pi}\frac{1}{\sqrt{2\widetilde{E}(p)}}\,\left[b^i(p)\,u(p)+d^{i \dagger}(-p)\,v(-p)\right]e^{ipz},
\eeq
where the factor $\sqrt{2\widetilde{E}(p)}$ is deliberately inserted in the integration measure,
to keep our normalization convention compatible with the standard text~\cite{Peskin:1995ev}. Here
$\widetilde{E}(p)$ can be interpreted as the energy carried by the dressed quark.
The spinor wave functions $u$, $v$ are parameterized as~\footnote{Note here the parametrization of the
dressed quark spinor wave functions differs from the preceding literature~\cite{Bars:1977ud,Kalashnikova:2001df},
where the Dirac-Pauli representation for $\gamma$-matrices were adopted.}
\begin{align}
u(p)=\sqrt{\widetilde{E}(p)}\,T(p)
\begin{pmatrix}
1\\
1
\end{pmatrix}
,\quad
v(-p)=\sqrt{\widetilde{E}(p)}\,T(p)
\begin{pmatrix}
1\\
-1
\end{pmatrix},
\end{align}
where $T(p)$ is a unitary $2\times 2$ matrix. In conformity with the convention adopted in
\cite{Peskin:1995ev}, the $u$, $v$ spinor wave functions carry the mass dimension of ${1\over 2}$.
Combining these two equations, one sees that the field expansion in \eqref{Psi:Fourier:expansion}
actually does not rely on the explicit form of $\widetilde{E}(p)$ at all,
but critically depends on the dressing function $T(p)$.

The quark vacuum state is defined to be
\beq
b^i({k})\left|0\right\rangle = d^i({k})\left|0\right\rangle =0.
\label{quark:vacuum}
\eeq
for all possible values of $k$.

Substituting the Fourier expansion of $\psi$ \eqref{Psi:Fourier:expansion},
into the Hamiltonian in \eqref{eq:H_orgnl}, and rearranging it into the normal-ordered form,
we can decompose the Hamiltonian into three pieces:
\beq
H = H_0+:H_2: +:H_4:
\label{Hamilton:three:parts:axial}
\eeq
which contain 0, 2 and 4 quark creation/annihilation operators accordingly,
\bseq
\label{Halmiton:normal:ordering:three:parts}
\bqa
\notag H_0 &=& N\int\! dz\! \int\!\frac{dp}{2\pi}\! \mathrm{Tr}\left[ \left(p\gamma^5\!+\!m\gamma^0\right)\Lambda_-(p)\!+\!\frac{\lambda}{2}\int\frac{dk}{\left(k\!-\!p\right)^2}
\Theta\left(|k\!-\!p|\!-\!\rho\right)\Lambda_+(p)\Lambda_-(k)\right],\\
\label{eq:H0:vacuum:energy}
\\
:\!H_2\!: &=&\int dp\ \mathrm{Tr}\left[\,\Xi(p)\Lambda_+(p){b^i}^\dagger(p)b^i(p)+\Xi(p)\Omega_-(p){b^i}^\dagger(p){d^i}^\dagger(-p)\right.
\nn\\
\label{eq:H2:dressed:quark}
 && \left. +\Xi(p)\Omega_+(p)d^i(-p)b^i(p)-\Xi(p)\Lambda_-(p){d^i}^\dagger(-p)d^i(-p)\right],\\
:\!H_4\!: &=&-{g_s^2 \over 2} \sum_{a}\iint_{t=\mathrm{const}}\!\!\!\!  dzdz'\,
:\psi^\dagger(z)T^a\psi(z)\,\widetilde{G}^{(2)}_\rho(z-z')\,\psi^\dagger(z')T^a\psi(z'):.\label{eq:H_4}
\eqa
\eseq
where the matrices $\Xi$, $\Lambda_\pm$ and $\Omega_\pm$ are defined as
\bseq
\bqa
&& \Xi(p)=p\gamma^5+m\gamma^0+\sum_i\frac{\lambda}{2}\int\frac{dk}{2\pi(p-k)^2}\Theta(|p-k|-\rho)\left(\Lambda_+(k)-\Lambda_-(k)\right),
\\
&& \Lambda_\pm(k)= T(k)\frac{1\pm \gamma^0}{2} T^\dagger(k),\\
&& \Omega_\pm(k)= T(k) {\gamma^0 \pm 1 \over 2} \gamma^z T^\dagger(k).
\eqa
\eseq
$H_0$ describes the vacuum energy. Let us first focus on the single dressed quark sector represented by $:H_2:$.
We are seeking a possible solution of $T(p)$ such that $:H_2:$ has a diagonalized form
in the basis of quark annihilation and creation operators:
\beq
\label{H2:diagonalized:form}
:H_2: =\sum_i \int_{-\infty}^\infty {dk\over 2\pi} \widetilde{E}_i(k)
\left({b^i}^\dagger(k) b^i(k)+{d^i}^\dagger(k)d^i (k)\right),
\eeq
where $i$ is the color index, and
$\widetilde{E}_i(k)$ denotes the energy of the dressed quark with momentum $k$.
To proceed, one parameterizes the $T(p)$ as~\cite{Bars:1977ud}
\beq
\label{Tp:theta:angle}
T(p)=\exp\left[-\frac{1}{2}\theta(p)\gamma^z\right],
\eeq
where $\theta(p)$ is called the Bogliubov-chiral angle, which is an odd function of
$p$~\cite{Bars:1977ud,Kalashnikova:2001df}.
As elucidated in Ref.~\cite{Shifman:2012zz}, $T(p)$ is reminiscent of the Foldy-Wouthuysen transformation that
decouples the positive and negative energy degrees of freedom in Dirac field,
and $\theta(p)$ play the role of the Foldy-Wouthuysen angle.

By the parametrization specified in \eqref{Tp:theta:angle},
diagonalization of (\ref{eq:H2:dressed:quark}) leads to
two coupled equations for $\theta(p)$ and $\widetilde{E}(p)$, respectively:
\bseq
\label{mass:gap:eqnarray}
\bqa
 && \widetilde{E}(p) \cos\theta(p) = m + {\lambda\over 2}
 \int \!\! {dk \over (k- p)^2} \Theta\left(|k-p|-\rho\right) \cos\theta(k),
\\
&& \widetilde{E}(p) \sin\theta(p) = p + {\lambda\over 2}
 \int \!\! {dk\over (k- p)^2} \Theta\left(|k-p|-\rho\right) \sin\theta(k).
\eqa
\eseq

After some plain linear algebra on two equations in (\ref{mass:gap:eqnarray}),
we finally arrive at the nonlinear equation for $\theta(p)$:
\beq
\label{formal:mass:gap:eq}
p\cos\theta(p)-m\sin\theta(p) = \frac{\lambda}{2} \lim_{\rho\to 0^+} \int_{-\infty}^{+\infty}\!\!
{dk\over (p-k)^2} \Theta\left(|k-p|-\rho\right) \sin\left[\theta(p)-\theta(k)\right],
\eeq
which is nothing but the celebrated {\it mass-gap equation}~\cite{Bars:1977ud}.
Note the limit $\rho\to 0^+$ just serves the standard Cauchy principal value prescription.
Examining the gap equation \ref{formal:mass:gap:eq},
the interpretation of $\theta(p)$ as the Foldy-Wouthuysen angle becomes
transparent if the interaction term, which is directly responsible for dressing the bare quark,
can be temporarily turned off.
The angle $\theta(p)$  plays a vital role for generating a nonvanishing quark vacuum condensate.
Practically speaking, the Bogoliubov-chiral angle can only be solved numerically,
even in the chiral limit.

In passing, we stress that the mass-gap equation (\ref{formal:mass:gap:eq})
can be obtained from another quite different perspective.
Rather than diagonalize $:\!H_2\!:$, one can take a closer look at
the vacuum energy constant. One can rewrite (\ref{eq:H0:vacuum:energy}) as
\beq
 {\mathcal E}_{\rm vac}[\theta(p)] = N \int \!\!\frac{dp}{2\pi}\left\{
 -m\cos \theta(p)-p\sin\theta(p)+\frac{\lambda}{2}\int \!\! dk\,\Theta(|k-p|-\rho)
 \frac{1-\cos[\theta(k)-\theta(p)]}{2(k-p)^2}\right\},
\label{vac:energy:density:H0}
\eeq
where ${\cal E}_{\rm vac} = H_0/L$ is the vacuum energy density, with $L$
the length of the spatial interval.

Minimizing Eq.~\eqref{vac:energy:density:H0} with respect to
$\theta(p)$, one can readily obtain a variational equation, which exactly
reproduce Eq.~(\ref{formal:mass:gap:eq})~\cite{Kalashnikova:2001df}.
Note that the true vacuum is no longer chiral invariant, and a nonzero quark condensate arises in
the chiral limit, which signals the spontaneous chiral symmetry breaking in
large-$N$ limit of ${\rm QCD}_2$~\cite{Li:1987hx}.

Once the Bogoliubov angle $\theta(p)$ is known, one can then infer
the dispersive law for a dressed quark:
\beq
\label{dispersion:relation}
\widetilde{E}(p)= m\cos\theta(p)+p\sin\theta(p)
+{\lambda \over 2} \int_{-\infty}^{+\infty}
{dk\over (p-k)^2} \Theta\left(|k-p|-\rho\right) \cos\left[\theta(p)-\theta(k)\right].
\eeq
It is straightforward to see that, the energy carried by the dressed quark
blows up for all values of momentum, $\widetilde{E}(p) \to {\lambda\over \rho}$,
after the IR regulator is removed.
This symptom is in sharp contrast to the regular dispersive law obtained in \cite{Bars:1977ud,Li:1987hx},
where the PV scheme is used to regularize the IR divergence there.
As a consequence, the free Hamiltonian in the dressed quark sector in
(\ref{H2:diagonalized:form}) is ill-defined, due to its sensitivity to the IR cutoff.
Nevertheless, this is a harmless and tolerable nuisance, since the colored object such as
dressed quark need not be affiliated with any physical significance.

For future usage, it is convenient to define the {\it regularized} dressed quark energy,
$E(p)$:
\beq
\label{regular:dispersion:relation}
E(p) \equiv \widetilde{E}(p)-\frac{\lambda}{\rho}= m\cos\theta(p)+p\sin\theta(p)
+\frac{\lambda}{2} \dashint_{-\infty}^{+\infty} \frac{dk}{(p-k)^2} \cos\left[\theta(p)-\theta(k)\right],
\eeq
where $\dashint$  denotes the PV scheme as specified in (\ref{PV:pres:defined:this:work}).
It is straightforward to see that $E(p)$ is an even function of $p$, and remain finite for all
finite $p$. Nevertheless, being a colored object,  the dispersive relation for a dressed quark,
no matter $\widetilde{E}(p)$ or $E(p)$, clearly violates Lorentz covariance.

\subsection{Bosonization}

In order to derive the bound state equation, we must take the interaction part of the Hamiltonian,
$:H_4:$, into account.
In parallel with the bosonization procedure for the LF Hamiltonian,
here we introduce the following  color-singlet compound operators analogous to (\ref{Four:Bosonic:operators}):
\bseq
\bqa
&& M(p,q) \equiv \frac{1}{\sqrt{N}}\sum_i d^{i}_{-p} b^{i}_{q},\qquad M^\dagger(p,q) \equiv \frac{1}{\sqrt{N}}\sum_i b^{i\dagger}_{q}d^{i\dagger}_{-p},
\\
&& B(p,q) \equiv \frac{1}{\sqrt{N}}\sum_i b^{i\dagger}_{p}b^{i}_{q},\qquad D(p,q) \equiv \frac{1}{\sqrt{N}} \sum_id^{i\dagger}_{-p}d^{i}_{-q}.
\eqa
\label{equal:time:quant:Bosonization}
\eseq

The commutation relations among $M$, $M^\dagger$, $B$ and $D$ in the large $N$ limit
are
\bseq
\bqa
&& \left[M\left(k_1,p_1\right),M^\dagger\left(k_2,p_2\right)\right]=
\left(2\pi\right)^2\delta(k_1\!-\!k_2)\delta(p_1\!-\!p_2)+\,\mathcal{O}\left(\frac{1}{\sqrt{N}}\right),
\\
&& \left[M\left(k_1,p_1\right),B\left(k_2,p_2\right)\right]=
\left[M^\dagger\!\left(k_1,p_1\right),B\left(k_2,p_2\right)\right]=\mathcal{O}\left(\frac{1}{\sqrt{N}}\right),
\\
&&\left[M\left(k_1^+,p_1^+\right),D\left(k_2,p_2\right)\right]=\left[M^\dagger\!\left(k_1,p_1\right),D\left(k_2,p_2\right)\right]=
\mathcal{O}\left(\frac{1}{\sqrt{N}}\right),
\\
&& \left[B\left(k_1,p_1\right),B\left(k_2,p_2\right)\right]=\left[D\,\left(k_1,\,p_1\right),
D\left(k_2,\,p_2\right)\right]=\mathcal{O}\left(\frac{1}{\sqrt{N}}\right),
\\
&& \left[B\left(k_1,p_1\right),D\left(k_2,p_2\right)\right]=\,0,
\eqa
\label{eq:MBD:commutation:equal:time}
\eseq
which are very similar to their light-cone counterparts \eqref{eq:MBD_cmmtr}.

Due to the confinement nature of ${\rm QCD}_2$, the same consideration
that leads to (\ref{Replace:B:D}) can also be applied here, {\it i.e.},
not all compound operators in
(\ref{equal:time:quant:Bosonization}) are independent.
In fact, one finds that~\cite{Kalashnikova:2001df}
\bseq
\begin{align}
B(p,p')=& \frac{1}{\sqrt{N}}\int_{-\infty}^{+\infty} \frac{dq}{2\pi} M^\dagger(q,p)M(q,p'),
\\
D(p,p')=& \frac{1}{\sqrt{N}}\int_{-\infty}^{+\infty} \frac{dq}{2\pi} M^\dagger(p,q)M(p',q).
\end{align}
\label{B:D:not:independent}
\eseq

Here we follow similar steps as what lead to (\ref{eq:LF_H}) in light-cone quantization.
Substituting the Fourier expansion of $\psi$, (\ref{Psi:Fourier:expansion}), into the Hamiltonian
(\ref{Halmiton:normal:ordering:three:parts}), then expressing everything in terms of the bosonic compound
operators introduced in (\ref{equal:time:quant:Bosonization}), eliminating $B, \,D$ in line with
(\ref{B:D:not:independent}), and only keeping terms at leading order in $1/N$,
the $:H_2:$ and $:H_4:$ pieces in (\ref{Hamilton:three:parts:axial}) read
\bseq
\begin{align}
:H_2: &= \iint\frac{dPdp}{(2\pi)^2}\,(\widetilde{E}(p)+\widetilde{E}(P-p))M^\dagger(p-P,p)M(p-P,p),
\label{eq:H2:MMdagger}\\
:H_4: &= -\frac{\lambda}{8\pi^2} \int dP \iint \frac{dp\, dk} {(p-k)^2}  \Theta\left({|p-k|-\rho}\right) \left\{2C(p,k,P)M^\dagger(p-P,p)M(k-P,k)\right.,
\nn\\
&\left.+S(p,k,P)\left[M(p,p-P)M(k-P,k)+M^\dagger(p,p-P)M^\dagger(k-P,k)\right]\right\}.
\label{eq:H4:MMdagger}
\end{align}
\label{H2:H4:MMdagger}
\eseq
where the function $S$ and $C$ are defined as~\cite{Bars:1977ud}
\bseq\label{C:S:functions:def}
\begin{align}
C\left(p,k,P \right) =& \cos\frac{\theta(p)-\theta(k)}{2}\cos\frac{\theta(P-p)-\theta(P-k)}{2},
\\
S\left(p,k,P \right) =& \sin\frac{\theta(p)-\theta(k)}{2}\sin\frac{\theta(P-p)-\theta(P-k)}{2}.
\end{align}
\eseq

\subsection{Bogoliubov transformation, diagonalization, and Bars-Green equations}
\label{Bogoliubov:diagonalization:BG:eqn}

The Hamiltonian $:H_2:+:H_4:$ in (\ref{H2:H4:MMdagger}) is not yet in the
diagonalized form. Parametrically,  it bears the specific structure:
\beq
H \sim H_0 + A M^\dagger M + B (M^\dagger M^\dagger + M M),
\eeq
which is reminiscent of the Hamiltonian for the  dilute weakly-interacting Bose gas~\cite{Schwabl:book}.
The familiar strategy of diagonalizing this type of Hamiltonian is through the
Bogoliubov transformation~\cite{Schwabl:book}:
\bseq
\label{Bogoliubov}
\bqa
&& m = u M + v M^\dagger,
\label{m:in:term:of:M:Mdagger}\\
&& m^\dagger = u M^\dagger + v M,
\\
&& u^2-v^2=1. \label{Bogoliubov:condition}
\eqa
\eseq

For our problem at hand, we can generalize \eqref{Bogoliubov}
by introducing two sets of operators $m_n$ and $m_n^\dagger$ ($n=0,1,\dots$),
which are the counterparts of the $m$ and $m^\dagger$ in (\ref{Bogoliubov}),
as the linear combination of the $M$ and $M^\dagger$ operators~\cite{Kalashnikova:2001df}:
\bseq
\label{eq:m:mdagger:M:Mdagger:equal:time}
\begin{align}
&m_n(P)=\int_{-\infty}^{+\infty}\frac{dq}{\sqrt{2\pi |P|}}\left[M(q-P,q)\,\varphi^+_n(q,P)+M^\dagger(q,q-P)\,\varphi^-_n(q,P)\right]
\\
&m_n^\dagger(P)=\int_{-\infty}^{+\infty}\frac{dq}{\sqrt{2\pi |P|}}\left[M^\dagger(q-P,q)\,\varphi^+_n(q,P)+M(q,q-P)\,\varphi^-_n(q,P)\right]
\label{mn:dagger:definition}\\
&M(q-P,q) = \sqrt{\frac{2\pi}{|P|}}\sum_{n=0}^\infty \left[m_n(P)\varphi_n^+(q,P)-m_n^\dagger(-P) \varphi_n^-(q-P,-P)\right]
 \\
&M^\dagger(q-P,q) =\sqrt{\frac{2\pi}{|P|}} \sum_{n=0}^\infty \left[m_n^\dagger(P)\varphi_n^+(q,P)-m_n(-P) \varphi_n^-(q-P,-P)\right],
\end{align}
\eseq
where $m_n(P)$ and $m_n^\dagger(P)$ will be interpreted as the annihilation and
creation operators for the $n$-th mesonic state carrying spatial momentum $P$.
The functions $\varphi_n^+(q,P)$ and $\varphi_n^-(q,P)$ play the role of Bogoliubov coefficients $u$ and $v$ in
(\ref{m:in:term:of:M:Mdagger}).

Similar to (\ref{LC:m:mdagger:commutation}) in the LF case, here we again postulate that the mesonic
annihilation and creation operators, $m_n$ and $m_n^\dagger$, obey
the canonical commutation relations:
\bseq
\bqa
&& \left[m_n(P),m^\dagger_m(P')\right]=2\pi\,\delta_{n m}\,\delta(P-P'),
\\
&& \left[m_n(P),m_m(P')\right]=\left[m_n^\dagger(P),m_m^\dagger(P')\right]=0.
\eqa
\label{Equal:time:m:mdagger:commutation}
\eseq
In order to satisfy these commutation
relations, the Bogoliubov functions $\varphi^n_\pm$ must obey the following
orthogonality and completeness conditions~\footnote{We stress that our normalization conditions
differ from those in \cite{Kalashnikova:2001df} by a factor of $|P|$, because
we demand that $\varphi_+^n(xP,P)$ remains dimensionless in conformity to the 't Hooft wave function $\phi(x)$,
which turns to be particularly convenient in comparing quasi and light-cone distributions.
Nevertheless, by adopting this convention, we are no longer capable of
studying the bound-state solutions in the rest frame ($P=0$) as was done in \cite{Li:1987hx}.}:
\bseq
\begin{align}
&\int_{-\infty}^{+\infty}{dp}\left[\varphi_+^n(p,P)\,\varphi_+^m(p,P)-\varphi_-^n(p,P)\,\varphi_-^m(p,P)\right]=|P|\delta^{nm}
\\
&\int_{-\infty}^{+\infty}{dp}\left[\varphi_+^n(p,P)\,\varphi_-^m(p-P,-P)-\varphi_-^n(p,P)\,\varphi_+^m(p-P,-P)\right]=0,
\\
&\sum_{n=0}^\infty\left[\varphi_+^n(p,P)\,\varphi_+^n(q,P)-\varphi_-^n(p-P,-P)\,\varphi_-^n(q-P,-P)\right]=|P|\delta(p-q)
\\
&\sum_{n=0}^\infty\left[\varphi_+^n(p,P)\,\varphi_-^n(q,P)-\varphi_-^n(p-P,-P)\,\varphi_+^n(q-P,-P)\right]=0,
\end{align}
\eseq
Note the relative minus sign in the last two equations reflects the
characteristic of the Bogoliubov transformation, as specified in
(\ref{Bogoliubov:condition})~\cite{Kalashnikova:2001df}.

We wish to diagonalize the axial-gauge Hamiltonian in the basis of $m_n$ and $m^\dagger_n$ operators.
Applying the Bogoliubov transformation (\ref{eq:m:mdagger:M:Mdagger:equal:time})
into (\ref{H2:H4:MMdagger}), we aim to obtain the intended form
\beq
\label{diagonalized:form:H:axial:gauge}
H=H_0^\prime + \int\frac{dP}{2\pi}\sum_n P_n^0 m_n^\dagger(P)m_n(P)+{\cal O}(1/\sqrt{N}),
\eeq
where $P_n^0 = \sqrt{M_n^2+ P^2}$, with $M_n$ the mass of the $n$-th mesonic state.
The interaction terms involving three or more mesons are suppressed by powers of $1/\sqrt{N}$,
which is completely immaterial to the theme of this work, so will be neglected.

The shifted vacuum energy in (\ref{diagonalized:form:H:axial:gauge}) is given by
\bqa
&& H_0^\prime = H_0+\sum_n\int dz\int\frac{dPdp}{2\pi |P|}\left\{\vphantom{\dashint^{+}_{\infty}}\left[\tilde{E}(p)+\tilde{E}(P-p)\right]\varphi_-^n(p,P)\,\varphi_-^n(p,P)\right.
\nn\\
&& -\lambda\int_{-\infty}^{+\infty}\frac{dk}{(p-k)^2}\Theta\left(|p-k|-\rho\right)\left[C(p,k,P)\varphi_-^n(p,P)\,\varphi_-^n(p,P)\right.
\\
&& \left.\left.-S(p,k,P)\left(\varphi_+^n(p,P)\,\varphi_-^n(k,P)+\varphi_-^n(p,P)\,\varphi_+^n(k,P)\right)
\right]\vphantom{\dashint^{+}_{\infty}}\right\},
\nn
\eqa
where $H_0$ is given in \eqref{eq:H0:vacuum:energy}.

We define the meson vacuum state $\vert \Omega \rangle$ by the condition
\beq
m_n(P)\vert \Omega \rangle = 0,
\eeq
for all $n$ and $P$. Consequently,
a single meson state can be constructed via
\beq
\vert P^0_n, P \rangle = \sqrt{2P_n^0}m^\dagger_n(P) | \Omega \rangle.
\eeq
Note the true vacuum state $\vert \Omega \rangle$  is highly nontrivial in the equal-time quantization.
This is clearly illustrated by \eqref{mn:dagger:definition}, according to which there are actually
two very different mechanisms to create a meson. First is by creating a pair of quark and antiquark,
no matter the vacuum is trivial or not.
The other mechanism is by removing a pair of quark and antiquark from the vacuum.
This is possible only if a flurry of correlated multi quark-antiquark constantly popping out of the vacuum,
plausibly a consequence of the nonzero quark condensate~\footnote{Note the true vacuum
$\vert \Omega \rangle$  is different from the quark vacuum $|0 \rangle$ defined in
(\ref{quark:vacuum}). It is supposed that they are connected by a unitary operator $S$~\cite{Bicudo:1989sh},
whose explicit form is unknown yet.
In the following sections, we will always use the true (``mesonic'') vacuum when computing
QCD matrix elements}.

After applying the Bogoliubov transformation to (\ref{H2:H4:MMdagger}),
in order to achieve the diagonalized form of (\ref{diagonalized:form:H:axial:gauge}), we have to enforce
the coefficients of operators $m^\dagger_n m_r$ ($n\neq r$), $m^\dagger_n m^\dagger_r+ m_n m_r$ to
vanish. After some algebra, we end up with two following equations:
\bqa
\label{BG:bound-state:eqs:IR:divergent}
&& \left[\widetilde{E}(p)+\widetilde{E}(P-p)\mp P_n^0)\right]\,\varphi^\pm_n(p,P)=
\nonumber\\
&& \lambda \int_{-\infty}^{+\infty} \frac{dk}{(p-k)^2}\Theta\left(|p-k|-\rho\right)\left[C(p,k,P)\,\varphi^\pm_n(k,P)-S(p,k,P)\,\varphi^\mp_n(k,P)\right].
\eqa

This pair of equations is not particularly convenient to use, since both $\widetilde{E}$ and the integrals are
sensitive to the IR cutoff $\rho$.
Miraculously, one can absorb the divergent $\frac{\lambda}{\rho}$ piece in $\widetilde{E}$ into the
cutoff-dependent integral, so that the modified integral becomes regular in the $\rho\to 0+$ limit.
After some manipulation, the axial-gauge bound state equations (\ref{BG:bound-state:eqs:IR:divergent}) can
be rewritten as
\bseq
\label{Final:BG:bound-state:eqs}
\begin{align}
&& \left(E(p)\!+\!E(P\!-\!p)\!-\! P^0 \right) \varphi_+(p,P)=\! \lambda \dashint_{-\infty}^{\infty}\!\! {dk\over (p-k)^2}\left[C(p,k,P)\,\varphi_+(k,P)-S(p,k,P)\varphi_-(k,P)\right],
\nn\\\\
&& \left(E(p)\!+\!E(P\!-\!p)\!+\! P^0 \right)\varphi_-(p,P)=\! \lambda \dashint_{-\infty}^{\infty} \!\! {dk\over (p-k)^2} \left[C(p,k,P)\,\varphi_-(k,P)-S(p,k,P)\varphi_+(k,P)\right],
\nn\\
\end{align}
\eseq
where $E(p)$ is the regularized dressed quark energy defined in ~\eqref{regular:dispersion:relation}.
Note that here we recover the principal value prescription as introduced in (\ref{PV:pres:defined:this:work})~\footnote{One
certainly can also use the equivalent PV prescriptions as specified
in (\ref{TWO:popular:PV:schemes}). Another practically useful prescription is the
{\it subtraction} scheme~\cite{Li:1987hx},
that is, for a test function $f(y)$ which is regular at $y=x$, one has
$\pvint\:\: dy {f(y)\over (x-y)^2} \equiv \int dy \,
{1 \over (x-y)^2} \left[f(y)\!-\!f(x)-(y-x)f'(x)\right]$.}.

Equations~(\ref{Final:BG:bound-state:eqs}) are the very bound-state equations in ${\rm QCD}_2$ in axial gauge,
first derived by Bars and Green back in 1978~\cite{Bars:1977ud}. For this reason, these equations will be
referred to as Bars-Green equations.
Consequently, the Bogoliubov-type functions $\varphi^n_\pm$ can now be interpreted as
the bound-state wave functions, or simply called Bars-Green wave functions.

A crucial feature of ${\rm QCD}_2$ in axial gauge is that, it preserves Poincar\'{e} invariance
in physical sector in a highly nontrivial way.
Notice the dispersive law for a colored object like dressed quark, which is
encoded in (\ref{dispersion:relation}) and (\ref{regular:dispersion:relation}),
is clearly not Lorentz covariant. However, as far as the color-singlet meson is concerned,
one is ensured to recover the standard dispersion relation dictated by special relativity. Specifically speaking,
irrespective of the Lorentz frame where the Bars-Green equations (\ref{Final:BG:bound-state:eqs}) are tackled,
one always ends up with the {\it identical} meson spectra, where the energy of the $n$-th mesonic state is always found
to satisfy $P_n^0 = \sqrt{M_n^2+ P^2}$.
It is important to emphasize that, in order to preserves Poincar\'{e} invariance,
the Bogoliubov angle $\theta(p)$ and the backward-motion component of Bars-Green wave functions,
$\phi^n_-$, appear to play a indispensable role in (\ref{Final:BG:bound-state:eqs}).
Thus, the 't Hooft model in axial gauge represents a rare example
that one knows exactly how to consistently boost a relativistic bound-state wave function
in the equal-time quantization.

A specific consequence of Poincar\'{e} invariance is that,
when the meson is viewed the IMF, that is, in the $P\to \infty$ limit,
one would still obtain the identical mesonic mass spectra.
In this specific Lorentz frame, $\theta(p)\to {\pi\over 2}$ as $p\to \infty$, and the $C$, $S$ functions in (\ref{C:S:functions:def})
reduce to simple step functions, one can show that the Bars-Green equations simply reduce to the 't Hooft equation.
Consequently, in the IMF, the $\phi^n_+$ component of the Bars-Green wave function reduces to
the 't Hooft light-cone wave functions, and the $\phi^n_-$ component fades away at a rate $\propto 1/P_z^2$~\cite{Jia:2017uul}.

The first numerical solution of Bars-Green equations was conducted by Li and collaborators
in late 1980s, yet only for {\it stationary} ($P_z=0$) mesons~\cite{Li:1987hx}. Very recently,
the Bars-Green equations (\ref{Final:BG:bound-state:eqs}), for the first time, were
solved numerically for an arbitrary moving frame for a variety of quark mass~\cite{Jia:2017uul},
thus explicitly establishing the Poincar\'{e} invariance of the 't Hooft model in axial gauge.
In particular, the authors of \cite{Jia:2017uul} concretely observe the tendency that,
when the meson gets more and more boosted, the $\phi_+$ component does converge to the 't Hooft wave function,
while the $\phi_-$ component quickly vanishes.  Moreover, some other physical quantities, such as
quark condensate, meson decay constant were also numerically investigated in different moving frame,
and prove to be Lorentz invariant~\cite{Jia:2017uul}. These studies unequivocally confirm the key role played by
the chiral angle $\theta(p)$ and $\phi^n_-$ to preserve the Poincar\'{e} invariance.

\section{Building PDF and quasi-PDF out of bound-state wave functions}
\label{Sec:PDF:Quasi:PDF}

Parton distribution functions (PDFs) count essentially the number density of a specific species of partons inside a parent hadron
that carry a specific momentum fraction of the hadron, $x$. Unexaggeratingly speaking,
they form the most indispensable inputs for yielding predictions for any high energy collision experiments
involving hadron beams.
In this section, starting from the operator definitions of light-cone PDFs and quasi-PDFs, we are going to
reexpress them in terms of the light-cone wave functions and Bars-Green wave functions for the ${\rm QCD}_2$ in the large $N$ limit.
We will also see that the Bogoliubov angle $\theta(p)$ will explicitly enter the expression for the quasi-PDFs.
This may offer some useful insight on the nature of quasi-PDF in $3+1$-dimensional QCD.
In this section, we assume the meson is moving along the positive $\hat{z}$ axis, so $P>0$.

\subsection{Light-cone PDF}
\label{LC:PDF:Subsection}

Collins and Soper have given a gauge-invariant operator definition for the PDFs~\cite{Collins:1981uw}.
One can readily adapt their definition to ${\rm QCD}_2$.
According to  \cite{Collins:1981uw}, the quark light-cone PDF in ${\rm QCD}_2$
is defined as the nonlocal light-cone correlators sandwiched between two equal-momentum mesons,
which are the $n$-th mesonic states carrying the light-cone momentum $P^+$:
\beq
q_n(x)= \int_{-\infty}^{+\infty} \frac{d\xi^-}{4\pi} e^{-i xP^+ \xi^-}
\left\langle P_n^-, P^+\left| \bar{\psi}(\xi^-) \gamma^+ \mathcal{W}\left[\xi^-,0\right] \psi(0)\right| P_n^-, P^+\right\rangle_C,
\label{CS:def:LC:PDF}
\eeq
where $P^-_n = M_n^2/(2P^+)$, with $M_n$ the meson mass of the $n$-th state.
Here $\psi$ denotes quark Dirac field, and
\beq
\mathcal{W}[\xi^-,0]=\mathcal{P}\left[\exp(-ig_s \int _0^{\xi^-}d\eta^- A^+(\eta^-))\right]
\eeq
is the gauge link connecting the two quark fields, inserted to ensure the gauge invariance of the PDF.
$x=k^+/P^+$ is the light-cone momentum fraction carried by the parton with respect to that of the meson.
By construction, the light-cone PDF in (\ref{CS:def:LC:PDF}) is boost invariant along the $z$ direction.
The subscript $C$ in (\ref{CS:def:LC:PDF}) indicates the disconnected contribution such as
\beq
 \left\langle P_n^-, P^+\left|\right. P_n^-, P^+\right\rangle \int_{-\infty}^{+\infty} \frac{d\xi^-}{4\pi} e^{-i xP^+ \xi^-}
  \left\langle {0} \left| \bar{\psi}(\xi^-) \gamma^+ \mathcal{W}\left[\xi^-,0\right] \psi(0) \right| {0} \right\rangle
\label{disconnected}
\eeq
should be discarded when calculating the forward matrix element
(\ref{CS:def:LC:PDF})~\cite{Collins:2011zzd}.

PDF has a most transparent probabilistic interpretation in the light-front quantization framework~\cite{Collins:2011zzd}.
Moreover, being a gauge-invariant quantity, the simplest way to proceed is to impose the light-cone gauge $A_-=0$ in (\ref{CS:def:LC:PDF}),
so that the gauge link can be dropped.
For simplicity, we will adopt the LF quantization and light-cone gauge in ${\rm QCD}_2$,
as was comprehensively described in Section~\ref{Sec:LC_H_method}.

The presence of $\gamma^+$ in (\ref{CS:def:LC:PDF}) implies that only the $\psi_R$ component (``good'' component)
of the Dirac field $\psi$ is projected out. Applying the Fourier expansion of the
$\psi_R$ as in \eqref{PsiL:Fourier:expansion},
and replacing the meson state by acting $m^\dagger_n(P^+)$ on the vacuum,
and the light-cone PDF in (\ref{CS:def:LC:PDF}) then becomes
\begin{align}
\label{eqn:PDF_exp}
\notag q_n(x) =&\int_{-\infty}^{+\infty} \frac{d\xi^-}{4\pi}e^{-ixP^+\xi^-}\iint {dk_1^+ dk_2^+\over 4\pi^2}\left\langle {0}\left|m_n(P^+)\sqrt{2P^+} \left[b^\dagger(k^+_1)e^{ik_1^+ \xi^-}\right.\right.\right.
\\
&\left.\left.\left.+d(k_1^+) e^{-ik_1^+ \xi^-} \right]\left[b(k_2^+)+d^\dagger(k_2^+)\right] \sqrt{2P^+}m_n^\dagger(P^+) \right| {0} \right\rangle_C.
\end{align}
Replacing the combinations $b^\dagger(k_1^+) b(k_2^+) $, $b^\dagger(k_1^+) d^\dagger(k_2^+)$, $d(k_1^+)b(k_2^+)$ and $d(k_1^+)d^\dagger(k_2^+)$ by the bosonic operators  $B$, $M^\dagger$, $M$, and $D$ as
in \eqref{Four:Bosonic:operators} correspondingly, and rewriting $B$ and $D$ in terms of $M$, $M^\dagger$
according to \eqref{Replace:B:D},
then eliminating $M$, $M^\dagger$ in favor of $m_n$ or $m_n^\dagger$ in line with \eqref{m:connected:to:M}, we end up with
the vacuum matrix element of the product of a string of mesonic creation and annihilation operators.
Discarding the disconnected piece (\ref{disconnected}), which arises from the commutator between $m_n$ and $m_n^\dagger$, we
obtain the intended LC PDF for the $n$-th mesonic state:
\beq
q_n(x) = \varphi_n(x)^2,
\label{LC:PDF:expression:fin}
\eeq
where $\varphi_n(x)$ is the 't Hooft light-cone wave function associated with the $n$-th mesonic state.
This result confirms what was obtained in \cite{Burkardt:2000uu} using simpler method.
Hearteningly, the light-cone PDF in 't Hooft model looks exceedingly simple.

It is also straightforward to account for the anti-quark distribution,
by extending the support of $x$ in (\ref{LC:PDF:expression:fin}) from $0\leq x \leq1$
to $-1\le x \le 1$:
\beq
q_n(x) =\epsilon(x) \varphi_n(|x|)^2,
\eeq
with the sign function $\epsilon(x)$ equal to 1 for positive $x$,
and equal to $-1$ for negative $x$.
Since we are considering only the flavor-neutral meson, the antiquark PDF is obviously identical
to the quark PDF.

\subsection{quasi-PDF}
\label{Sec:qs_PDF_def}

quasi-PDF was recently introduced by Ji as a proxy to facilitate the extraction of the light-cone PDF
from Euclidean lattice QCD~\cite{Ji:2013dva,Ji:2014gla}.
It is defined as the equal-time spatially-nonlocal correlation functions
sandwiched between two equal-momentum hadrons.
Although quasi-PDF is obviously not boost-invariant, its profile is expected to converge to the
light-cone PDF in the IMF.
Analogous to the definition of the quasi-PDF in realistic QCD~\cite{Ji:2013dva},
the quasi-PDF in ${\rm QCD}_2$ can be defined as the following forward matrix element,
with the external hadron taken to be the $n$-th mesonic state with the spatial momentum $P$:
\beq
\tilde{q}_n (x, P)=\int_{-\infty}^{+\infty}\frac{dz}{4\pi}\,e^{ixP z}\,\left\langle P_n^0, P\left|\bar{\psi}(z)\,\gamma^z\mathcal{W}\left[z,0\right]
\psi(0)\right|P_n^0,P\right\rangle_C,
\label{Def:quasi:PDF}
\eeq
where $P_n^0=\sqrt{P^2+M_n^2}$, $x=k/P$ is the spatial momentum fraction carried by the parton
with respect to that of the meson. Unlike the light-cone PDF,
the range of $x$ is unconstrained, $-\infty < x < \infty$.
The space-like gauge link
\beq
\mathcal{W}[z,0]=\mathcal{P}\left[\exp(-ig_s \int _0^{z}d z' A^z(z^\prime))\right]
\eeq
has been inserted in (\ref{Def:quasi:PDF}) to ensure gauge invariance of the quasi-PDF.

Similar to (\ref{CS:def:LC:PDF}),
the subscript $C$ in (\ref{Def:quasi:PDF}) again indicates that only the connected contributions are included.
Therefore, the disconnected piece,
\begin{align}
\left\langle P_n^0,P\left|\right. P_n^0,P\right\rangle\int_{-\infty}^{+\infty}\frac{dz}{4\pi}\,e^{ixP z}\,\left\langle {\Omega}\left|\bar{\psi}(z)\,\gamma^z\mathcal{W}\left[z,0\right] \psi(0)\right|{\Omega}\right\rangle
\label{Discarded:equal:time}
\end{align}
should be discarded when calculating the matrix element affiliated with the quasi-PDF.

Since quasi-PDF is time-independent, it is natural to study its property in the equal-time quantization.
Moreover, it is most convenient to compute the quasi-PDF is in the axial gauge $A^z=0$ in (\ref{Def:quasi:PDF}),
so that one can neglect the gauge link.
For simplicity, in this subsection, we will stay with the equal-time quantization and work with axial
gauge gauge in ${\rm QCD}_2$,
closely following the quantization
procedure detailed in section~\ref{Sec:H_Axl}.

We proceed by the bosonization procedure similar to computing the light-cone PDF in Section~\ref{LC:PDF:Subsection}.
First replacing the meson state by acting $m^\dagger_n(P)$ on the true vacuum $\vert \Omega\rangle$,
our task then becomes to compute the vacuum matrix element.
Conducting the Fourier expansion of the Dirac field $\psi$
in accordance with \eqref{Psi:Fourier:expansion},
expressing the product of two quark annihilation and creation operators in terms of
$B$, $M^\dagger$, $M$, and $D$ as introduced
in \eqref{equal:time:quant:Bosonization}, followed by rewriting $B$, $D$ as the convolution integral
between $M$, $M^\dagger$ according to \eqref{B:D:not:independent},
then trading $M$, $M^\dagger$ for the meson annihilation and creation operators $m_n$ or $m_n^\dagger$
in line with \eqref{eq:m:mdagger:M:Mdagger:equal:time}, we end up with
the vacuum matrix element of the product of a string of meson annihilation and creation operators.
Repeatedly applying the commutation relations \eqref{Equal:time:m:mdagger:commutation}, and
discarding the disconnected piece (\ref{Discarded:equal:time}), we are finally capable of
expressing the quasi-PDF as
\bqa
\tilde{q}_n(x,P) &=& {P_n^0 \over P}\, \sin \theta(xP)\,\bigg[\left(\varphi^n_+(xP,P)\right)^2+
\left(\varphi^n_-(xP,P)\right)^2
\nn\\
&+ & \left(\varphi^n_+(-xP,P)\right)^2
+\left(\varphi^n_-(-xP,P)\right)^2 \bigg].
\label{Quasi:PDF:in:term:of:BG:wave:func}
\eqa
The explicit occurrences of the Bogliubov angle $\theta(p)$, and the
backward-moving component of the Bars-Green wave functions $\varphi_-^n$,
make the quasi-PDF a much more complicated object than the light-cone PDF.
It is reassuring to see that, in the IMF, {\it i.e.}, in the $P\to \infty$ limit,
where $P_n^0\to P$, $\theta(xP)\to {\pi\over 2}\epsilon(x)$,
and $\phi_-^n$ dies away,
the quasi-PDF in \eqref{Quasi:PDF:in:term:of:BG:wave:func} does recover
the light-cone PDF in \eqref{LC:PDF:expression:fin}!

Eq.~\eqref{Quasi:PDF:in:term:of:BG:wave:func} is one of the key achievements of this paper.
We have successfully constructed the quasi-PDF in terms of the basic building block of ${\rm QCD}_2$
in axial gauge, the chiral angle and the Bars-Green wave functions.
We are wondering whether this reduction pattern, at least to some extent,
can also be carried over to the realistic QCD.

Charge conjugation symmetry imposes the following relation for the $\tilde{q}(x)$:
\begin{align}
\tilde{q}\left(-x,P\right)= -\tilde{q}\left(x,P\right).
\label{Charge:conjugation}
\end{align}
Reassuringly, the quasi-PDF as specified in \eqref{Quasi:PDF:in:term:of:BG:wave:func},
indeed obeys this relation.

It is worth mentioning here, the definition of quasi-PDF is by no means unique.
In principle, one can construct an infinite number of gauge-invariant quasi-PDFs,
all of which are equally legitimate provided that all of them can reduce to the light-cone PDF in IMF.
It can be said that all the legal definitions of quasi-PDF form a universality class~\cite{Hatta:2013gta}.
In Appendix~\ref{Sec:g0_gz}, we will numerically compare two different definitions of quasi-PDF,
the one just considered in this Section, versus the other defined
by replacing $\gamma^z$ in \eqref{Def:quasi:PDF} with $\gamma^0$.

\section{Building LCDA and quasi-DA out of bound-state wave functions}
\label{Sec:LCDA:Quasi:DA}

For hard exclusive reactions involving hadrons, it is the light-cone distribution amplitude (LCDA), rather than the PDF,
that directly enters the QCD factorization theorem~\cite{Lepage:1980fj}. Therefore, LCDAs represent the
fundamental nonperturbative inputs in order to describe the hard exclusive QCD processes.

Analogous to Section~\ref{Sec:PDF:Quasi:PDF}, we will in this section
express the LCDA and quasi-DA of a flavor-neutral meson in ${\rm QCD}_2$,
in terms of its bound-state wave functions.

\subsection{LCDA}

In line with Refs.~\cite{Lepage:1980fj,Radyushkin:1977gp}, one defines
the LCDA of a flavor-neutrual meson in ${\rm QCD}_2$ as
\begin{align}
&\Phi_{n}(x)=\frac{1}{f^{(n)}}\int_{-\infty}^{+\infty} {d\xi^- \over 2\pi}\,e^{-i\left(x-\tfrac{1}{2}\right) P^+\xi^-}
\,\left\langle P_{n}^-, P^+\left|\bar{\psi}\left(\tfrac{\xi^-}{2}\right) \mathcal{W}\left[\tfrac{\xi^-}{2},-\tfrac{\xi^-}{2}\right]\gamma^+\gamma_5\psi\left(-\tfrac{\xi^-}{2}\right)\right|0\right\rangle,\label{Def:LC:DA}
\end{align}
where $\mathcal{W}$ is the
light-like gauge link similar to what is introduced in \eqref{CS:def:LC:PDF}.
$f^{(n)}$ denotes the decay constant of the $n$-th mesonic state, defined through~\cite{Callan:1975ps,Jia:2017uul}
\beq
\left\langle n,\, P \left\vert \bar{\psi} \gamma^\mu \gamma^5 \psi
\right\vert \Omega \right\rangle
= f^{(n)} {P^\mu \over \sqrt{2P^0}}.
\eeq

Analogous to Section~\ref{LC:PDF:Subsection}, it is most transparent to study the LCDA in LF quantization
supplemented with further imposing the light-cone gauge.
The expressions of the meson decay constants are particularly simple in LF quantization~\cite{Callan:1975ps},
\begin{align}
f^{(n)}=  \begin{cases}
    \sqrt{\frac{N}{\pi}}\int_0^1 dx\, \varphi_{n}\left(x\right)\;  &\text{even } n, \\
    0\;  &\text{odd } n, \\
  \end{cases}
 \end{align}
which are particularly simple.
Due to the parity consideration, the decay constants of all the $n$-odd flavor-neutral mesons vanish.
Therefore, we will concentrate on the LCDAs of those $n$-even mesonic states.

Following the essentially same bosonization techniques
that lead to the light-cone PDF in Section~\ref{LC:PDF:Subsection},
we find the LCDA for the $n$-even mesonic states to be
\beq
\Phi_{2n}\left(x\right)=\frac{1}{f^{(2n)}}\sqrt{\frac{N}{\pi}}\varphi_{2n}(x),
\label{LCDA:in:term:of:tHooft:wf}
\eeq
which is simply proportional to the 't Hooft wave function.
Since the decay constant scales as $\sqrt{N}$,
the LCDAs thereby assume finite value in the $N\to \infty$ limit.
As a matter of fact, the LCDAs are subject to the normalization condition by construction:
\beq
\int^{1}_{0} \!\!dx \, \Phi_{2n}(x)= 1.
\label{Normalization:LCDA}
\eeq

\subsection{quasi-DA}
\label{SubSec:quasi:DAs}

Analogous to the quasi-DA introduced in ${\rm QCD}_4$~\cite{Ji:2013dva,Ji:2014gla}, here
we choose to define the quasi-DA in ${\rm QCD}_2$ in its canonical form:
\beq
 \widetilde{\Phi}_{n}(x, P)=\frac{1}{f^{(n)}}\int_{-\infty}^{+\infty}\frac{dz}{2\pi}\,e^{i\left(x-\tfrac{1}{2}\right)P z}\,\left\langle P_{n}^0,P\left|\bar{\psi}\left(\tfrac{z}{2}\right) \mathcal{W}\left[\tfrac{z}{2},-\tfrac{z}{2}\right]\gamma^z\gamma_5\psi\left(-\tfrac{z}{2}\right)\right|\Omega\right\rangle,
\label{Quasi:DA:Definition}
\eeq
where $\mathcal{W}$ is the space-like gauge link as introduced in \eqref{Def:quasi:PDF}.

Employing essentially the same bosonization procedure, which leads to the analytic expression for
quasi-PDF in Section~\ref{Sec:qs_PDF_def}, we finally find that the
quasi-DAs of those $n$-even mesonic states can be formulated as
\beq
\widetilde{\Phi}_{2n}(x,P)= {1\over f^{(2n)}} \sqrt{N\over \pi} \sqrt{P^0\over P} \,\sin {\theta(xP)+\theta(P-xP) \over 2}
\Big[
\varphi^{2n}_+(xP,P)+\varphi^{2n}_-(xP,P)
\Big],
\label{quasi:DA:in:term:of:Bars:Green}
\eeq
where $\varphi^{2n}_\pm$ denote the Bars-Green wave functions associated with the $2n$-th excited mesonic state.
The explicit form of the decay constant $f^{(n)}$ in axial gauge has also been worked out~\cite{Jia:2017uul},
which looks considerably more complicated than the LF quantization case:
\begin{align}
&f^{(n)}=\begin{cases}
\sqrt{\frac{N P^0}{\pi P}}\int_{-\infty}^{\infty}dx\, \sin \frac{\theta\left(xP\right)+\theta\left(P-xP\right)}{2}\left[\varphi^{n}_+\left(xP,P\right)+\varphi^{n}_-\left(xP,P\right)\right]\;&\text{even }n,\\
0\;&\text{odd }n.
\\
\end{cases}\label{eq:dcy_cnstnt}
\end{align}
Note the Bars-Green wave functions and the Bogoliubov angle conspire in a nontrivial manner
so that $f^{(n)}$ is independent of the Lorentz frame.

Comparing \eqref{quasi:DA:in:term:of:Bars:Green} and \eqref{eq:dcy_cnstnt},
one sees that, by construction the quasi-DAs also obey a very simple normalization condition:
\beq
\int^{\infty}_{-\infty} \!\!dx \, \widetilde{\Phi}_{2n}(x,P)= 1.
\label{Normalization:Quasi:DA}
\eeq

Reassuringly, in the IMF ($P\rightarrow \infty$), one readily verifies that,
the quasi-DA in \eqref{quasi:DA:in:term:of:Bars:Green} does recover the LCDA
as given in (\ref{LCDA:in:term:of:tHooft:wf}).

\section{Numerical results of quasi-PDF and DA}
\label{Sec:numerical}

Based on the analytic expressions for the light-cone and quasi distributions worked out
in the preceding sections,
we are going to present a comparative study for quasi distributions and their light-cone counterparts in this section.
We consider four types of lowest-lying flavor-neutral mesons: chiral (massless) pion ($\pi_\chi$), physical pion $\pi$,
a fictitious ``strangeonium'' $s\bar{s}$, and charmonium, varying the quark masses according to the
recipe described in Ref.~\cite{Jia:2017uul}.
For the light-cone and quasi-PDFs, we also consider the first
excited states associated with these four meson species.

\begin{figure}[htbp]
 	\centering
 	\includegraphics[width=1.0\textwidth]{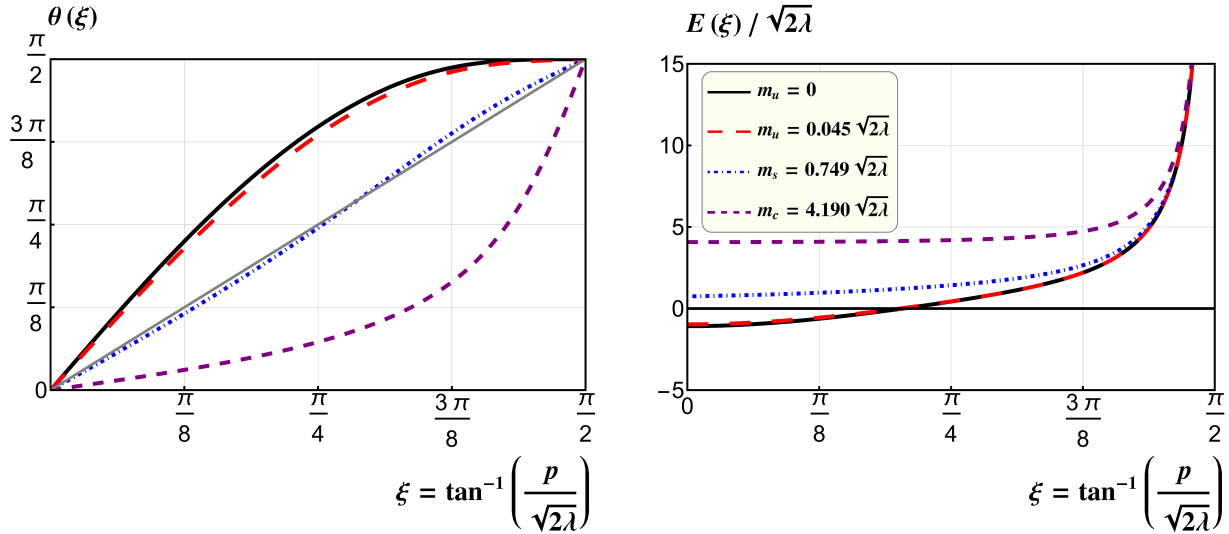}
 \caption{Bogoliubov-chiral angle $\theta(p)$, and the regularized dressed quark energy, $E(p)$,
 as functions of $\xi=\tan^{-1}\left({p\over \sqrt{2\lambda}}\right)$, with different quark masses as
 specified in Table~\ref{tab:masses}.
 Most curves are taken from Ref.~\cite{Jia:2017uul}, except those for the charm quark.
 \label{theta_omega}}
 \end{figure}

The 't Hooft coupling $\lambda=0.18/\pi\;\rm{GeV}^2$ is taken to coincide with
the value of the string tension in realistic $\rm{QCD}_4$~\cite{Burkardt:2000uu}.
The quark masses (in units of $\sqrt{2\lambda}$) are tuned in such a way that the ground-state
meson masses coincide with the realistic meson masses of $\pi_\chi$, $\pi$ and $c\bar c$,
while the mass of the $s$ quark is determined by demanding that
the Bogoliubov angle $\theta(p)$ as a function of $\xi=\tan^{-1} (p/\sqrt{2\lambda})$ is
closest to a straight line~\cite{Jia:2017uul}.
The numerical solutions of 't Hooft equation, mass-gap equation and Bars-Green equations have already
been presented comprehensively in Ref.~\cite{Jia:2017uul}, and we refer the interested reader to that paper
for technical details.
Here we will directly present our numerical results. For the sake of clarity,
the profiles of the Bogoliubov angle $\theta(p)$ and the dispersion relation $E(p)$, which are
affiliated with the aforementioned quark masses, are depicted in Fig.~\ref{theta_omega}.

The quark masses and the corresponding meson mass spectra (for simplicity, we only include the ground state $n=0$ and
the first excited state $n=1$) are listed in Table~\ref{tab:masses}.
\begin{table}[H]
\begin{centering}
\begin{tabular}{|c|c|c|c|c|}
\hline
$m_{q}$ & $u/d$ & $u/d$ & $s$ & $c$\tabularnewline
\hline
\hline
 & 0 & 0.045 & 0.749 & 4.190\tabularnewline
\hline
$M_n $ & $\pi_{\chi}$ & $\pi$ & $s\bar{s}$ & $c\bar{c}$\tabularnewline
\hline
$n=0$ & 0 & 0.41 & 2.18 & 9.03\tabularnewline
\hline
$n=1$ & 2.43 & 2.50 & 3.72 & 10.08\tabularnewline
\hline
\end{tabular}
\par\end{centering}
\begin{centering}
\caption[Caption for LOF]{Quark masses and the corresponding meson mass spectra, where only the ground state and the first excited state are included.\protect\footnotemark}
\label{tab:masses}
\par\end{centering}
\end{table}\footnotetext{In Ref.~\cite{Jia:2017uul}, the charm quark mass is ``erroneously''
take to be $m_c = 4.23\sqrt{2\lambda}$. In this work, we take $m_c$ to be $4.19 \sqrt{2\lambda}$,
which is tuned to reproduce the center-of-gravity mass of the lowest-lying charmonia,
$M_{\rm C.O.G}=\frac{1}{4} M_{\eta_c}+\frac{3}{4} M_{J/\psi}$, associated with the real world.}

\begin{figure}
  \begin{centering}
  \includegraphics[clip,width=\textwidth]{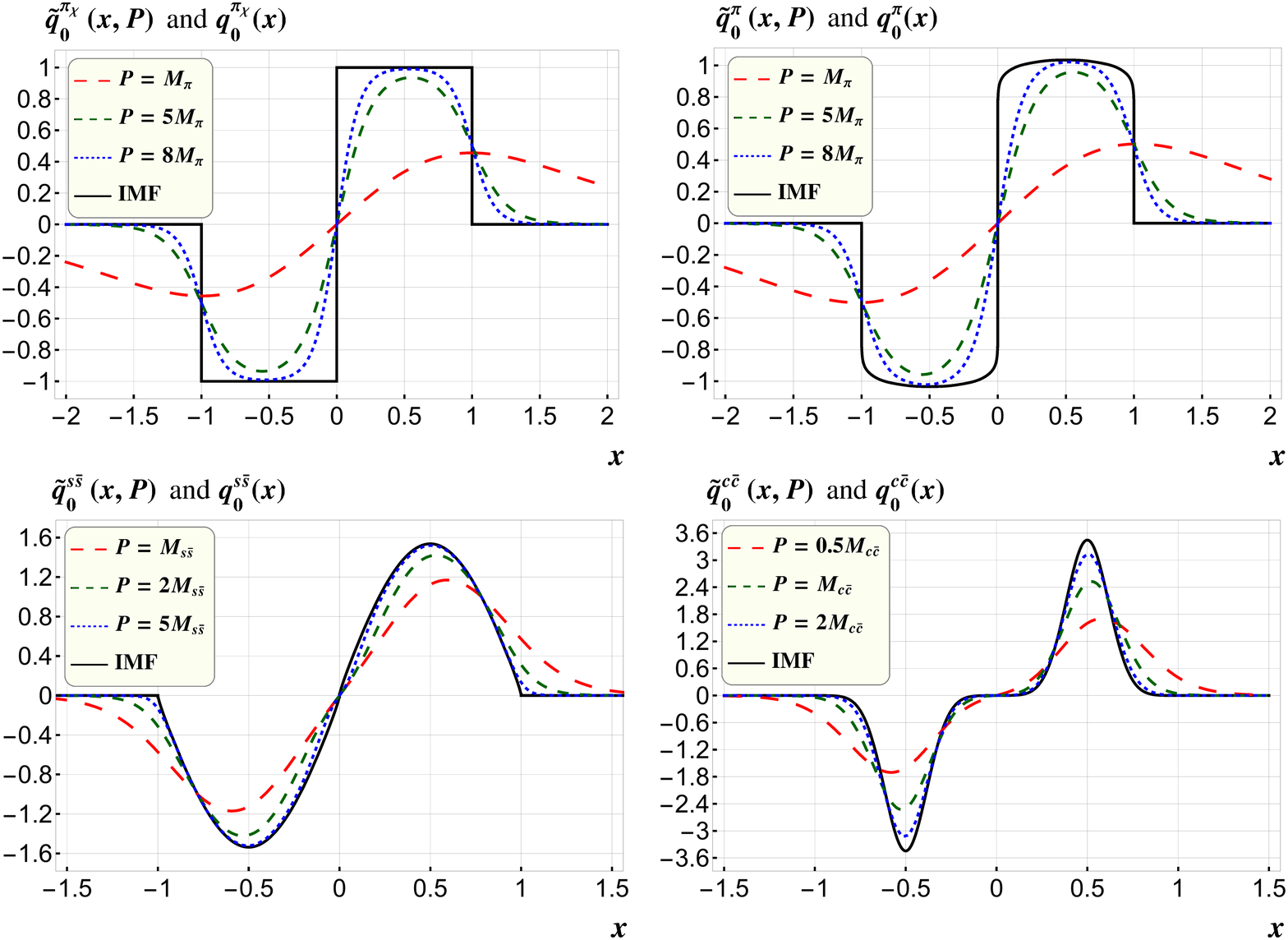}
  \end{centering}
  \caption{Quark light-cone and quasi-PDFs for the chiral pion, physical pion,
  lowest-lying strangeonium and charmonium.
  The momentum of chiral pion is in unit of the physical pion mass.
  \label{fig:PDF}}
  \end{figure}

\begin{figure}
  \begin{centering}
  \includegraphics[clip,width=\textwidth]{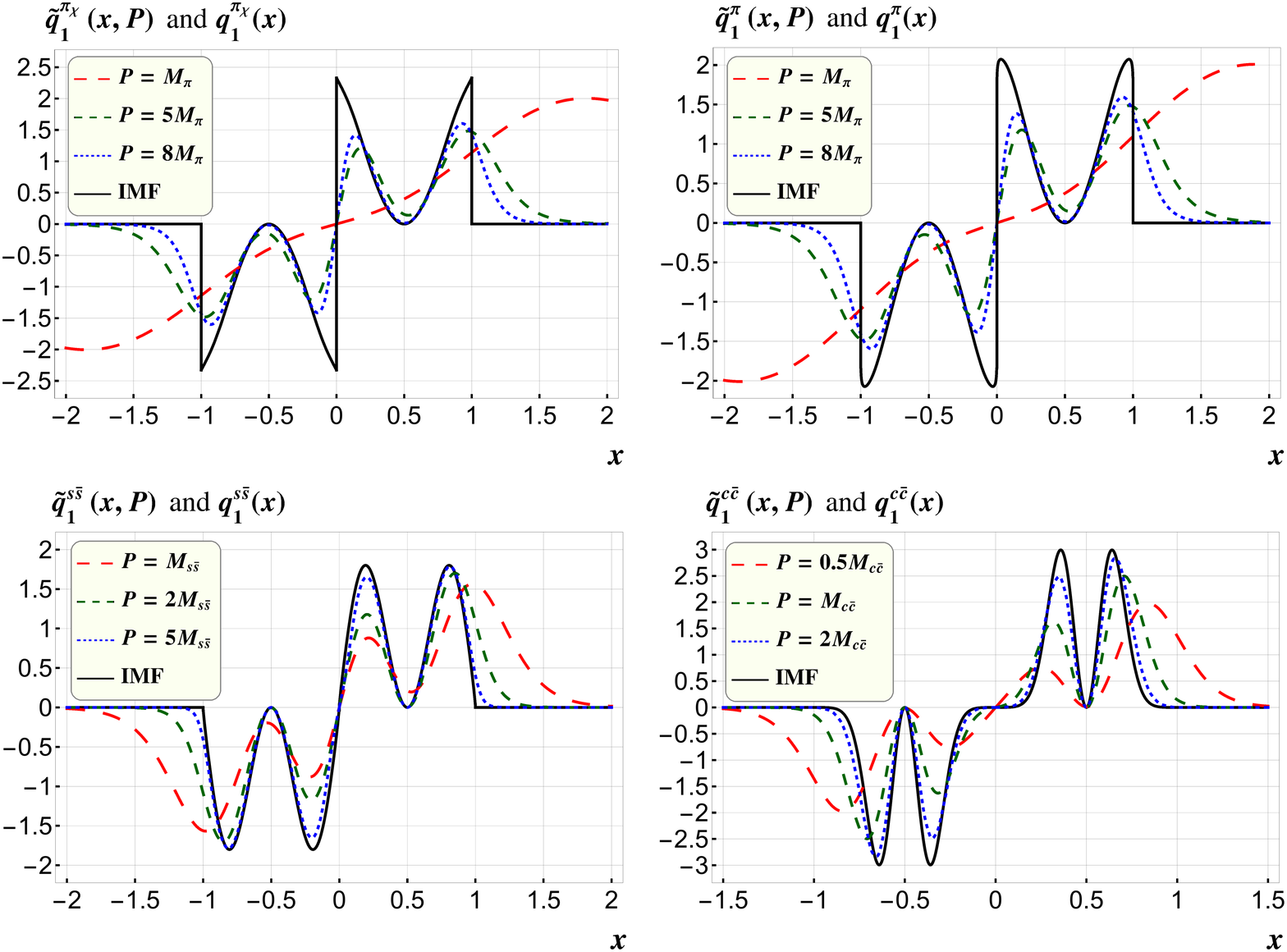}
  \end{centering}
  \caption[justification=raggedright]{Quark light-cone and quasi-PDFs for the $1$st excited state
  corresponding to four different quark masses as specified in Table~\ref{tab:masses}.
  The meson momenta are in units of the ground-state mass for each quark specifies.
  \label{fig:PDFn1}}
  \end{figure}

\begin{figure}
    \begin{centering}
    \includegraphics[clip,width=\textwidth]{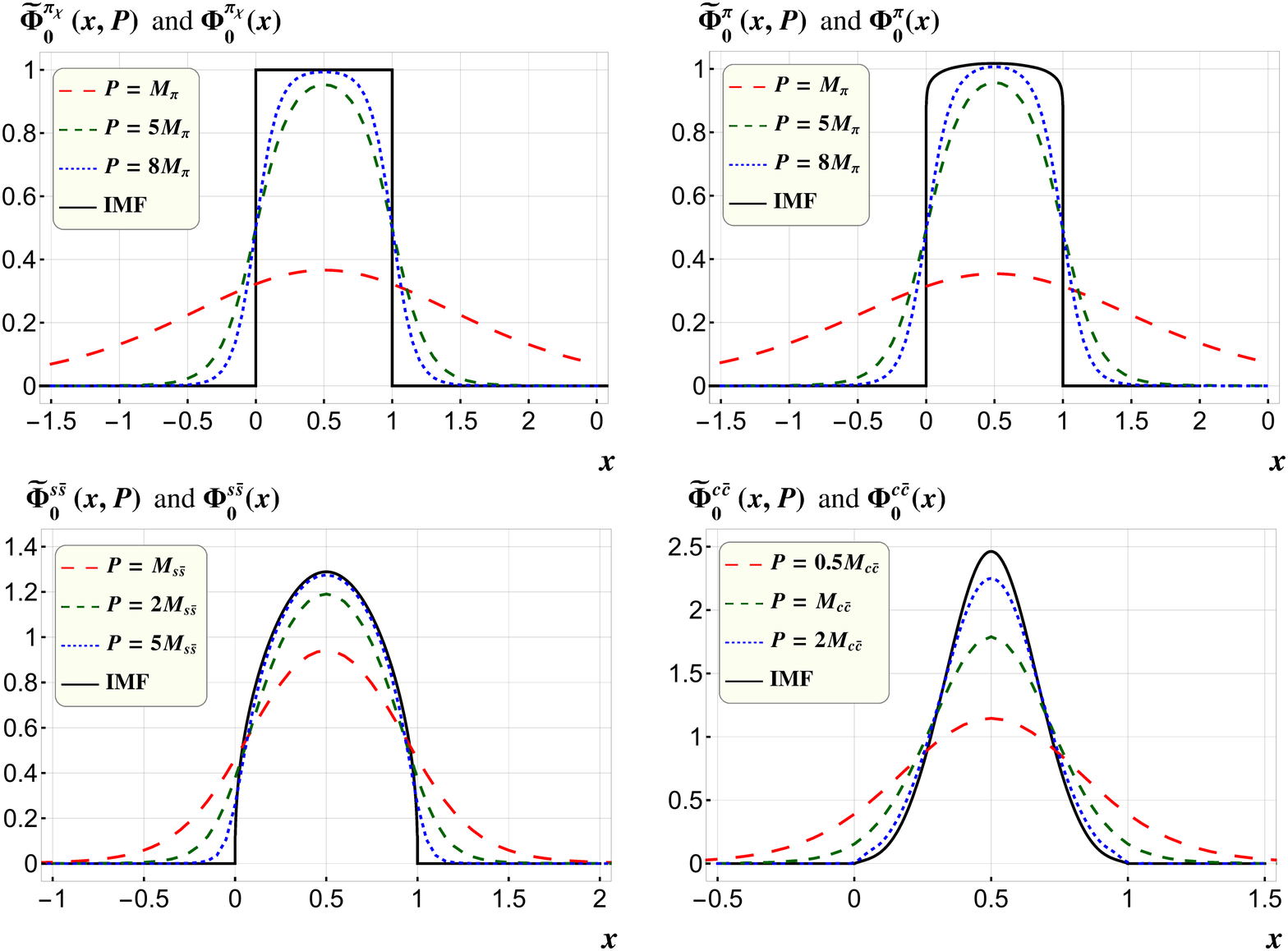}
    \end{centering}
    \caption[justification=RaggedLeft]{LCDAs and quasi-DAs for the chiral pion, pion, lowest-lying
    strangeonium and charmonium. The momentum of chiral pion is in unit of the physical pion mass.
    \label{fig:DA}}
    \end{figure}

In light of the numerically available 't Hooft and Bars-Green wave functions~\cite{Jia:2017uul},
as well as \eqref{Quasi:PDF:in:term:of:BG:wave:func} and \eqref{quasi:DA:in:term:of:Bars:Green},
we calculate the quasi distributions of those mesons in several different reference frames.
The light-cone distributions are also juxtaposed for comparison.
The numerical results of light-cone and quasi-PDFs for lowest-lying mesons are shown in Fig.~\ref{fig:PDF},
and those for the $1$-st excited state in Fig.~\ref{fig:PDFn1},
while the numerical results for the LCDAs and quasi-DAs of the ground-state mesons
are presented in Fig.~\ref{fig:DA}.

From Fig.~\ref{fig:PDF}, \ref{fig:PDFn1}, and \ref{fig:DA}, one clearly observes the
general tendency, that irrespective of the meson species,
the quasi distributions are indeed converging to their light-cone counterparts,
as the meson gets more and more boosted.

An interesting observation is that, the quasi distributions of heavier mesons ($s\bar s$, $c\bar c$) appear to
converge to the light-cone distributions in a {\it faster} pace than those of lighter mesons ($\pi_\chi$, $\pi$).
To quantify this assertion, let us introduce the ratio $r \equiv P_n /M_n$.
For light mesons, as exemplified by the physical pion, even when boosted to $r=8$,
there still exists considerable difference between the shapes of the light-cone and quasi distributions;
on the other hand, heavy mesons tends to exhibit a rather different pattern.
When $r=5$ for the $s\bar{s}$ meson, or when $r=2$ for the $c\bar c$ meson~\footnote{We stop at $r=2$ for
charmonium, mainly because the technical challenge about numerical instability of boosting a heavy meson quickly becomes insurmountable.},
the quasi distributions already coincide with the light-cone distributions
to a decent degree.

The correlation between convergence behavior of quasi distributions under boost
and the hadron species has already been noticed in lattice simulations in realistic ${\rm QCD}_4$~\cite{Chen:2017mzz,Lin:2017ani,Chen:2017lnm,Zhang:2017bzy,Chen:2017gck}.
There it is found that, somewhat counter-intuitively,
nucleon's quasi-PDF approaches its light-cone PDF in a much {\it faster} pace than
the quasi-DA of pion approaches its LCDA.

Were it not a sheer accident, it will be valuable if the 't Hooft model
can offer some insight in unravelling this curious correlation pattern observed in real world.

We end this section by commenting on a simple fact concerning the first excited state.
As can be seen in Fig.~\ref{fig:PDFn1}, one observes a zero at $x= \pm {1\over 2}$ for
various light-cone PDFs in the $n=1$ state, but not for the corresponding quasi-PDFs.
This can be understood from the angle of the charge conjugation symmetry~\cite{Jia:2017uul}:
\begin{equation}
 \varphi^{n} (1-x) = (-1)^n\varphi^{n}(x),\qquad \tilde\varphi^{n}_\pm \left(1-x,P\right) = (-1)^n\tilde\varphi_\pm\left(x,P\right),
 \end{equation}
which implies that the light-cone as well as Bars-Green wave functions of the
$n$-odd states must have a zero at $x= {1\over 2}$. Nevertheless,
as one can see from \eqref{Quasi:PDF:in:term:of:BG:wave:func},
due to the presence of $\sin\theta(xP)$ as well as the different arguments arising in
four types of squared Bars-Green functions, the quasi-PDFs in finite reference frame
no longer possess the simple odd symmetry under $x\leftrightarrow 1-x$.

\section{One-loop perturbative calculations for light-cone and quasi distributions and IR divergences}
\label{Sec:Matching}

So far, we have been completely concentrating on the nonperturbative aspects of the
light-cone and quasi distributions in ${\rm QCD}_2$.
In this Section, we will instead switch the gear,
to address some theoretical issues within the confine of perturbation theory.

The key insight underlying LaMET in ${\rm QCD}_4$ is that quasi distributions exhibit the same IR behavior as their light-cone counterparts
in the leading power of $1/P^z$~\cite{Ji:2013dva}.
Because of this peculiar feature, there arises a factorization theorem that connects the
quasi and light-cone PDFs~\cite{Ji:2013dva,Xiong:2013bka,Ji:2014gla}:
\begin{align}
\tilde{q}(x,P^z,\Lambda)= \int^{1}_{-1} \frac{dy}{|y|}\, Z\left(\frac{x}{y},\frac{\Lambda}{P^z},\frac{\mu}{P^z}\right)q(y,\mu)+
\mathcal{O}\left(\frac{\Lambda_{\rm{QCD}}^n}{P_z^n},\frac{M^r}{P_z^r}\right).
\label{Matching:4:dim:Quasi}
\end{align}
where $\Lambda$ represents a UV cutoff in the transverse momentum space associated with quasi-PDF, and
$\mu$ is the renormalization scale associated with the light-cone PDF. The neglected terms represent the higher twist corrections.
The factorization theorem \eqref{Matching:4:dim:Quasi} states that the $Z$ factor takes into account the difference between the
UV regimes of the light-cone and quasi-PDFs, which is thus amenable to perturbation theory owing to the asymptotic freedom of QCD.
Through the one-loop order, the $Z$ factor affiliated with the quark PDF can be expressed as
\begin{align}
Z\left(\xi,\frac{\Lambda}{P^z},\frac{\mu}{P^z}\right)
= \delta(\xi-1)+ {\alpha_s\over \pi} Z^{(1)}\left(\xi,\frac{\Lambda}{P^z},\frac{\mu}{P^z}\right)+\cdots.
\end{align}
The order-$\alpha_s$ coefficient can be computed by the {\it perturbative matching} procedure,
\begin{align}
Z^{(1)}\left(\xi,\frac{\Lambda}{P^z},\frac{\mu}{P^z}\right) = \tilde{q}^{(1)}\left(\xi,{P^z},\Lambda\right)-{q}^{(1)}\left(\xi,{\mu}\right),
\end{align}
where the physical hadron has been replaced by a single quark,
$\tilde{q}^{(1)}$ and ${q}^{(1)}$ signify the corresponding quasi and light-cone PDF associated with this ``fictitious'' hadron,
accurate to the order-$\alpha_s$.
In four spacetime dimension, due to the severe UV divergence emerging from the transverse momentum integration,
the limit of $P^z\rightarrow\infty$ and $\Lambda\rightarrow\infty$ generally
do not commute~\cite{Xiong:2013bka}.
It is this very noncommutativity that leads to a nontrivial matching factor in realistic QCD.

In this section, we will calculate the one-loop corrections to the light-cone and quasi distributions in $d=2$ spacetime dimension.
The major motif of such computation is to verify one of the backbone of LaMET, that quasi and light-cone distributions
indeed possess the same IR behavior in the leading order in $1/P^z$, even in ${\rm QCD}_2$.
Recall that the gauge coupling in $\rm{QCD}_2$ carries a positive mass dimension,
thereby the 't Hooft model is a superrenormalizable theory.
Therefore, the (almost) absence of UV divergences in loop diagrams~\footnote{One exception is the
perturbative correction to the quark condensate $\langle \bar{\psi}\psi \rangle$ in ${\rm QCD}_2$,
which receives a logarithmic UV divergence from the one-loop tadpole diagram, and can be eliminated
through additive renormalization~\cite{Burkardt:1995eb}.}
nullifies the aforementioned non-commutativity, thus we do not expect a nontrivial $Z$ factor to arise.
On the other hand, ${\rm QCD}_2$ has much more severe IR divergences than its 4-dimensional cousin, so
it is interesting to explicitly examine the IR behavior of light-cone and quasi distributions.

At first sight, it may appear attractive to utilize dimensional regularization (DR)
to regularize the IR divergence.
Nevertheless, apart from automatically preserving Lorentz and gauge invariance, this popular regularization scheme
is not suited for our purpose. First, we will encounter severe power IR divergences, which are simply absent in DR,
but are actually what we desire to see.
More importantly, DR in $1+1$-dimensional theory has some intrinsic drawback.
When working in $2-2\epsilon$ dimensions, we have artificially introduced some fictitious
transverse degrees of freedom, which might
lead to some pathological behavior when taking the $\epsilon\to 0$ limit in the end.

In formulating the bound-state equation in Hamiltonian approach in Sections~\ref{Sec:LC_H_method} and
\ref{Sec:H_Axl}, we have used an infrared momentum cutoff to
regularize the IR divergence.
In this Section, we will again employ this ``physical'' IR cutoff, which turns out to be convenient and less confusing.
It is worth mentioning that the large $N$ limit is no longer required
in this Section.

\subsection{Light-cone and quasi-PDFs to one-loop order}
\label{SubSec:PDF:Quasi:PDF}

\begin{figure}
\centering
\begin{minipage}[c]{\textwidth}
\includegraphics[scale=0.9]{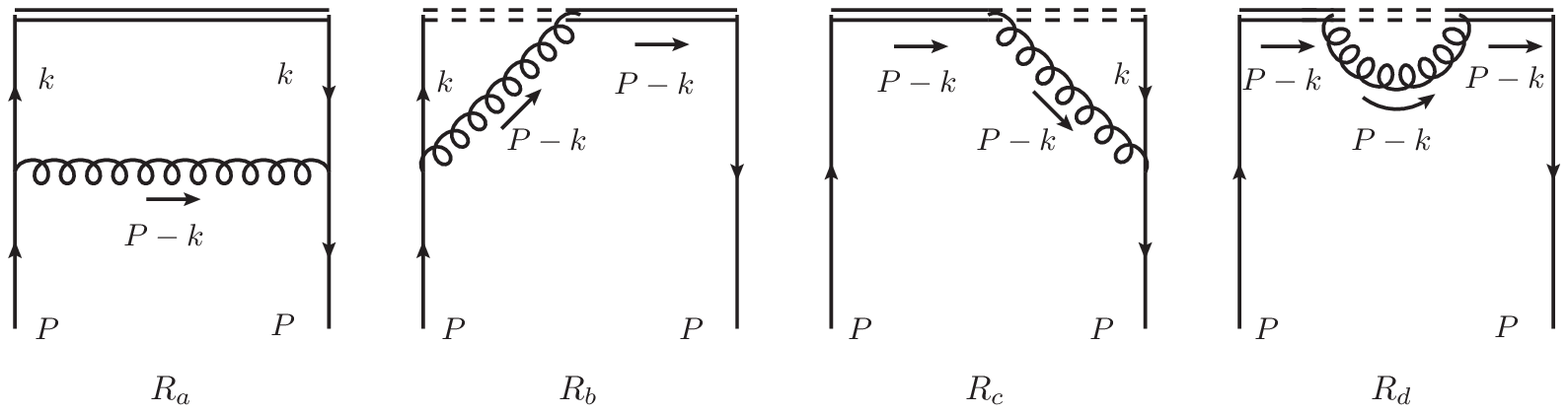}
\vspace{0.3cm}
\end{minipage}
\noindent
\centering
\begin{minipage}[c]{\textwidth}
\includegraphics[scale=0.9]{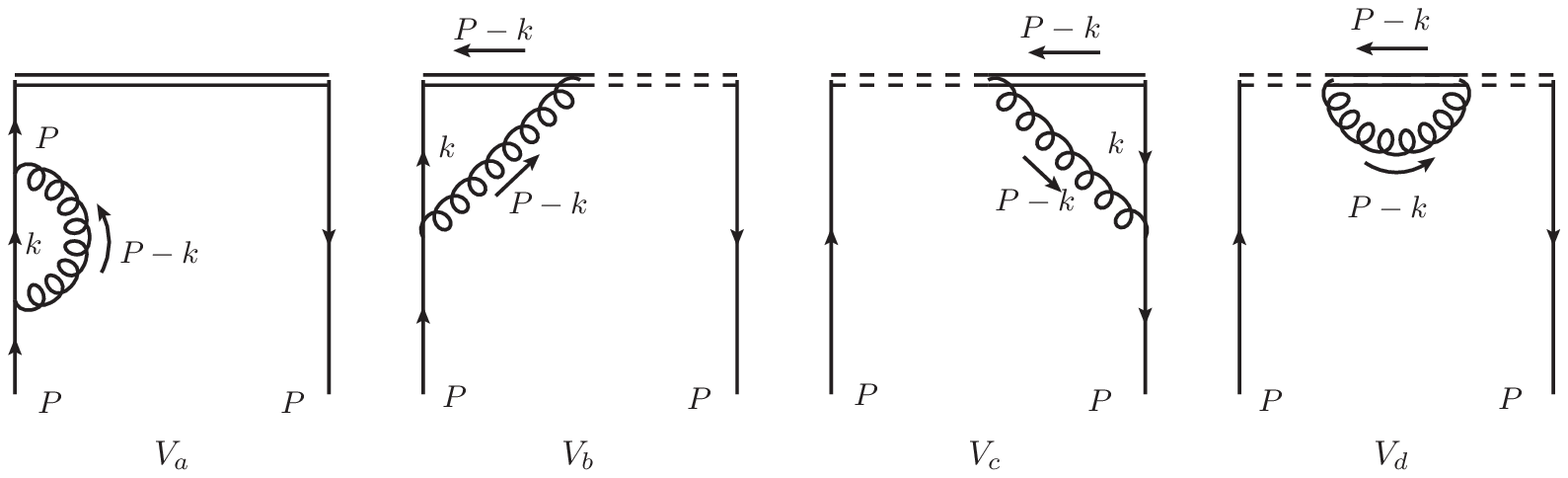}
\end{minipage}
\caption{One-loop Feynman diagrams for quark light-cone and quasi-PDF. The upper row corresponds to the real corrections $\mathcal{Q}^R$,
while the lower row corresponds to the virtual corrections $\mathcal{Q}^V$ which are proportional to $\delta(x-1)$. The Feynman diagrams show the one-loop corrections to the process of a quark with momentum $P$ splitting to a quark with momentum $k$ ($R_a$), the quark interacting with gauge link though a gluon exchange ($R_{b,c}$, $V_{b,c}$), the quark's self energy ($V_a$) and gauge link's self-interaction ($R_d$, $V_d$).  The double line represents the gauge link in PDF and quasi-PDF definition in \eqref{CS:def:LC:PDF} and \eqref{Def:quasi:PDF}, while the dashed double line represents the gauge link on which no net momentum flows. For simplicity, we have also omitted the wave function renormalization on the quark leg in the final state.}
\label{fig:PDF_1loop_RealVirtual}
\end{figure}

In computing the quark light-cone and quasi-PDFs, we replace a physical meson by a single quark.
For technical simplicity, in this subsection, we will no longer stay with the non-covariant gauge,
rather conduct all the calculation in Feynman gauge. The one-loop Feynman diagrams for quark PDF
are shown in Fig.~\ref{fig:PDF_1loop_RealVirtual}.

At this stage, we will treat the one-loop corrections for light-cone and quasi-PDF
in a unified manner. Following the Feynman rules for PDF (quasi-PDF) and gluon-gaugue link interaction term in Ref.~\cite{Collins:2011zzd}, the contributions from real correction diagrams are~\cite{Ishikawa:2017faj,Ji:2015jwa}
\bseq
\begin{align}
  &\mathcal{Q}^R_{a}\left(x,n\right)=\frac{-ig_{s}^{2}}{2}\int \!\! {d^{2}k \over (2\pi)^{2}} \,\frac{\bar{u}\left(P\right)\gamma^{\mu}\left(k\!\!\!/+m\right) n\!\!\!/ \left(k\!\!\!/+m\right)\gamma_{\mu}u\left(P\right)}
  {\left(k^{2}-m^{2}+i\epsilon\right)^{2}\left[\left(P-k\right)^{2}+i\epsilon\right]}\delta\left(n\cdot k-xn\cdot P\right),
\\
&\mathcal{Q}^R_{b+c}\left(x,n\right)= -ig_{s}^2 \int\!\! {d^{2}k \over (2\pi)^{2}} \,\frac{\bar{u}\left(P\right)n\!\!\!/ \left(k\!\!\!/+m\right)n\!\!\!/ u\left(P\right)}{\left(k^{2}-m^{2}+i\epsilon\right)\left[\left(P-k\right)^{2}+i\epsilon\right]}\frac{\delta\left(n\cdot k-xn\cdot P\right)}{n\cdot\left(P-k\right)},
\\
&\mathcal{Q}^R_{d}\left(x,n\right)= {-ig_{s}^{2}\over 2} \int \!\! {d^{2}k \over (2\pi)^{2}}  \,\frac{n^{2}\bar{u}\left(P\right) n\!\!\!/ u\left(P\right)}{\left(P-k\right)^{2}+i\epsilon}\frac{\delta\left(n\cdot k-xn\cdot P\right)}{\left[n\cdot\left(P-k\right)\right]^2}.
\end{align}
\label{one:loop:real:PDF}
\eseq
and the contributions from virtual correction diagrams are~\cite{Ishikawa:2017faj,Ji:2015jwa}
\bseq
\begin{align}
  &\mathcal{Q}_{a}^{V}\left(x,n\right)=\delta Z_{F}\delta\left(1-x\right)=-\delta\left(1-x\right)\left.\frac{d\, \Sigma\left(P\right)}{dP\!\!\!\!/}\right|_{P\!\!\!\!/=m},\\
&\mathcal{Q}_{b+c}^{V}\left(x,n\right)=-\delta\left(x-1\right)\int dx\,\mathcal{Q}^R_{b+c}\left(x,n\right),\\
&\mathcal{Q}_{d}^{V}\left(x,n\right)=-\delta\left(x-1\right)\int dx\,\mathcal{Q}^R_{d}\left(x,n\right),
\end{align}
\label{one:loop:virtual:PDF}
\eseq
where $\Sigma(P)$ represents the quark self energy, and $Z_F$ denotes the quark wave function renormalization constant.

The above expressions in \eqref{one:loop:real:PDF} and \eqref{one:loop:virtual:PDF}
can be adapted to both light-cone or quasi-PDF,
depending on the specific choice of the reference vector $n^\mu$.
For the former, $n^\mu$ is chosen to be the null vector $n^\mu=n^\mu_{\rm LC}$ such that
$n_{\mathrm{LC}}^2=0$, $k\cdot n_{\mathrm{LC}}=k^{+}$, and
$\gamma\cdot n_{\mathrm{LC}} =\gamma^{+}$;
while for the latter, the reference vector is taken to be $n^\mu_z$,
such that $n_{z}^{2}=-1$, $k \cdot n_z =k^z$, and $\gamma \cdot n_z = \gamma^z$.

To compute the one-loop corrections to the light-cone PDF,
it is convenient to switch to the light-cone coordinate,
so one can write $P\cdot k=P^{+}k^{-}+P^{-}k^{+}$.
The integration over $k^{+}$ can be trivially carried out using the $\delta$
function, while the $k^{-}$ integration is performed via the method of residue.
Summing all the real correction diagrams, we find a null result:
\begin{align}
\notag q^R\left(x\right)&=\mathcal{Q}^R_{a}\left(x,n_{\mathrm{LC}}\right)+\mathcal{Q}^R_{b+c}\left(x,n_{\mathrm{LC}}\right)+\mathcal{Q}^R_{d}\left(x,n_{\mathrm{LC}}\right)\\
&=\begin{cases}
\left(\frac{-xg_{s}^{2}C_F}{\pi m^{2}\left(1-x\right)^{3}}\right)_a+\left(\frac{xg_{s}^{2}C_F}{\pi m^{2}\left(1-x\right)^{3}}\right)_{b+c}+\left(0\right)_d & 0<x<1\\
0 & \mathrm{Otherwise}
\end{cases}=0,
\end{align}
where the subscript $a$, $b+c$, $d$ denote the contributions from
$R^a$, $R^b+R^c$ and $R^d$ in Fig.~\ref{fig:PDF_1loop_RealVirtual}, respectively.

In computing the virtual corrections \eqref{one:loop:virtual:PDF}, we have employed
the momentum fraction $\eta$ as an IR regulator, imposed on the $k^{+}$ integration:
\beq
\int_{0}^{P^{+}}dk^{+}\Rightarrow P^{+}\int_{0}^{1}dx \, \Theta(1-x-\eta),
\label{LC:pert:theory:cutoff}
\eeq
where $x=k^+/P^+$ and $\eta\rightarrow 0^+$.

The sum of virtual diagrams also vanishes:
\begin{align}
\notag q^{\!V}\!\!\left(x\right) & = \mathcal{Q}^{\!V}\!\!\left(x,n_\mathrm{LC}\right)= \left[\!\delta\left(\!x\!-\!1\!\right)\frac{g_{s}^{2}C_F}{2\pi m^{2}}\left(\!\frac{1}{\eta^{2}}\!-\!\frac{2}{\eta}\!+\!1\!\right)\!\right]_a
\!\!+\!\!\left[\!-\delta\left(x\!-\!1\right)\frac{g_{s}^{2}C_F}{2\pi m^{2}}\left(\frac{1}{\eta^{2}}\!-\!\frac{2}{\eta}\!+\!1\right)\!\right]_{b+c}\!\!+\!\!\left(0\right)_d\\
&= 0.
\label{total:virtual}
\end{align}

Piecing all terms together, the light-cone PDF at one-loop level is
\begin{equation}
\label{LC:PDF:one:loop:total}
q\left(x\right)= q^R(x) + q^V(x)=0.
\end{equation}

The vanishing light-cone PDF in two dimensions is not surprising.
The one-loop quark PDF can be interpreted as the probability of
a parent quark splitting into a daughter quark plus a on-shell gluon.
However, there is no physical gluon in two dimensions because the
lacking of transverse degree of freedom.
Consequently, the one-loop light-cone PDF (splitting function) vanishes.

For the quasi-PDF, we stay with the ordinary coordinates and the reference two-vector $n$ is chosen as the space-like unit vector $n_z$.
For the real correction diagrams,
first performing the $k^z$ integration using the $\delta$
function, then integrating over
$k^{0}$ via the method of residues, we obtain
\begingroup
\allowdisplaybreaks
\bseq
\begin{align}
\notag \tilde{q}^R_{a}\left(x,P^{z}\right)=&\mathcal{Q}^R_{a}\left(x,n_{z}\right)\\
\notag =&\frac{g_{s}^{2}C_Fm^{2}}{4\pi}\left\{ \frac{xP^{z}\left[2P^{0}\sqrt{m^{2}+x^{2}P_{z}^{2}}+2m^{2}+x(x+1)P_{z}^{2}\right]}{\left(m^{2}+x^{2}P_{z}^{2}\right){}^{3/2}\left(P^{0}\sqrt{m^{2}+x^{2}P_{z}^{2}}+m^{2}+xP_{z}^{2}\right)^{2}}\right.\\
&\left.-\frac{x}{\left|x-1\right|P_{z}^{2}\left[\left|x-1\right|P^{0}+(x-1)P^{z}\right]^{2}}\right\},\\
\notag\tilde{q}^R_{b+c}\left(x,P^{z}\right)=&\mathcal{Q}^R_{b+c}\left(x,n_{z}\right)\\
\notag=&\frac{g_{s}^{2}C_F}{4\pi}\left\{ \frac{P^{0}\left|x-1\right|-P^{z}(x+1)}{P_{z}^{2}(x-1)\left|x-1\right|\left[P^{0}\left|x-1\right|+P^{z}(x-1)\right]}\right.\\
&\left.-\frac{P^{0}\sqrt{m^{2}+x^{2}P_{z}^{2}}+m^{2}-xP_{z}^{2}}{P^{z}(x-1)\sqrt{m^{2}+x^{2}P_{z}^{2}}\left(P^{0}\sqrt{m^{2}+x^{2}P_{z}^{2}}+m^{2}+xP_{z}^{2}\right)}\right\},\\
\tilde{q}^R_{d}\left(x,P^{z}\right)=&\mathcal{Q}^R_d\left(x,n_z\right)=\frac{g_{s}^{2}C_F}{4\pi P_{z}^{2}\left|x-1\right|^{3}}.
\end{align}
\eseq
\endgroup
It is straightforward to check that, in the IMF, the one-loop corrections to the quasi-PDF do
approach their light-cone counterparts,
in a diagram-by-diagram basis.

Next we turn to the virtual corrections for the quasi-PDF.
Each individual virtual one-loop diagram yields the following contribution, respectively:
\bseq
\begin{align}
\tilde{q}_{a}^{V}\left(x,P^{z}\right)= & \delta(x-1)\frac{g_{s}^{2}C_F}{2\pi m^{2}}\left(\frac{P_{0}^{2}+P_{z}^{2}}{2P_{z}^{2}\eta^{2}}-\frac{P_{0}^{2}+P_{z}^{2}}{P^{z}P^{0}\eta}+1\right),
\\
\notag\tilde{q}_{b+c}^{V}\left(x,P^{z}\right)=&Q_{b+c}^{V}\left(x,n_{z}\right)
\\
=&-\delta\left(x-1\right)\frac{g_{s}^{2}C_F}{2\pi m^{2}}\left[\frac{1}{\eta^{2}}-\frac{2P^{0}}{\eta P^{z}}+\frac{m^{4}\tanh^{-1}\left(\frac{P^{z}}{P^{0}}\right)+P^{0}P^{z}\left(m^{2}+P_{0}^{2}\right)}{P_{0}^{3}P^{z}}\right],
\\
\tilde{q}_{d}^{V}\left(x,P^{z}\right)=&Q_{d}^{V}\left(x,n_{z}\right)=-\delta(x-1)
\frac{g_{s}^{2}C_F}{4\pi P_{z}^{2}\eta^{2}},
\end{align}
\eseq
where the quadratic IR singularity emerges in each diagram.

In computing the virtual corrections for quasi-PDF, analogous to \eqref{LC:pert:theory:cutoff} in the light-cone case,
we again utilize the momentum fraction variable $\eta$ as an IR regulator,
imposed on the $k^z$ integration:
\beq
\int_{0}^{P^{z}}dk^{z}\Rightarrow P^{z}\int_{0}^{1}dx \, \Theta(1-x-\eta),
\label{equal:time:pert:theory:cutoff}
\eeq
where $x=k^z/P^z$ and $\eta\rightarrow 0^+$.

Summing up $\tilde{q}^R_{a,b+c,d}$ and $\tilde{q}^V_{a,b+c,d}$, we obtain the
complete one-loop corrections to the quasi-PDF:
\beq
\tilde{q}\left(x,P^{z}\right)  =  \frac{g_{s}^{2}m^{2}}{4\pi P^{z}}\left[\frac{1}{\left(x-1\right)\left(m^{2}+P_{z}^{2}x^{2}\right)^{3/2}}\right]_{+}+
 \frac{g_{s}^{2}}{2\pi\eta P^{0}P^{z}}\delta\left(x-1\right),
\label{Quasi:PDF:one:loop:total}
\eeq
which is suppressed by at least one inverse power of $P^z$.
Note the linear IR divergence is still present,
but accompanied with a ${\cal O}(1/P_z^2)$ suppression factor.

Comparing \eqref{LC:PDF:one:loop:total} with \eqref{Quasi:PDF:one:loop:total}, we verify that, to the one-loop order,
both the light-cone and quasi-PDFs do share the same IR behavior at the leading power in $1/P^z$,
which is simply zero.

Integrating $\tilde{q}(x,P^z)$ in \eqref{Quasi:PDF:one:loop:total}
over entire range of $x$ generates the one-loop correction to the vector current $\bar\psi \gamma^z\psi$.
This non-vanishing integral indicates that the vector current no longer conserves,
with the extent of violation of $\mathcal{O}\left(g_s^2/P_z^2\right)$, clearly a
higher twist effect.
We suspect that the breaking of vector current conservation may
originate from the fact that the momentum cut-off in $k^z$ integration likely
violates Lorentz invariance.
To check this assumption, we also recalculate the one-loop correction to the vector current in DR,
and confirm that the current conservation holds at one-loop order.
Despite this nuisance, in order to be consistent with the rest of the paper,
we will still stick to the soft momentum cutoff as a viable IR regulator.

In passing, it is worth mentioning that, in the matching between the
light-cone and quasi generalized parton distribution functions (GPDs) in four spacetime dimensions,
a similar pattern has also been observed: for the $E$-type GPD,
the light-cone and quasi GPD differ in IR at one-loop order
only by a higher-twist term~\cite{Ji:2015qla,Xiong:2015nua}.

As anticipated, due to UV finiteness of ${\rm QCD}_2$ at one-loop level, the matching between quasi-PDF
and light-cone PDF turns out to be trivial, at least to this perturbative order,
thereby the corresponding $Z$ factor is simply $\delta\left(\xi-1\right)$.

It is also illuminating to trace the origin of the $1/P^z$-suppressed scaling behavior
of quasi-PDF (hence the vanishing light-cone PDF)
from another angle, {\it i.e.},  from the time-ordered perturbation theory (TOPT),
or often referred to as the old-fashioned perturbation theory.
From \eqref{eq:H_orgnl}, one can split the axial-gauge ${\rm QCD}_2$ Hamiltonian
into the free and the interaction parts:
\begin{subequations}
\bqa
&& H = H_{0}+ H_{\rm int},
\\
&& H_0 = \int \!\!dz \: \psi^\dagger(z)\,\left(-i \gamma^5 \partial_z +m\gamma^0\right)\,\psi(z),
\\
&& H_{\rm int} = \int\!\!\!\!\!\int dz\,dz^\prime \: {\cal H}_{\rm int}(z,z^\prime) =  - {g_s^2\over 2} \sum_a\int \!\! dz\,dz'\,\psi^\dagger(z)T^a\psi(z)\,\widetilde{G}^{(2)}_\rho(z-z')\,\psi^\dagger(z')T^a\psi(z').
\nn\\
\eqa
\label{Axial:H:decomposition}
\end{subequations}

It is convenient to conduct the TOPT calculation for the partonic quasi distributions
in the $A^z=0$ gauge, where the gauge links ${\cal W}$ in the quasi-PDF in \eqref{Def:quasi:PDF}
and quasi-DA in \eqref{Quasi:DA:Definition} simply disappear.
Through the second order in $g_s$, it turns out that the quasi-PDF in
\eqref{Def:quasi:PDF} and quasi-DA in \eqref{Quasi:DA:Definition} can be recast into the equivalent
TOPT format, each of which consists of
two distinct time-ordering between $\bar{\psi}\gamma^z\psi$ and $H_{\rm int}$:
\bseq
\bqa
\tilde{q}\left(x,P^{z}\right) &=&
\int\!\!\!\!\!\int \! dz_1 dz_2\,\int {dz\over 4\pi}\, e^{ix P^z z} \bigg\{ \left\langle P  \right| \bar\psi(z)\gamma^z\psi(0)
{1\over P^0-H_0}
\mathcal{H}_{\rm int}(z_1,z_2)\left| P \right\rangle_C
\nn \\
&+& \left\langle  P \right| \mathcal{H}_{\rm int}\left(z_1,z_2\right)
{1\over P^0-H_0} \bar\psi(z)\gamma^z\psi(0)\left|  P  \right\rangle_C \bigg\},
\label{eqn:TOPT_PDF}
\\
\widetilde{\Phi} \left(x,P^z\right) &=&
\int\!\!\!\!\!\int \! dz_1 dz_2\, \int {dz\over 2\pi}\, e^{i\left(x-\tfrac{1}{2}\right)P^z z}
\bigg\{\,\left\langle p, P-p\right\vert \bar\psi\left(\tfrac{z}{2}\right)\gamma^z\psi\left(-\tfrac{z}{2}\right) {1\over P^0-H_0}
\mathcal{H}_{\rm int}(z_1,z_2)\vert 0 \rangle_C
\nn \\
&+& \left\langle p, P-p\right\vert  \mathcal{H}_{\rm int}(z_1,z_2){1\over P^0-H_0}\bar\psi\left(\tfrac{z}{2}\right)\gamma^z\psi\left(-\tfrac{z}{2}\right)
\vert 0 \rangle_C \bigg\},
\label{eqn:TOPT_DA}
\eqa
\label{TOPT:two:time-ordering}
\eseq
where $H_0$ appearing in the energy denominator refers to the free part of the Hamiltonian,
and $\mathcal{H}_{\rm int}$  represents the instantaneous Coulomb interaction, both of which are defined
in \eqref{Axial:H:decomposition}.

\begin{figure}[H]
\begin{centering}
\includegraphics{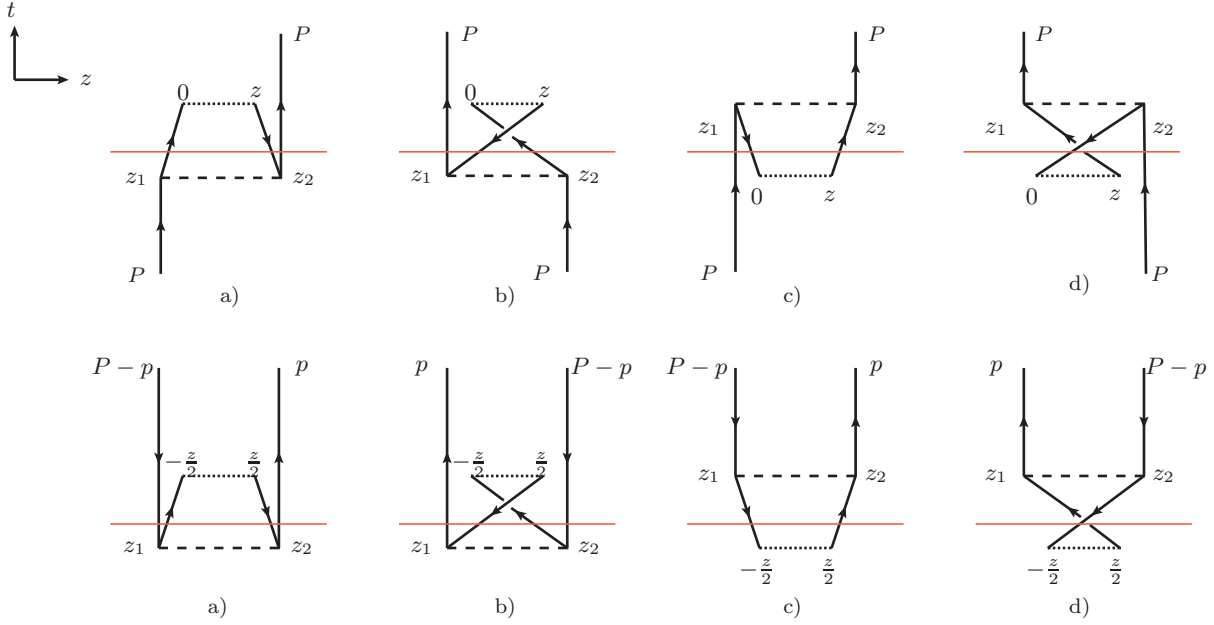}
\end{centering}
\caption{Various TOPT diagrams that are responsible for the ${\cal O}(g_s^2)$ real correction contributions to
quasi-PDF (upper row) and quasi-DA (lower row).
The dashed line represents the instantaneous color Coulomb potential.
The horizontal (red) solid line specifies the allowed intermediate on-shell partonic states.}
\label{fig:TOPT}
\end{figure}

As before, we first replace the external hadronic states in \eqref{eqn:TOPT_PDF} by
an on-shell quark with 2-momentum $P^\mu=(P^0,P^z)$.
We proceed by inserting a complete set of eigenstates of $H_0$
immediately left to ${1\over P^0-H_0}$ in \eqref{eqn:TOPT_PDF}.
To obtain a nonvanishing result,
the viable intermediate states are inevitably composed of three free particles,
$q\bar{q}q$, which turns out to contribute to the real corrections for
the quasi-PDF~\footnote{If the intermediate states only consist of
the single quark $q$, the matrix elements in \eqref{eqn:TOPT_PDF} then correspond to the virtual correction
to the quasi-PDF. For simplicity, we will not bother to consider this piece of contribution.}.
We then compute the matrix elements of $\bar{\psi}\gamma^z\psi$ and $H_{\rm int}$ separately,
by contracting the field operators with the external partonic states in all possible way.
Integrating over the spatial variables $z_1$, $z_2$ and $z$, we then end with
the product of several momentum-conserving $\delta$-function.
One finally can write down all the order-$g_s^2$ contributions to the quasi-PDF.
Each individual contribution corresponds to a particular way of
contracting field operators and external states, which is schematically represented by the those TOPT
diagrams in Fig.~\ref{fig:TOPT}.

Since the intermediate states must contain three particles for quasi-PDF,
it is inevitable for the vacuum creation and annihilation vertices to arise in the TOPT diagrams,
as can be clearly seen from the top row of Fig.~\ref{fig:TOPT}.
Obviously, it is the resulting large energy denominator that is responsible for the $1/P^z$-suppressed behavior~\cite{Weinberg:1966jm} of quasi-PDF.
The quasi-PDF eventually vanishes when viewed in the IMF,
which amounts to the vanishing light-cone PDF.


\subsection{LCDA and quasi-DA to one-loop order}\label{Sec:Matching_DA}

\begin{figure}[H]
\begin{centering}
\includegraphics[clip,width=\textwidth]{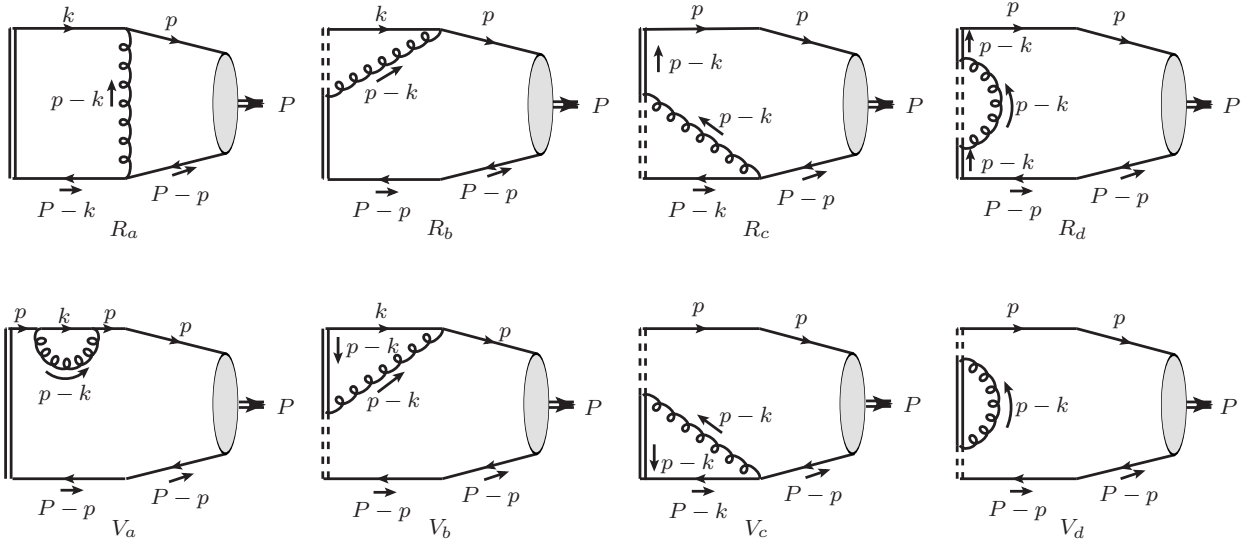}
\end{centering}
\caption{One-loop Feynman diagrams for the real (upper row) and virtual (lower row) corrections to DA of a fictitious meson with momentum $P$, the Feynman diagrams show the one-loop corrections to the amplitude of extracting a quark (anti-quark) with momentum $k$   ($P-k$) from a quark anti-quark pair with momentum $p$ ($P-p$) correspondingly. For simplicity, we have omitted the wave function renormalization diagram for the antiquark line.
Identical to the PDF case, the virtual corrections to the DA
take the form of $Z_F \delta(x-\tfrac{1}{2})$. The double line represents the gauge link in DA and quasi-DA definition \eqref{Def:LC:DA} and \eqref{Quasi:DA:Definition}, while the dashed double line represents the gauge link on which no net momentum flows.}
\label{fig:DA_1loop_RealVirtual}.
\end{figure}

To access the LCDA and quasi-DA in perturbation theory, we proceed to replace a meson by a color-singlet
$q\bar{q}$ pair. To justify perturbative expansion, in this subsection we assume the
weak coupling limit: $g_s \ll m$, has been taken.
The corresponding one-loop diagrams for DAs are shown in Fig.~\ref{fig:DA_1loop_RealVirtual}.

In four spacetime dimensions, the one-loop matching factor linking the LCDA and quasi-DA, is more involved than the one linking the light-cone PDF and quasi-PDF.
One needs to start with a more general momentum configuration $k^+ = xP^+$ and $p^+=yP^+$. Thus the matching factor $Z$ cannot be written as
a single-variable function~\cite{Ji:2015qla}, instead must depend on both $x$ and $y$.
Owing to the UV finiteness of ${\rm QCD}_2$, the matching factor is doomed to be trivial.
Therefore, for illustrative purpose, we will focus on one specific kinematic configuration of the external ``mesonic'' state,
$y={1\over 2}$, that is,
$p^+={P^+\over 2}$, so that the $q$ and $\bar{q}$ equally partition the fictitious meson's total momentum.
As a consequence, the DA becomes the function of $x$ only.

Following basically the same strategy as adopted in one-loop calculation for the
light-cone and quasi quark PDF, as described in Section~\ref{SubSec:PDF:Quasi:PDF},
we obtain the one-loop corrections to the LCDA, $\Phi(x)$, and the quasi-DA, $\widetilde{\Phi}(x,P^z)$:
\allowdisplaybreaks{
\bseq
\begin{align}
 & \Phi\left(x\right) =
\begin{cases}
-\left[\frac{4g_s^2 C_F x (1-x)}{\pi  m^2 (1-2 x)^4}\right]_{4+}+\frac{g_s^2C_F(1-4\eta)}{16\pi m^2\eta}\delta''\left(x-\frac{1}{2}\right) &0<x<1
\\
 0 &\mathrm{Otherwise}
\end{cases},
\label{eq:LCDA_1loop}
\\
& \widetilde{\Phi}\left(x,P^z\right)=
\left\{\frac{g_s^2C_F}{\pi  \left(1-2x\right)^2 P^z}\left[-\frac{2 P^0 \left(m^2+ x(1-x)  P_z^2\right)}{m^2 (1-2 x)^2 P_z^2}+\frac{1}{\sqrt{m^2+(1-x)^2 P_z^2}}\right.\right.
\nn \\
\notag &\left.\left.\times\frac{x (x-1) P_z^2-x P^0\sqrt{m^2+(x-1)^2
   P_z^2}\!+\!2 m^2 (x-1)}{
   -P^0\sqrt{m^2+(x-1)^2 P_z^2}+2 m^2-x
   P_z^2+P_z^2}\right]\!+\!\left(x\rightarrow 1\!-\!x\right)\right\}_{4+}
\\
   &-\frac{g_s^2C_F}{16 \pi  m^2 \left(P^z\right)^3}\delta''\left(x-\tfrac{1}{2}\right)\left[4 P^z \left(P_0^2-m^2\right)-\frac{P_0^3}{\eta }\right],
\label{eq:qsDA_1loop}
\end{align}
\label{Quasi:DA:one:loop:order}
\eseq
}
where a ``4-plus'' prescription has been introduced.
This prescription is understood in a distributive sense, which is defined as
\bqa
\int_{x_1}^{x_2} dx  \left[\frac{g\left(x\right)} {\left|x-\tfrac{1}{2}\right|^4} \right]_{4+}\!\! f(x) & = &
\int_0^1 dx  \frac{g\left(x\right)} {\left|x-\tfrac{1}{2}\right|^4} \bigg[f(x)\!-\!f\left(\tfrac{1}{2}\right)- f'\left(\tfrac{1}{2}\right)\left(x\!-\!\tfrac{1}{2}\right)
\nn \\
&-& \frac{f''\left(\tfrac{1}{2}\right)}{2!}\left(x-\tfrac{1}{2}\right)^2 - \frac{f'''\left(\tfrac{1}{2}\right)}{3!}\left(x-\tfrac{1}{2}\right)^3 \bigg],
\eqa
where $f(x)$ is any smooth functions that are regular at $x={1\over 2}$.
We further assume $g(x)$ is symmetric under the exchange $x\leftrightarrow 1-x$, so that $g'\left({1\over 2}\right)=0$.
The integration boundaries are $x_1=0$, $x_2=1$ for LCDA, and
$x_1=-x_2=-\infty$ for quasi-DA, respectively.
In \eqref{eq:LCDA_1loop} and \eqref{eq:qsDA_1loop}, we have employed some distribution identities
to express the DAs in terms of these ``4-plus'' distributions.
More details about those identities can be found in Appendix~\ref{Sec:distribution:identities}.

In contrast to \eqref{LC:PDF:one:loop:total} and \eqref{Quasi:PDF:one:loop:total} in the PDF case,
one sees that, to the one-loop order,
the quasi-DA contains some leading-twist pieces that are not suppressed by powers of $1/P^z$.
As anticipated, boosting \eqref{eq:qsDA_1loop} to IMF, one readily recovers \eqref{eq:LCDA_1loop}.
The difference between quasi-DA and LCDA is certainly of the higher twist origin, of
the order $g_s^2/P_z^2$.

Examining \eqref{eq:LCDA_1loop} and \eqref{eq:qsDA_1loop}, reassuringly,
we do observe that both LCDA and quasi-DA possess the identical linear IR singularity,
$\propto {g_s^2\over m^2 \eta} \delta''\left(x-\frac{1}{2}\right)$.

It is again elucidating to see why the quark DA, in contrast to the quark PDF,
contains a leading twist term, from the angle of time-ordered perturbation theory.
Similar what is done to quasi-PDF, we also insert 
a complete set of eigenstates of $H_0$
immediately left to ${1\over P^0-H_0}$ in \eqref{eqn:TOPT_DA}.
Unlike the case of quasi-PDF, here 
the allowed intermediate states can be either $q\bar{q}$ or $q\bar{q}q\bar{q}$, in order to obtain
a nonvanishing results for the real corrections to  quasi-DA.
Computing both matrix elements involving $\bar{\psi}\gamma^z\psi$ and $H_{\rm int}$,
exhausting all possible contractions between Dirac field operators and the external partonic states,
integrating over the spatial variables $z_1$, $z_2$ and $z$, we  
finally end up with all the order-$g_s^2$ contributions to the quasi-DA.
Each contribution specifies a particular way of
contracting field operators and external states, which are represented by the those TOPT
diagrams in lower row of Fig.~\ref{fig:TOPT}.

In contrast to the case for quasi-PDF, apart from tetra-quark states, the $q\bar{q}$ two-particle states
also constitute the legitimate
intermediate states. As a result, the corresponding TOPT diagrams, {\it e.g.} Fig.~\ref{fig:TOPT}~$c)$, $d)$
in the lower row, are absent of the vacuum creation and annihilation vertices,
therefore freed from suppression by large energy denominator 
Consequently, the leading scaling behavior of the quasi-DA in the large momentum limit is
$g_s^2/P_z^0$, which leads to a nonvanishing LCDA when viewed in the IMF.

We now conclude this Section. By explicitly working out the one-loop corrections to
quark PDF and DA in ${\rm QCD}_2$, we have firmly established the validity of the cornerstone of LaMET,
{\it viz.}, the partonic quasi and light-cone distributions do share the identical IR behavior a
t the leading power in $1/P^z$.  The one-loop correction to the DA
appears to constitute a more nontrivial example than
the PDF.

\section{Summary and outlook}\label{Sec:Summary}

In this paper, we have carried out a comprehensive study of two important classes of
meson's parton distributions, the PDF and DA, in the context of
large-$N$ limit of $\rm{QCD}_2$.
Our approach is entirely based upon the first principles of QCD.
We have applied the Hamiltonian operator method as well as bosonization technique
to construct both light-cone and quasi distributions out of the basic building blocks,
that is, the 't Hooft wave function for the former, Bars-Green wave functions and the
Bogoliubov angle for the latter.
In a sense, equations~\eqref{Quasi:PDF:in:term:of:BG:wave:func} and \eqref{quasi:DA:in:term:of:Bars:Green}
are the key formulae of this work. Unlike their four-dimensional counterparts,
which can only be accessed by numerical lattice simulation in Euclidean spacetime,
we have directly probed the quasi distributions in Minkowski spacetime, and have developed a
thorough understanding about what they are made of in the two-dimensional case.

We justify the 't Hooft wave function as the valid light-cone Fock state wave function of the hadron.
Consequently, the quark PDF and LCDA can be directly built out of 't Hooft wave function, in an exceedingly
simple manner.
On the contrary, in the equal-time quantization,
a pair of Bars-Green wave functions alone is not sufficient to
express the quasi distributions, and one must supplement another important ingredient, the
Bogoliubov-chiral angle, which may be viewed as characterizing the nonperturbative nature of the vacuum.

We have presented a comparative numerical study between
light-cone PDFs and quasi-PDFs, as well as between LCDAs and quasi-DAs, for a variety of meson species.
It is straightforward to see from \eqref{Quasi:PDF:in:term:of:BG:wave:func} and \eqref{quasi:DA:in:term:of:Bars:Green} that,
the quasi distributions do converge to their light-cone counterparts in the IMF.
We also numerically verified the tendency that, the quasi distributions do approach
their light-cone counterparts, when the meson gets more and more boosted.
We have also observed an interesting pattern, that light meson's quasi distributions in general
approach the light-cone distributions in a slower rate compared with the heavy mesons under boost.
This somewhat counterintuive pattern is qualitatively consistent what is observed in lattice simulations in
realistic four-dimensional QCD~\cite{Zhang:2017bzy,Chen:2018fwa}.

Within the realm of perturbation theory, we have also investigated the one-loop corrections to
the light-cone and quasi distributions in ${\rm QCD}_2$,
yet abandoning the large $N$ limit.
We have verified the backbone of LaMET in this novel theoretical setting, that the IR behaviors of
quasi and light-cone distributions are identical at the leading power in $1/P^z$.
It is theoretically interesting, since ${\rm QCD}_2$ has more severe IR divergence than ${\rm QCD}_4$.
We do witness how the linear IR divergences in LCDA and quasi-DA agree with each in ${\rm QCD}_2$.
Nevertheless, since ${\rm QCD}_2$  is a super-renormalizable theory,
the matching $Z$ factor linking the light-cone with quasi distributions
turns out to be trivial.

Equipped with the bosonization method, we are capable of computing virtually all the nonperturbative
gauge-invariant matrix elements in the 't Hooft model.
For instance, besides quasi-PDF, we are also able to compute the lattice cross section~\cite{Ma:2017pxb}
as well as pseudo PDF~\cite{Radyushkin:2017cyf,Orginos:2017kos}, which have been advocated as viable competitors of quasi-PDF,
presumed to be more efficient to extract the light-cone PDF.
There is no principle difficulty to perform a similar study for these alternative options of parton distributions as in this work.
To some extent, ${\rm QCD}_2$ may be viewed as an ideal and
fruitful theoretical laboratory, which can examine many interesting ideas
concerning a variety of parton distributions.

\begin{acknowledgments}

We thank LiuJi Li for participating in the early stage of this work. We are grateful to Xiangdong Ji and
Jianhui Zhang for useful discussions.
The work of Y.~J., S.-R. Liang and R.~Y. is supported in part by the National Natural Science Foundation of China under Grants No.~11475188,
No.~11621131001 (CRC110 by DFG and NSFC), by the IHEP Innovation Grant under contract number Y4545170Y2,
and by the State Key Lab for Electronics and Particle Detectors.
The work of X.-N.~X. is supported by the Deutsche Forschungsgemeinschaft (Sino-German CRC 110).
\end{acknowledgments}

\appendix

\section{Alternative definitions of quasi-PDF and DA: $\gamma^z$ versus $\gamma^0$}
\label{Sec:g0_gz}

As mentioned in Sec.~\ref{Sec:qs_PDF_def}, one is free to invent different operator definitions for quasi distributions,
all of which are legitimate provided that they can reduce to the correct light-cone distributions in IMF.
It is said that they then form a universality class~\cite{Hatta:2013gta}.
The difference among them must be suppressed by powers of $1/P^z$.

In this Appendix, we wish to critically compare two simplest definitions for quasi-PDF:
\bseq
\bqa
\tilde{q}_{\gamma^z}(x,P)&=&\int_{-\infty}^{+\infty}\frac{dz}{4\pi}\,e^{-ixP^{z}z}\,\left\langle P_n^0,P|\bar{\psi}(z)\,\gamma^{z}\mathcal{W}\left[z,0\right]\psi(0)|P_n^0,P\right\rangle _{C},
\\
\tilde{q}_{\gamma^0}(x,P)&=&\int_{-\infty}^{+\infty}\frac{dz}{4\pi}\,e^{-ixP^{z}z}\,\left\langle P_n^0,P|\bar{\psi}(z)\,\gamma^{0}\mathcal{W}\left[z,0\right]\psi(0)|P_n^0,P\right\rangle _{C}.
\eqa
\eseq
The first canonical definition follows from \eqref{Def:quasi:PDF}, which has already
been investigated in the main text.
The second definition is new, which we are going to explore.
The subscript ``$C$'' again implies that only the connected part of the matrix element is retained.

Through the operator approach and the Bogoliubov transformation,
the functional forms of two different
definitions of quasi-PDFs in terms of $\tilde\varphi_\pm\left(x,P\right)$ and $\theta$ angle can be worked out,
\bseq
\bqa
\tilde{q}_{n,\gamma^z}(x,P) &\!=\! & {\frac{P_n^0}{P}}\sin \theta(xP)\!\! \left[\left(\varphi^n_+(xP,P)\right)^2\!+\!
\left(\varphi^n_-(xP,P)\right)^2 \!+\!\left(\varphi^n_+\left(-xP,P\right)\right)^2
\!+\!\left(\varphi^n_-\left(-xP,P\right)\right)^2\right],
\nn\\
\label{quasi:PDF:gammaz}
\\
\tilde{q}_{n,\gamma^0}(x,P) &\!=\!&{\frac{P_n^0}{P}} \left[\left(\varphi^n_+(xP,P)\right)^2\!-\!
\left(\varphi^n_-(xP,P)\right)^2\!+\!\left(\varphi^n_-\left(-xP,P\right)\right)^2\!-\!
\left(\varphi^n_+\left(-xP,P\right)\right)^2\right].
\nn\\
\label{quasi:PDF:gamma0}
\eqa
\eseq
Here \eqref{quasi:PDF:gammaz} simply duplicates \eqref{Quasi:PDF:in:term:of:BG:wave:func}.
Absence of the factor $\sin\theta$ in \eqref{quasi:PDF:gamma0}
may account for why the new quasi-PDF approaches the light-cone PDF in a faster pace than the canonical one.
It is curious to know whether this reason has any connection to the realistic ${\rm QCD}_4$.

\begin{figure}[h]
\begin{centering}
\includegraphics[clip,width=\textwidth]{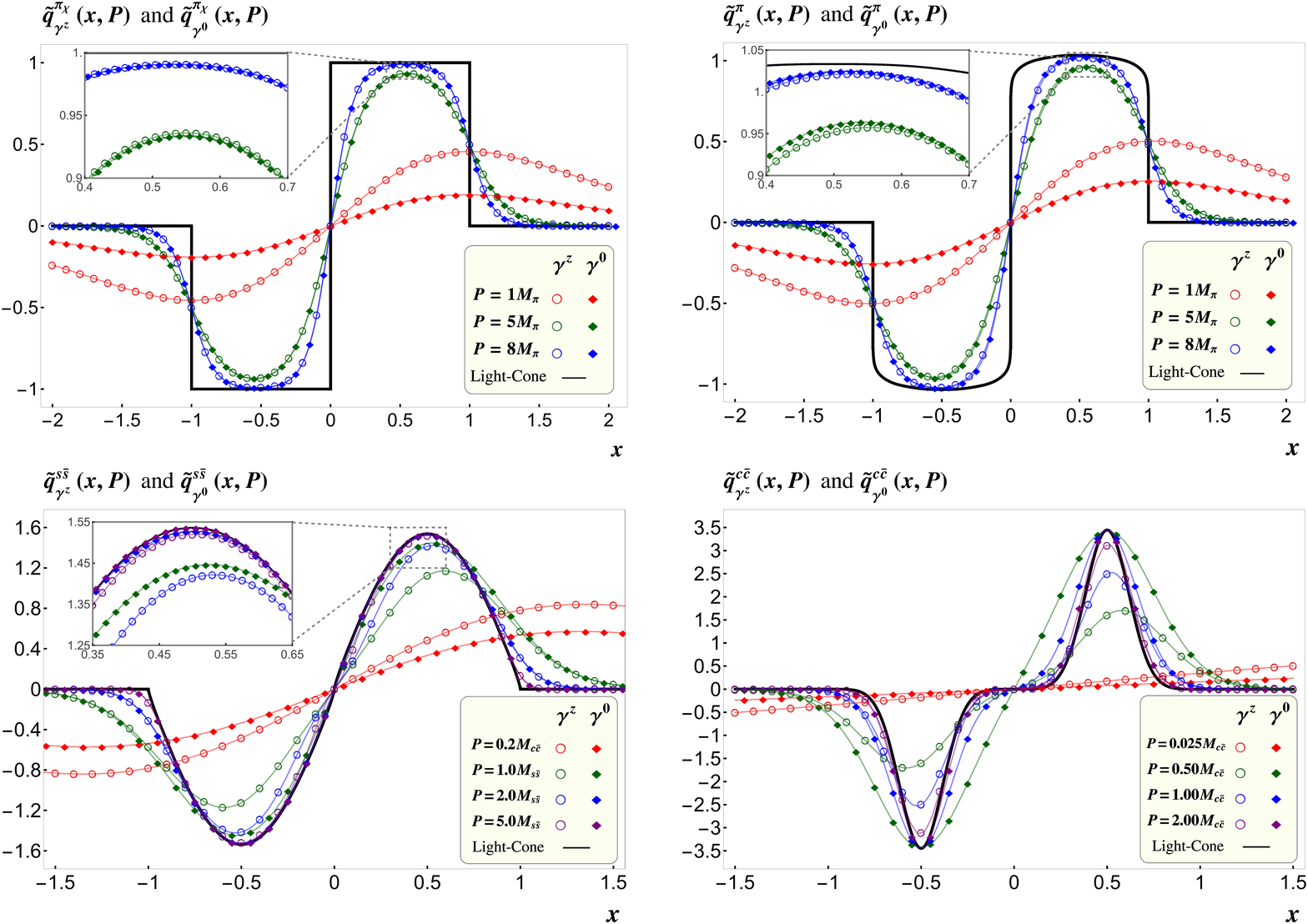}
\end{centering}
\caption{Comparison between two versions of the quark quasi-PDF defined with $\gamma^z$ (open circles) and
$\gamma^0$ (filled diamond), for four different ground-state mesons as specified
in Table~\ref{tab:masses}.}\label{fig:PDF_gz_g0}
\end{figure}

In Fig.~\ref{fig:PDF_gz_g0}, we juxtapose two versions of quasi-PDFs viewed from different reference frames,
for the four different specifies of ground-state mesons.
From the plots, we clearly see the tendency that both versions of quasi-PDFs would converge to the correspondng
light-cone PDF in IMF.
However, they evolve quite differently under Lorentz boost.
When the meson momentum is small,
$\tilde{q}_{\gamma^0}$ appears to converge with a considerably slower pace than $\tilde{q}_{\gamma^z}$;
nevertheless, when the meson momentum get large, $\tilde{q}_{\gamma^0}$ appears to converge faster than $\tilde{q}_{\gamma^z}$.
If this pattern persists in ${\rm QCD}_4$,
one may be persuaded that $\tilde{q}_{\gamma^0}$ is perhaps a more favourable choice for lattice simulation
than $\tilde{q}_{\gamma^z}$.

From Fig.~\ref{fig:PDF_gz_g0}, one can also observe that how the evolution patterns of two different quasi-DAs depend on
the quark mass. For each meson species, one might be interested in the critical threshold point
of the momentum-to-mass ratio, $r_{\rm crt}= P^z_{\rm crt}/M$, after which the
$\tilde{q}_{\gamma^0}$ starts to have a better convergence behavior than $\tilde{q}_{\gamma^z}$.
For lighter mesons ($\pi_\chi$, $\pi$), the critical $r$ values
are quite large, $r_{\rm crt}^{\pi_\chi}$,
$r_{\rm crt}^{\pi}$ are about 5.
In contrast, for heavier mesons ($s\bar s$, $c\bar c$), the critical $r$ values are rather small,
$r_{\rm crt}^{s\bar s} \approx 0.2$, and $r_{\rm crt}^{c\bar c} \approx 0.025$.

Next we turn to the quasi-DAs.
Like the quasi-PDF case, we also intend to compare two simplest definitions for quasi-DA:
\bseq
\begin{align}
\widetilde{\Phi}_{2n,\gamma^z}(x, P)&=\frac{1}{f^{(2n)}}\int_{-\infty}^{+\infty}\frac{dz}{2\pi}\,e^{i\left(x-\tfrac{1}{2}\right)P z}\,\left\langle P_{2n}^0,P\left|\bar{\psi}\left(\tfrac{z}{2}\right) \mathcal{W}\left[\tfrac{z}{2},-\tfrac{z}{2}\right]\gamma^z\gamma_5\psi\left(-\tfrac{z}{2}\right)\right|\Omega\right\rangle,\\
\widetilde{\Phi}_{2n,\gamma^0}(x, P)&=\frac{1}{f^{(2n)}}\int_{-\infty}^{+\infty}\frac{dz}{2\pi}\,e^{i\left(x-\tfrac{1}{2}\right)P z}\,\left\langle P_{2n}^0,P\left|\bar{\psi}\left(\tfrac{z}{2}\right) \mathcal{W}\left[\tfrac{z}{2},-\tfrac{z}{2}\right]\gamma^0\gamma_5\psi\left(-\tfrac{z}{2}\right)\right|\Omega\right\rangle,
\end{align}
\eseq

\begin{figure}[h]
\begin{centering}
\includegraphics[clip,width=\textwidth]{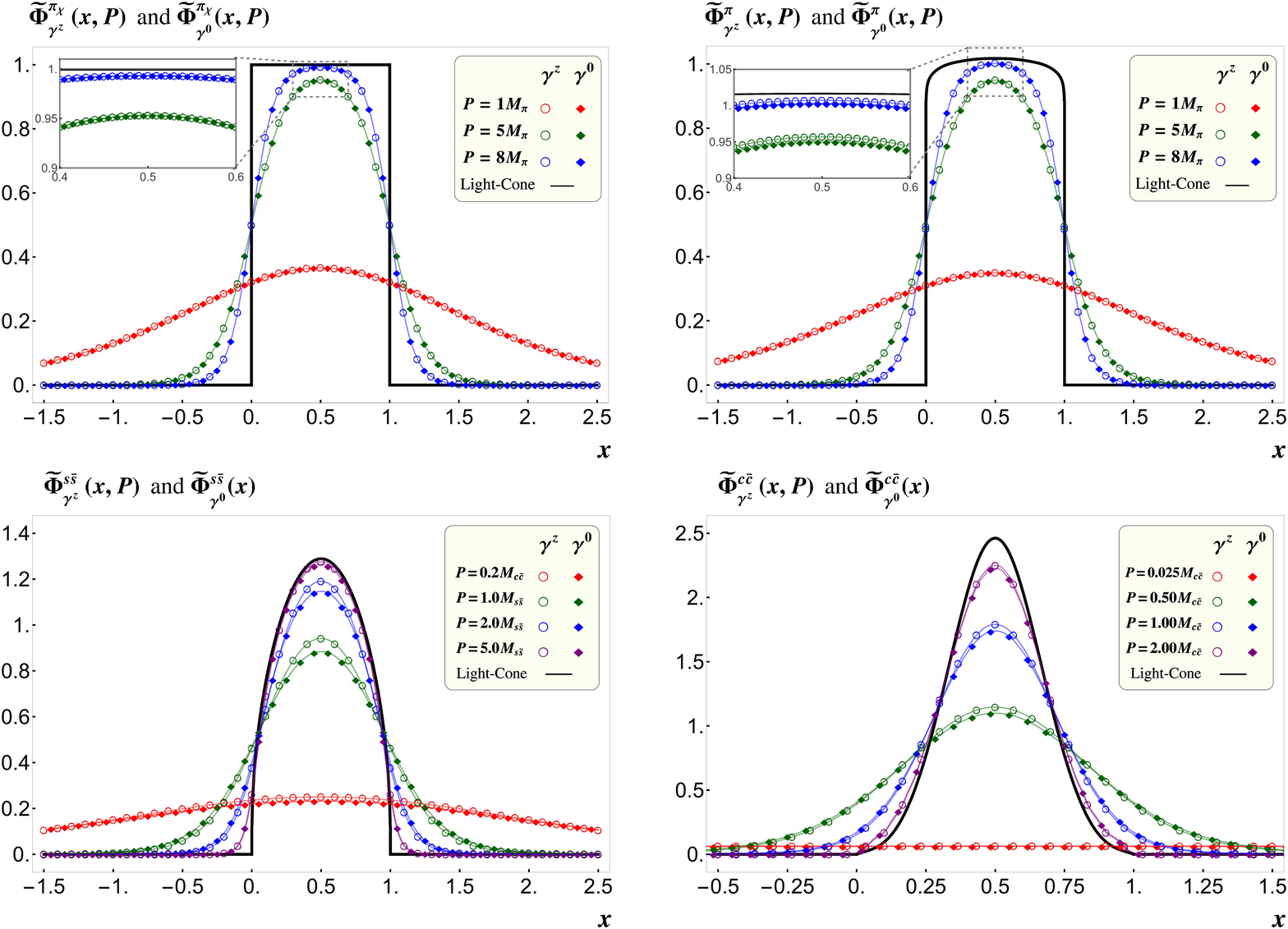}
\end{centering}
\caption{Comparison between two versions of the quasi-DA defined with $\gamma^z$ (open circles) and
$\gamma^0$ (filled diamond), for four different ground-state mesons as specified
in Table~\ref{tab:masses}.}\label{fig:DA_gz_g0}
\end{figure}

Utilizing the operator approach together with the Bogoliubov transformation, these two different versions of
quasi-DA can be expressed as
\bseq
\bqa
\widetilde{\Phi}_{2n,\gamma^z}\left(x,P\right)= &&{1\over f^{(2n)}} \sqrt{N\over \pi} \sqrt{P^0\over P} \sin {\theta(xP)+\theta(P-xP)\over 2}\left[\varphi^{2n}_+\left(xP,P\right)+\varphi^{2n}_-\left(xP,P\right)\right],
\nn\\
\label{Quasi:DA:gammaz}
\\
\widetilde{\Phi}_{2n,\gamma^0}\left(x,P\right)= &&{1\over f^{(2n)}} \sqrt{N\over \pi} \sqrt{P^0\over P} \cos \frac{\theta\left(xP\right)-\theta\left(P-xP\right)}{2}\left[\varphi^{2n}_+\left(xP,P\right)-\varphi^{2n}_-\left(xP,P\right)\right],
\nn\\
\label{Quasi:DA:gamma0}
\eqa
\eseq
where $f^{(2n)}$ is the decay constant of the $2n$'th mesonic state,
one of whose explicit expressions has been given in \eqref{eq:dcy_cnstnt}.
From Eq.~(3.10) of Ref.~\cite{Jia:2017uul}, one can find another equivalent expression of $f^{(2n)}$,
from which one immediately sees that $\widetilde{\Phi}_{2n,\gamma^0}$ in \eqref{Quasi:DA:gamma0}
also obeys the normalization condition \eqref{Normalization:Quasi:DA}.

For the canonical quasi-DA, we have actually duplicated
\eqref{quasi:DA:in:term:of:Bars:Green} for \eqref{Quasi:DA:gammaz}.
The new quasi-DA assumes the form of \eqref{Quasi:DA:gamma0}.
One readily sees that both types of quasi-DAs approach the LCDA in the IMF, therefore they belong to
the same universality class.
Nevertheless, it is not straightforward to see
which trigonometric function, $\sin[(\theta(xP)+\theta(P-xP))/2]$ or $\cos[(\theta(xP)-\theta(P-xP))/2]$,
approaches unity in a faster pace as $P\to \infty$. Therefore, just by inspection of the
analytical form, it is difficult to judge which definition of quasi-DAs bears better convergence behavior.

From Fig.~\ref{fig:DA_gz_g0}, we see that two versions of quasi-DAs for chiral pion are identical.
This can be readily proved, since the analytical expressions for
Bars-Green wave functions for $\pi_\gamma$ are exactly known~\cite{Kalashnikova:2001df,Jia:2017uul}.
For other massive mesons, it turns out that $\widetilde \Phi_{\gamma^z}$ has always better convergence
behavior than $\widetilde \Phi_{\gamma^0}$, irrespective of the
velocity of the boosted frame.
The difference between these two quasi-DAs are always insignificant.

\section{Distribution identities encountered in momentum cutoff IR regularization}
\label{Sec:distribution:identities}

In this Appendix, we collect some useful distribution identities that enable us to rewrite the LCDA and quasi-DA
in terms of the ``4-plus'' distributions in \eqref{Quasi:DA:one:loop:order}.
The validity of the following identities can be examined by picking up an arbitrary test function $f(x)$.
Rather than specialize to the ``$4$-plus'' distribution,
for the sake of generality, here we will introduce a ``$n$-plus'' distribution
that would appear in a loop calculation implementing the IR momentum cutoff.
For the light-cone loop integral, the ``$n$-plus'' distribution is defined as
\begin{align}
\int_0^1 dx\, \left[\frac{g\left(x\right)} {\left|\tfrac{1}{2}-x\right|^n}\right]_{n+}f(x)
 \equiv \int_0^1 dx\, \frac{g\left(x\right)} {\left|\tfrac{1}{2}-x\right|^n}\left[f(x)-\sum_{i=0}^{n}\frac{1}{i!} f^{(i)}\left(\frac{1}{2}\right)\left(x-\frac{1}{2}\right)^{i}\right].
\label{n:plus:LC}
\end{align}
For our purpose, we assume $g(x)$ is symmetric under the exchange $x\leftrightarrow 1-x$.

In Section~\ref{Sec:Matching}, we have adopted a soft momentum fraction $\eta$ as the IR regulator
in the loop integration using light-cone coordinates (see \eqref{LC:pert:theory:cutoff}).
We often encounter the following type of integral:
\bqa
&& \lim_{\eta\rightarrow 0}\int_0^1 dx\, \Theta\left(\left|x-\tfrac{1}{2}\right|-\eta\right) \frac{g\left(x\right)f(x) } {\left|\tfrac{1}{2}-x\right|^n}
\\
&&  = \int_0^1 dx \left\{\vphantom{\sum_{i}^{\left[\tfrac{n}{2}\right]}}  \left[\frac{g\left(x\right)}
 {\left|\tfrac{1}{2}-x\right|^n} \right]_{n+} f(x)
+f(x)\sum_{i=0}^{\left[\tfrac{n}{2}\right]}\frac{\delta^{(2i)}\left(x-\frac{1}{2}\right)}{\left(2i\right)!}\lim_{\eta\rightarrow 0}\int_0^1dy \, \Theta\left(\left|y-\tfrac{1}{2}\right|-\eta\right)\frac{g(y)\left|y-\tfrac{1}{2}\right|^{2i}}{\left|y-\tfrac{1}{2}\right|^n} \right\}.
\nn
\eqa
In the second line, this integral has been rewritten in terms of the ``$n$-plus'' distribution together with
a series of product of $\delta$-function and some integrals in the $\eta\to 0$ limit.
Since we only consider the DA of the flavor-neutral mesons,
we have dropped the odd number of derivatives of $\delta$ function, because those terms do not contribute
when $g(x)$ is symmetric under $x\leftrightarrow1-x$, since the odd number of derivatives of $g(x)$ vanishes at $x=1/2$.

The ``$n$-plus'' distribution in \eqref{n:plus:LC} can be readapted to a convolution integral
with unrestricted domain, $-\infty<x<\infty$, which is relevant to the loop calculation for the quasi-DAs:
\beq
\int_{-\infty}^{+\infty} dx\, \left[\frac{g\left(x\right)} {\left|\tfrac{1}{2}-x\right|^n}\right]_{n+}f(x)
 =\int_{-\infty}^{+\infty} dx\, \frac{g\left(x\right)} {\left|\tfrac{1}{2}-x\right|^n}\left[f(x)-\sum_{i=0}^{n}\frac{1}{i!} f^{(i)}\left(\frac{1}{2}\right)\left(x-\frac{1}{2}\right)^{i}\right]
\eeq

In Section~\ref{Sec:Matching}, we have also adopted a soft momentum fraction $\eta$ as the IR regulator
in the loop integration using ordinary coordinates (see \eqref{equal:time:pert:theory:cutoff}).
We often confront the following type of integrals:
\bqa
 &&\lim_{\eta\rightarrow 0}\int_{-\infty}^{+\infty}  dx\, \Theta\left(\left|x\!-\!\tfrac{1}{2}\right|-\eta\right) \frac{g\left(x\right)f(x) } {\left|\tfrac{1}{2}-x\right|^4}
\\
 && = \int_{-\infty}^{+\infty} \!\!\!\! dx \left\{\!\!\vphantom{\sum_{i}^{\left[\tfrac{n}{2}\right]}}  \left[\frac{g\left(x\right)} {\left|\tfrac{1}{2}-x\right|^n} \right]_{n+} \!\!\!f(x)
\!+\!f(x)\sum_{i=0}^{\left[\tfrac{n}{2}\right]}\frac{\delta^{(2i)}\!\!\left(\!x-\frac{1}{2}\!\right)}{\left(2i\right)!}\lim_{\eta\rightarrow 0}\int_{-\infty}^{+\infty}\!\!\!  dy \, \Theta\left(\left|y-\tfrac{1}{2}\right|-\eta\right)\frac{g(y)\left|y-\tfrac{1}{2}\right|^{2i}}{\left|y-\tfrac{1}{2}\right|^n} \right\}
\nn
\label{n:plus:quasi}
\eqa
In the second line, we again have dropped the odd number of derivatives of $\delta$ function.
In this identity, it is necessary to assume that the test function $f(x)$ falls off
sufficiently fast as $|x|\rightarrow\infty$.



\begin{thebibliography}{100}

\bibitem{Jia:2017uul}
  Y.~Jia, S.~Liang, L.~Li and X.~Xiong,
  JHEP {\bf 1711}, 151 (2017)
  [arXiv:1708.09379 [hep-ph]].


\bibitem{Detmold:2001dv}
  W.~Detmold, W.~Melnitchouk and A.~W.~Thomas,
  Eur.\ Phys.\ J.\ direct {\bf 3}, no. 1, 13 (2001)
  [hep-lat/0108002].


\bibitem{Hagler:2007xi}
  Ph.~H{\"a}gler {\it et al.} [LHPC Collaboration],
  Phys.\ Rev.\ D {\bf 77}, 094502 (2008)
  [arXiv:0705.4295 [hep-lat]].


\bibitem{Musch:2011er}
  B.~U.~Musch, P.~Hagler, M.~Engelhardt, J.~W.~Negele and A.~Schafer,
  Phys.\ Rev.\ D {\bf 85}, 094510 (2012)
  [arXiv:1111.4249 [hep-lat]].


\bibitem{Alexandrou:2015qia}
  C.~Alexandrou, M.~Constantinou, S.~Dinter, V.~Drach, K.~Hadjiyiannakou, K.~Jansen, G.~Koutsou and A.~Vaquero,
  JHEP {\bf 1506}, 068 (2015)
  [arXiv:1501.03734 [hep-lat]].


\bibitem{Braun:2015lfa}
  V.~M.~Braun, S.~Collins, M.~G\"ockeler, P.~P\'erez-Rubio, A.~Sch\"afer, R.~W.~Schiel and A.~Sternbeck,
  PoS QCDEV {\bf 2015}, 009 (2015)
  [arXiv:1510.07429 [hep-lat]].


\bibitem{Ji:2013dva}
  X.~Ji,
  Phys.\ Rev.\ Lett.\  {\bf 110}, 262002 (2013)
  [arXiv:1305.1539 [hep-ph]].


\bibitem{Ji:2014gla}
  X.~Ji,
  Sci.\ China Phys.\ Mech.\ Astron.\  {\bf 57}, 1407 (2014)
  [arXiv:1404.6680 [hep-ph]].


\bibitem{Xiong:2013bka}
  X.~Xiong, X.~Ji, J.~H.~Zhang and Y.~Zhao,
  Phys.\ Rev.\ D {\bf 90}, no. 1, 014051 (2014)
  [arXiv:1310.7471 [hep-ph]].


\bibitem{Ma:2014jla}
  Y.~Q.~Ma and J.~W.~Qiu,
  arXiv:1404.6860 [hep-ph].


\bibitem{Ishikawa:2017faj}
  T.~Ishikawa, Y.~Q.~Ma, J.~W.~Qiu and S.~Yoshida,
  Phys.\ Rev.\ D {\bf 96}, no. 9, 094019 (2017)
  [arXiv:1707.03107 [hep-ph]].


\bibitem{Ji:2017oey}
  X.~Ji, J.~H.~Zhang and Y.~Zhao,
  Phys.\ Rev.\ Lett.\  {\bf 120}, no. 11, 112001 (2018)
  [arXiv:1706.08962 [hep-ph]].


\bibitem{Stewart:2017tvs}
  I.~W.~Stewart and Y.~Zhao,
  Phys.\ Rev.\ D {\bf 97}, no. 5, 054512 (2018)
  [arXiv:1709.04933 [hep-ph]].


\bibitem{Alexandrou:2017huk}
  C.~Alexandrou, K.~Cichy, M.~Constantinou, K.~Hadjiyiannakou, K.~Jansen, H.~Panagopoulos and F.~Steffens,
  Nucl.\ Phys.\ B {\bf 923}, 394 (2017)
  [arXiv:1706.00265 [hep-lat]].


\bibitem{Lin:2014zya}
  H.~W.~Lin, J.~W.~Chen, S.~D.~Cohen and X.~Ji,
  Phys.\ Rev.\ D {\bf 91}, 054510 (2015)
  [arXiv:1402.1462 [hep-ph]].


\bibitem{Chen:2016utp}
  J.~W.~Chen, S.~D.~Cohen, X.~Ji, H.~W.~Lin and J.~H.~Zhang,
  Nucl.\ Phys.\ B {\bf 911}, 246 (2016)
  [arXiv:1603.06664 [hep-ph]].


\bibitem{Alexandrou:2016jqi}
  C.~Alexandrou, K.~Cichy, M.~Constantinou, K.~Hadjiyiannakou, K.~Jansen, F.~Steffens and C.~Wiese,
  Phys.\ Rev.\ D {\bf 96}, no. 1, 014513 (2017)
  [arXiv:1610.03689 [hep-lat]].


\bibitem{Alexandrou:2015rja}
  C.~Alexandrou, K.~Cichy, V.~Drach, E.~Garcia-Ramos, K.~Hadjiyiannakou, K.~Jansen, F.~Steffens and C.~Wiese,
  Phys.\ Rev.\ D {\bf 92}, 014502 (2015)
  [arXiv:1504.07455 [hep-lat]].


\bibitem{Zhang:2017bzy}
  J.~H.~Zhang, J.~W.~Chen, X.~Ji, L.~Jin and H.~W.~Lin,
  Phys.\ Rev.\ D {\bf 95}, no. 9, 094514 (2017)
  [arXiv:1702.00008 [hep-lat]].


\bibitem{Chen:2017mzz}
  J.~W.~Chen, T.~Ishikawa, L.~Jin, H.~W.~Lin, Y.~B.~Yang, J.~H.~Zhang and Y.~Zhao,
  Phys.\ Rev.\ D {\bf 97}, no. 1, 014505 (2018)
  [arXiv:1706.01295 [hep-lat]].


\bibitem{Lin:2017ani}
  H.~W.~Lin, J.~W.~Chen, T.~Ishikawa and J.~H.~Zhang,
  arXiv:1708.05301 [hep-lat].


\bibitem{Chen:2017lnm}
  J.~W.~Chen, T.~Ishikawa, L.~Jin, H.~W.~Lin, A.~Sch\"afer, Y.~B.~Yang, J.~H.~Zhang and Y.~Zhao,
  arXiv:1711.07858 [hep-ph].


\bibitem{Chen:2017gck}
  J.~W.~Chen {\it et al.},
  arXiv:1712.10025 [hep-ph].


\bibitem{Chen:2018xof}
  J.~W.~Chen, L.~Jin, H.~W.~Lin, Y.~S.~Liu, Y.~B.~Yang, J.~H.~Zhang and Y.~Zhao,
  arXiv:1803.04393 [hep-lat].


\bibitem{Ishikawa:2016znu}
  T.~Ishikawa, Y.~Q.~Ma, J.~W.~Qiu and S.~Yoshida,
  arXiv:1609.02018 [hep-lat].


\bibitem{Carlson:2017gpk}
  C.~E.~Carlson and M.~Freid,
  Phys.\ Rev.\ D {\bf 95}, no. 9, 094504 (2017)
  [arXiv:1702.05775 [hep-ph]].


\bibitem{Xiong:2017jtn}
  X.~Xiong, T.~Luu and U.~G.~Mei\ss ner,
  arXiv:1705.00246 [hep-ph].


\bibitem{Chen:2016fxx}
  J.~W.~Chen, X.~Ji and J.~H.~Zhang,
  Nucl.\ Phys.\ B {\bf 915}, 1 (2017)
  [arXiv:1609.08102 [hep-ph]].


\bibitem{Monahan:2016bvm}
  C.~Monahan and K.~Orginos,
  JHEP {\bf 1703}, 116 (2017)
  [arXiv:1612.01584 [hep-lat]].


\bibitem{Monahan:2017hpu}
  C.~Monahan,
  Phys.\ Rev.\ D {\bf 97}, no. 5, 054507 (2018)
  [arXiv:1710.04607 [hep-lat]].


\bibitem{Wang:2017qyg}
  W.~Wang, S.~Zhao and R.~Zhu,
  Eur.\ Phys.\ J.\ C {\bf 78}, no. 2, 147 (2018)
  [arXiv:1708.02458 [hep-ph]].


\bibitem{tHooft:1973alw}
  G.~'t Hooft,
  Nucl.\ Phys.\ B {\bf 72}, 461 (1974).


\bibitem{Witten:1979kh}
  E.~Witten,
  Nucl.\ Phys.\ B {\bf 160}, 57 (1979).


\bibitem{Coleman:1985}
S.~R.~Coleman, {\it Aspects of Symmetry}, Cambridge University Press, 1985.
See Chapter 8, $1/N$.

\bibitem{tHooft:1974pnl}
  G.~'t Hooft,
  Nucl.\ Phys.\ B {\bf 75}, 461 (1974).


\bibitem{Burkardt:2000uu}
  M.~Burkardt,
  Phys.\ Rev.\ D {\bf 62}, 094003 (2000)
  [hep-ph/0005209].


\bibitem{Bars:1977ud}
  I.~Bars and M.~B.~Green,
  Phys.\ Rev.\ D {\bf 17}, 537 (1978).


\bibitem{Radyushkin:2017cyf}
  A.~V.~Radyushkin,
  Phys.\ Rev.\ D {\bf 96}, no. 3, 034025 (2017)
  [arXiv:1705.01488 [hep-ph]].


\bibitem{Orginos:2017kos}
  K.~Orginos, A.~Radyushkin, J.~Karpie and S.~Zafeiropoulos,
  Phys.\ Rev.\ D {\bf 96}, no. 9, 094503 (2017)
  [arXiv:1706.05373 [hep-ph]].


\bibitem{Ma:2017pxb}
  Y.~Q.~Ma and J.~W.~Qiu,
  Phys.\ Rev.\ Lett.\  {\bf 120}, no. 2, 022003 (2018)
  [arXiv:1709.03018 [hep-ph]].


\bibitem{Zhitnitsky:1985um}
  A.~R.~Zhitnitsky,
  Phys.\ Lett.\  {\bf 165B}, 405 (1985)
  [Sov.\ J.\ Nucl.\ Phys.\  {\bf 43}, 999 (1986)]
  [Yad.\ Fiz.\  {\bf 43}, 1553 (1986)].


\bibitem{Kikkawa:1980dc}
  K.~Kikkawa,
  Annals Phys.\  {\bf 135}, 222 (1981).


\bibitem{Nakamura:1981zi}
  A.~Nakamura and K.~Odaka,
  Phys.\ Lett.\  {\bf 105B}, 392 (1981).


\bibitem{Rajeev:1994tr}
  S.~G.~Rajeev,
  Int.\ J.\ Mod.\ Phys.\ A {\bf 9}, 5583 (1994)
  [hep-th/9401115].


\bibitem{Dhar:1994ib}
  A.~Dhar, G.~Mandal and S.~R.~Wadia,
  Phys.\ Lett.\ B {\bf 329}, 15 (1994)
  [hep-th/9403050].


\bibitem{Dhar:1994aw}
  A.~Dhar, P.~Lakdawala, G.~Mandal and S.~R.~Wadia,
  Int.\ J.\ Mod.\ Phys.\ A {\bf 10}, 2189 (1995)
  [hep-th/9407026].


\bibitem{Cavicchi:1993jh}
  M.~Cavicchi,
  Int.\ J.\ Mod.\ Phys.\ A {\bf 10}, 167 (1995)
  [hep-th/9401086].


\bibitem{Barbon:1994au}
  J.~L.~F.~Barbon and K.~Demeterfi,
  Nucl.\ Phys.\ B {\bf 434}, 109 (1995)
  [hep-th/9406046].


\bibitem{Itakura:1996bk}
  K.~Itakura,
  Phys.\ Rev.\ D {\bf 54}, 2853 (1996)
  [hep-th/9604032].


\bibitem{Kogut:1969xa}
  J.~B.~Kogut and D.~E.~Soper,
  Phys.\ Rev.\ D {\bf 1}, 2901 (1970).


\bibitem{Callan:1975ps}
  C.~G.~Callan, Jr., N.~Coote and D.~J.~Gross,
  Phys.\ Rev.\ D {\bf 13}, 1649 (1976).


\bibitem{Einhorn:1976uz}
  M.~B.~Einhorn,
  Phys.\ Rev.\ D {\bf 14}, 3451 (1976).


\bibitem{Brodsky:1997de}
  S.~J.~Brodsky, H.~C.~Pauli and S.~S.~Pinsky,
  Phys.\ Rept.\  {\bf 301}, 299 (1998)
  [hep-ph/9705477].


\bibitem{Hornbostel:1988ne}
  K.~Hornbostel,
  {\it The Application Of Light Cone Quantization To Quantum Chromodynamics In (1+1)-dimensions}, Ph.D. thesis,
Stanford Linear Accelerator Center, Stanford University Stanford,California 94309,December 1988


\bibitem{Lenz:1991sa}
  F.~Lenz, M.~Thies, K.~Yazaki and S.~Levit,
  Annals Phys.\  {\bf 208}, 1 (1991).


\bibitem{Mandelstam:1982cb}
  S.~Mandelstam,
  Nucl.\ Phys.\ B {\bf 213}, 149 (1983).


\bibitem{Leibbrandt:1987qv}
  G.~Leibbrandt,
  Rev.\ Mod.\ Phys.\  {\bf 59}, 1067 (1987).


\bibitem{Hadamard:1923}
J. Hadamard,
\textit{Lectures on Cauchy's problem in linear partial differential equations},
Dover, New York, 1923.

\bibitem{Kalashnikova:2001df}
  Y.~S.~Kalashnikova and A.~V.~Nefediev,
  Phys.\ Usp.\  {\bf 45}, 347 (2002)
  [Usp.\ Fiz.\ Nauk {\bf 172}, 378 (2002)]
  [hep-ph/0111225].


\bibitem{Peskin:1995ev}
M.~E.~Peskin and D.~V.~Schroeder,
{\it An Introduction to Quantum Field Theory}
(Addison-Wesley Publishing Company, Fifth (corrected) printing, 1997).


\bibitem{Shifman:2012zz}
M.~Shifman,
{\it Advanced Topics In Quantum Field Theory: A Lecture Course}
(Cambridge University Press, Cambridge, UK, 2012).


\bibitem{Li:1987hx}
  M.~Li, L.~Wilets and M.~C.~Birse,
  J.\ Phys.\ G {\bf 13}, 915 (1987).


\bibitem{Schwabl:book}
F.~Schwabl, {\it Advanced Quantum Mechanics}, Springer, 2005 (3rd ed.).
See Section 3.2.2, Bogoliubov Theory of the Weakly Ineracting Bose Gas.

\bibitem{Bicudo:1989sh}
  P.~J.~d.~A.~Bicudo and J.~E.~F.~T.~Ribeiro,
  Phys.\ Rev.\ D {\bf 42}, 1611 (1990).


\bibitem{Collins:1981uw}
  J.~C.~Collins and D.~E.~Soper,
  Nucl.\ Phys.\ B {\bf 194}, 445 (1982).


         \bibitem{Collins:2011zzd}
         J.~Collins,
         {\it Foundations of perturbative QCD},
         (Cambridge University Press, Cambridge, England, 2013).


\bibitem{Hatta:2013gta}
  Y.~Hatta, X.~Ji and Y.~Zhao,
  Phys.\ Rev.\ D {\bf 89}, no. 8, 085030 (2014)
  [arXiv:1310.4263 [hep-ph]].


\bibitem{Lepage:1980fj}
  G.~P.~Lepage and S.~J.~Brodsky,
  Phys.\ Rev.\ D {\bf 22}, 2157 (1980).


\bibitem{Radyushkin:1977gp}
  A.~V.~Radyushkin,
  [hep-ph/0410276].


\bibitem{Burkardt:1995eb}
  M.~Burkardt,
  Phys.\ Rev.\ D {\bf 53}, 933 (1996)
  [hep-ph/9509226].

\bibitem{Ji:2015jwa}
  X.~Ji and J.~H.~Zhang,
  Phys.\ Rev.\ D {\bf 92}, 034006 (2015)
  doi:10.1103/PhysRevD.92.034006
  [arXiv:1505.07699 [hep-ph]].


\bibitem{Ji:2015qla}
  X.~Ji, A.~Sch\"afer, X.~Xiong and J.~H.~Zhang,
  Phys.\ Rev.\ D {\bf 92}, 014039 (2015)
  [arXiv:1506.00248 [hep-ph]].


\bibitem{Xiong:2015nua}
  X.~Xiong and J.~H.~Zhang,
  Phys.\ Rev.\ D {\bf 92}, no. 5, 054037 (2015)
  [arXiv:1509.08016 [hep-ph]].
  

\bibitem{Weinberg:1966jm} 
  S.~Weinberg,
  Phys.\ Rev.\  {\bf 150}, 1313 (1966).
  doi:10.1103/PhysRev.150.1313


\bibitem{Chen:2018fwa}
  J.~W.~Chen, L.~Jin, H.~W.~Lin, Y.~S.~Liu, A.~Sch\"afer, Y.~B.~Yang, J.~H.~Zhang and Y.~Zhao,
  arXiv:1804.01483 [hep-lat].


\end{thebibliography}
\end{document}